%Paper: hep-th/9211038
%From: Emmanuel Guitter <guitter@amoco.saclay.cea.fr>
%Date: 09 Nov 92 16:56:23+0100

%%%%%%%%%%%%%%%%%%%%%%%%%%%%%%%%%%%%%%%%%%%%%%%%%%%%%%%%%%%%%%%%%%%%%%%%%%%%%
%
%  WARNINGS!!!!!!
%
% - This plain tex file needs harvmac.tex
%
% - Choose the option `big'. The option `little' + landscape mode
%   are not recommended.
%
% - You also need the font files amssym.def and amssym.tex.
%   If you don't have them, suppress the two corresponding lines
%      \input amssym.def
%      \input amssym.tex
%   10 lines below, and replace them by the line
%      \def\Bbb{\bf } \def\varsubsetneq{\ne }
%
%
%%%%%%%%%%%%%%%%%%%%%%%%%%%%%%%%%%%%%%%%%%%%%%%%%%%%%%%%%%%%%%%%%%%%%%%%%%%%
%\magnification=\magstep0
\input harvmac
%%%%%%%%%%%%%%%%%%%%%%%%%%%%%%%%%%%%%%%%%
% some additional math fonts            %
%%%%%%%%%%%%%%%%%%%%%%%%%%%%%%%%%%%%%%%%%
\input amssym.def
\input amssym.tex
%\def\Bbb{\bf } \def\varsubsetneq{\ne }
%%%%%%%%%%%%%%%%%%%%%%%%%%%%%%%%%%%%%%%%%
% Slight modifications of plain harvmac %
%%%%%%%%%%%%%%%%%%%%%%%%%%%%%%%%%%%%%%%%%
\def\writetoc{\immediate\openout\tfile=toc.tmp
   \def\writetoca##1{{\edef\next{\write\tfile{\noindent ##1
   \string\leaderfill {\noexpand\number\pageno} \par}}\next}}}
\def\appendix#1#2{\global\meqno=1\global\subsecno=0\xdef\secsym{\hbox{#1.}}
\bigbreak\bigskip\noindent{{\titlermsb Appendix #1.\ }{\bf #2}}
\message{(#1. #2)}
\writetoca{{\titlermsb Appendix } {\titlermsb #1.\ } {\bf #2}}\par
\nobreak\medskip\nobreak}
\def\newsec#1{\global\advance\secno by1\message{(\the\secno. #1)}
%\ifx\answ\bigans \vfill\eject \else \bigbreak\bigskip \fi  %if desired
\global\subsecno=0\eqnres@t\noindent{\titlermsb\the\secno. #1}
\writetoca{{\bf \secsym} {\titlermsb #1}}\par\nobreak\medskip\nobreak}
\def\eqnres@t{\xdef\secsym{\the\secno.}\global\meqno=1\bigbreak\bigskip}
\def\sequentialequations{\def\eqnres@t{\bigbreak}}\xdef\secsym{}
\global\newcount\subsecno \global\subsecno=0
\def\subsec#1{\global\advance\subsecno by1\message{(\secsym\the\subsecno. #1)}
\ifnum\lastpenalty>9000\else\bigbreak\fi
\noindent{\bf\secsym\the\subsecno. #1}\writetoca{\string\quad
{\bf \secsym\the\subsecno.} {\bf #1}}\par\nobreak\medskip\nobreak}
\font\titlermsb=cmbx7 \tfontsize \font\tay=eusb10
\font\tentit=cmmib10
\font\ninetit=cmmib9
\font\seventit=cmmib7

\font\fivetit=cmmib5
\newfam\titfam
\textfont\titfam=\tentit
\scriptfont\titfam=\seventit
\scriptscriptfont\titfam=\fivetit
\def\tit{\fam\titfam\tentit}
\def\CA{{\cal A}}

\def\CC{{\cal C}}
\def\CD{{\cal D}}
\def\CE{{\cal E}}
\def\CF{{\cal F}}
\def\CG{{\cal G}}
\def\CH{{\cal H}}

\def\CJ{{\cal J}}

\def\CM{{\cal M}}
\def\CN{{\cal N}}
\def\CO{{\cal O}}
\def\CP{{\cal P}}
\def\CQ{{\cal Q}}
\def\CR{{\cal R}}
\def\CS{{\cal S}}
\def\CT{{\cal T}}

\def\CV{{\cal V}}

\def\CZ{{\cal Z}}
%%%%%%%%%%%%%%%%%%%%
% personal macros %
%%%%%%%%%%%%%%%%%%%%
\def\aa{{\Bbb A}}
\def\cc{{\Bbb C}}
\def\pp{{\Bbb P}}
\def\nn{{\Bbb N}}
\def\rr{{\Bbb R}}
\def\rvec{{\bf \vec r}}
\def\rvect{{\bf \vec {\tilde r}}}
\def\kvec{{\bf \vec k}}
\def\qvec{{\bf \vec q}}
\def\Avril{Y}
\def\sT{{\scriptscriptstyle T}}
\def\sA{{\scriptscriptstyle A}}
\def\sB{{\scriptscriptstyle B}}
\def\sC{{\scriptscriptstyle C}}
\def\sD{{\scriptscriptstyle D}}
\def\sP{{\scriptscriptstyle P}}
\def\sI{{\scriptscriptstyle I}}
\def\sJ{{\scriptscriptstyle J}}
\def\sL{{\scriptscriptstyle L}}
\def\sK{{\scriptscriptstyle K}}
\def\sM{{\scriptscriptstyle M}}
\def\SJ{{\scriptstyle J}}
\def\SI{{\scriptstyle I}}
\def\SL{{\scriptstyle L}}
\def\SK{{\scriptstyle K}}
\def\SM{{\scriptstyle M}}
\def\ST{{\scriptstyle T}}
\def\SA{{\scriptstyle A}}
\def\SB{{\scriptstyle B}}
\def\SC{{\scriptstyle C}}
\def\SD{{\scriptstyle D}}
\def\SP{{\scriptstyle P}}
\def\cn{{}^c\kern -2.pt {\cal N}}
\def\Tay{{\hbox{\tay T}}}
\def\RR{\relax{\rm I\kern-.18em R}}
\def\II{\relax{\rm 1\kern-.24em I}}
\def\setminuso{\hbox{$/ \kern -3pt {}_o$}}
\def\setminusp{\hbox{$/ \kern -3pt {}_p$}}
\def\setminuspk{\hbox{$/ \kern -3pt {}_{\{p_k\}}$}}
\def\setminusw{\hbox{$/ \kern -3pt {}_w$}}
\def\setminuswjmin{\hbox{$/ \kern -3pt {}_{\omega^0_{J-1}}$}}
\def\setminusbullet{\hbox{$/ \kern -3pt {}_\bullet$}}
\def\veew{\hbox{$\kern 2.pt \vee \kern -2pt {}_\omega$}}
\def\veewj{\hbox{$\kern 2.pt \vee \kern -2pt {}_{\omega_\sJ}$}}
\def\veewjmin{\hbox{$\kern 2.pt \vee \kern -2pt {}_{\omega_{\sJ-1}}$}}
\def\veewprim{\hbox{$\kern 2.pt \vee \kern -2pt {}_{ \omega'}$}}
\def\veewzeroj{\hbox{$\kern 2.pt \vee \kern -2pt {}_{ \omega^0_\sJ}$}}
\def\veevv{\hbox{$\kern 2.pt \vee \kern -2pt {}_w$}}
\def\veevvj{\hbox{$\kern 2.pt \vee \kern -2pt {}_{w_j}$}}
\def\veegodot{\hbox{$\kern 2.pt \vee \kern -2pt {}_{G_\odot}$}}
\def\veeplus{\hbox{$\kern 2.pt \vee\kern -6.1pt \raise 6.5pt
\hbox{${\scriptscriptstyle \oplus}$}\kern 4.pt$}}
\def\encadre#1{\vbox{\hrule\hbox{\vrule\kern8pt\vbox{\kern8pt#1\kern8pt}
\kern8pt\vrule}\hrule}}
\def\encadrefantome#1{\vbox{\hbox{\kern1pt\vbox{\kern8pt#1\kern8pt}
\kern1pt}}}
\def\Encadre#1{\vcenter{\hrule\hbox{\vrule\kern8pt\vbox{\kern8pt#1\kern8pt}
\kern8pt\vrule}\hrule}}
\def\boxit#1{\vbox{\hrule\hbox{\vrule\kern1pt\vbox{\kern10pt\vbox{\hsize45pt
\noindent #1}\kern10pt}\kern1pt\vrule}\hrule}}
\def\noboxit#1{\vbox{\hbox{\kern1pt\vbox{\kern10pt\vbox{\hsize45pt
\noindent #1}\kern10pt}\kern1pt}}}
\def\Boxit#1{\vbox{\hrule\hbox{\vrule\kern1pt\vbox{\kern16pt\vbox{\hsize59pt
\noindent #1}\kern16pt}\kern1pt\vrule}\hrule}}
\def\matt{\left(\matrix{\Encadre{\hbox{$\displaystyle \matrix{
\boxit{$\ \ \rho^{2\nu}\Pi^{{\bf T}_1}$}&&&\noboxit{$\rho^{2\nu} {\cal O}
(\rho^{2\delta})$}
\cr
&\boxit{$\ \ \rho^{2\nu}\Pi^{{\bf T}_2}$}&&\cr
&&\ddots&\cr
\noboxit{$\rho^{2\nu} {\cal O}(\rho^{2\delta})$}&&&\boxit{$\,\rho^{2\nu}
\Pi^{{\bf T}_{m-1}}$}
\cr}$}}&\noboxit{$\rho^\nu {\cal O}(\rho^\delta)$}\cr
\noboxit{$\rho^\nu {\cal O}(\rho^\delta)$}
&\Boxit{$\Pi^{{\bf T}_m}+{\cal O}(\rho^{2\delta})$}\cr
}\right)}
\def\mattt{\left(\matrix{\Encadre{\hbox{$\displaystyle \matrix{
\boxit{$\ \ \ \Avril^{{\bf T}_1}$}&&&\noboxit{\ \ $ {\cal O}
(\rho^{2\delta})$}
\cr
&\boxit{$\ \ \ \Avril^{{\bf T}_2}$}&&\cr
&&\ddots&\cr
\noboxit{$\ \  {\cal O}(\rho^{2\delta})$}&&&\boxit{$\ \
\Avril^{{\bf T}_{m-1}}$}
\cr}$}}&\noboxit{$\ \ {\cal O}(\rho^\delta)$}\cr
\noboxit{$\ \  {\cal O}(\rho^\delta)$}
&\Boxit{$\Avril^{{\bf T}_m}+{\cal O}(\rho^{2\delta})$}\cr
}\right)}
\def\mattY{\left[\matrix{
\boxit{$\ \ \ \Avril^{{\bf T}_{\sJ,1}}$}&\noboxit{$\ \ 0$}
&\noboxit{$\ \ \ldots$}&\noboxit{$\ \ 0$} \cr
\noboxit{$\ \ 0$} &\boxit{$\ \ \ \Avril^{{\bf T}_{\sJ,2}}$}&
\noboxit{$\ \ \ldots$}&
\noboxit{$\ \ 0$}\cr\noboxit{$\ \ \vdots$}
&\noboxit{$\ \ \vdots$}&\noboxit{$\ \ \ddots$}&\noboxit{$\ \ \vdots$}\cr
\noboxit{$\ \ 0$}&\noboxit{$\ \ 0$}&\noboxit{$\ \ \ldots$}&\boxit{$\ \
\Avril^{{\bf T}_{\sJ,{j^{\rm max}}}}$}
\cr}\right]
}
\def\figcap#1#2{\nfig#1{#2} \noindent {Fig.\ }\xfig#1{:\ }{\sl #2}}
\def\figinsert#1#2#3{\vbox to #1{\vfill \vbox {\hsize=#2
\baselineskip=10pt {#3} \par}}}
%%%%%%%%%%%%%%%%%%%%%%%%%%%%%%%%%%%%%%%%%%%%%%%%%%%%%%%%%%%%%%%%%%%%%%%%%%%%
%\draftmode
\Title{SPhT/92-124\ \ hep-th/9211038}{
\vbox{
\vskip -2truecm
\centerline{Renormalization Theory for}\vskip2pt
\centerline{ Interacting Crumpled Manifolds}}
}

%For more complicated situations, substitute for {\it either\/} argument:
%\Title{\vbox{\baselineskip12pt\hbox{SPhT/92-124}\hbox{SLAC-PUB 88-001}
%		\hbox{photocopy at own risk}}}
%{\vbox{\centerline{This title is too long to fit}
%	\vskip2pt\centerline{comfortably on one line*}}}
%   \footnote{}{*optional footnote on title}

\centerline{Fran\c cois David\footnote{$^\dagger$}
{Member of C.N.R.S.},
Bertrand Duplantier{$^\dagger$}{} and Emmanuel Guitter}
\bigskip\centerline{Service de Physique Th\'eorique\footnote{$^\star$}{
Laboratoire de la Direction des Sciences de la Mati\`ere du Commissariat
\`a l'Energie Atomique}}
\centerline{C.E. Saclay}
\centerline{F-91191 Gif-sur-Yvette Cedex}
\vskip 1.truecm
\centerline{\bf Abstract}{
\ninerm
\textfont0=\ninerm
\font\ninemit=cmmi9
\font\sevenmit=cmmi7
\textfont1=\ninemit
\scriptfont0=\sevenrm
\scriptfont1=\sevenmit
\baselineskip=11pt
\bigskip
We consider a continuous
model of $D$-dimensional elastic (polymerized) manifold fluctuating in
$d$-dimensional Euclidean space, interacting with a single impurity via
an attractive or repulsive {\ninetit\char'016}-potential (but without
self-avoidance interactions).
Except for $D=1$ (the polymer case), this model cannot
be mapped onto a local field theory.
We show that the use of intrinsic distance geometry allows for a rigorous
construction of the high-temperature perturbative expansion and for analytic
continuation in the manifold dimension $D$.
We study the renormalization properties of the model for $0<D<2$,
and show that
for bulk space dimension $d$ smaller that the upper critical dimension
$d^{\star}=2D/(2-D)$, the perturbative
expansion is ultra-violet finite, while
ultraviolet divergences occur as poles at
$d=d^{\star}$.
The standard proof of perturbative renormalizability
for local field theories (the Bogoliubov Parasiuk Hepp theorem)
does not apply to this model.
We prove perturbative renormalizability to all orders by constructing a
subtraction
operator {\ninebf R} based on a generalization of the Zimmermann forests
formalism,
and which makes the theory finite at $d=d^{\star}$.
This subtraction operation corresponds to a renormalization of the coupling
constant of the model (strength of the interaction with the impurity).
The existence of a Wilson function, of an $\epsilon$-expansion
{\nineit \`a la} Wilson Fisher around
the critical dimension, of scaling laws for
$d<d^{\star}$ in
the repulsive case, and of non-trivial critical exponents of the
delocalization transition for $d>d^{\star}$ in
the attractive case is thus
established.
To our knowledge, this study provides the first proof of renormalizability
for a model of extended objects, and should be applicable to the study of
self-avoidance interactions for random manifolds.
\par
}
%if too many authors for abstract on same page, say   \vfill\eject\pageno0
\vskip .3in
\Date{11/92} %replace this line by \draft  for preliminary versions
	     %or specify \draftmode at some point

%if you want double-space, use e.g. \baselineskip=20pt plus 2pt minus 2pt
\hfuzz 1.pt
%%%%%%%%%%%%%%%%%%%%%%%%
%  TABLE OF CONTENT
%%%%%%%%%%%%%%%%%%%%%%%%
\vfill \eject
\listtoc
\writetoc
\vfill \eject
%%%%%%%%%%%%%%%%%%%%%%%%

\newsec{Introduction}

One general problem arising in statistical physics is the
understanding of the effect of interactions on the thermodynamical
properties of extended fluctuating geometrical objects. These objects may
be (1-dimensional) lines, like long linear macromolecules or polymers,
(2-dimensional) surfaces, like membranes or interfaces, or even
(3-dimensional) volumes, like gels.
The interactions involve in general two-body attractive or repulsive
forces, and one may in general reduce such problems into two different classes:
(i) either one deals with self-interactions between distinct points of the
same fluctuating object, or mutual interactions between several fluctuating
objects;
(ii) or one deals with the interaction of a single freely fluctuating
object with another non-fluctuating fixed object.
Case (i) includes
for instance self-avoiding polymers or membranes, polyelectrolytes and
charged gels, as well as the description of intersections of random walks.
Case (ii) includes the problems
of binding/unbinding of a long molecule or a membrane on a wall,
the wetting of an interface. One can also reduce to this class the problems
of unbinding of two membranes or interfaces, and that
of the steric repulsions between membranes in a lamellar phase.
\medskip
Among the many different generic situations one can think of, one case
is now well understood, namely that where the fluctuating objects are
only {\it one dimensional objects}.
Indeed, many problems in case (ii) can then be solved by simple analogy with
quantum mechanics, {\it i.e.} by use of a diffusion equation. The situation
is more complicated in case (i), a paradigm of which is the celebrated
problem of self-avoiding polymers. Still in this case, the use of perturbative
expansions and Renormalization Group techniques allows for explicit results
on the thermodynamics of these objects.
For instance, a self-avoiding polymer embedded in a $d$-dimensional
external space can be described by the continuous Edwards Hamiltonian
\ref\SirSam{S. F. Edwards, Proc. Phys. Soc. Lond. {\bf 85} (1965) 613.}
\ref\desClJan{J. des Cloizeaux and G. Jannink,
{\sl Polymers in Solution, their Modelling and Structure}, Clarendon
Press, Oxford (1990).}:
\eqn\Edwards{
{\cal H}={1\over 2} \int_0^S ds\, \Big({d\vec r\over ds}
\cdot{d\vec r\over ds}\Big) +\
{b\over 2}\int_0^S ds\int_0^S ds'\, \delta^{d}\big(\vec r(s)-\vec r(s')\big)
\ .}
This model can then be viewed as a 1-dimensional field theory,
with position field $\vec r (s)$ at abscissa $s$ along the chain of size $S$,
and with a {\it non-local} interaction term.
This field theory then has a formal perturbative
expansion in $b$: this point of view dates back the work of Fixman
\ref\Fixman{M. Fixman, J. Chem. Phys. {\bf 23} (1955) 1656.} and has
been developed by des Cloizeaux
\desClJan\
\ref\desCloiz{J. des Cloizeaux, J. Phys. France {\bf 42} (1981) 635.}.
The terms of this expansion are in general integrals over the
{\it internal} coordinates $s$ of the
interaction points and may diverge when these interaction points
come close to each other ($|s-s'|\to 0$).
The theory can then be regularized by analytic continuation in $d\ge 2$, and
the natural expansion parameter is then $b S^{2-d/2}$, hence large in
the thermodynamic limit $S\to\infty$ for $d<4$.
For dimensional reasons, the corresponding large distance divergences are
twinned with the short-distance divergences, and appear as
poles in $d$ at $d=4$. Within a double expansion
in $b$ and $\epsilon=4-d$, the structure of these poles is such that
the theory is {\it renormalizable} for $\epsilon \ge 0$. This
means that the poles at $\epsilon =0$ can actually be absorbed
into redefinitions
of the parameters of the model, and that a scaling limit is obtained
for the thermodynamical properties of the polymer when $\epsilon \ge 0$.
Still, a rigorous proof of
renormalizability requires the use of the famous equivalence of
the Edwards model with the ${\rm O}(n)$ model for $n=0$, that is a model
with a $n$-component field $\vec \Phi (\vec r)$ in the
$d$-dimensional {\it external} space, as shown by de~Gennes
\ref\deGennes{P.-G. de Gennes, Phys. Lett. {\bf 38A} (1972) 339.}.
 From this different point of view, which was the first to be developed in the
70's, the self-avoiding polymer problem is seen as a
%This leads to a different point of view, where
$d$-dimensional {\it local} field theory, that is a theory
with local interactions, and amenable to the standard renormalization group
treatments for critical phenomena
\ref\DombGreen{{\sl Phase Transitions and Critical Phenomena}, Vol. {\bf 6},
eds. C. Domb and M. S. Green, Academic Press, New York (1976).}
\ref\Zinn{J. Zinn-Justin, {\sl Quantum Field Theory and Critical Phenomena},
Clarendon Press, Oxford (1989).}.
Again, this field theory can be studied via a perturbative expansion,
the terms of which may diverge when two {\it external} interaction points
$\vec r$ and ${\vec r\,}'$ come close to each other
($|\vec r -{\vec r}\,' |\to 0$).
Now the general renormalization scheme for local
field theories applies and ensures (perturbative) renormalizability,
from which one deduces a posteriori the renormalizability of the direct
approach ``\`a la des Cloizeaux"
\ref\BenMa{M. Benhamou and G. Mahoux, J. Phys. France {\bf 47} (1986) 559.}
\ref\BDdir{B. Duplantier, J. Phys. France {\bf 47} (1986) 569.}
\ref\SW{L. Sch\"afer and T. A. Witten, J. Phys. France {\bf 41} (1980) 459.}.
This equivalence with a local field theory also holds for
1-dimensional problems in case (ii), and methods of perturbative field theory
can also be applied in this case. Although they are in general
more complicated than the simple diffusion equation, they
give comparable results (see \ref\DWFreed{J. F. Douglas, S.-Q. Wang and
K. F. Freed, Macromolecules {\bf 19} (1986) 2207; see also
M. K. Kosmas, Makromol. Chem., Rapid Commun. {\bf 2} (1981) 563.}).
\medskip
Beside the perturbative framework, one should notice that rigorous
non-perturbative results have been obtained for the Edwards model and
related models: the mathematical construction of the measure on random
paths associated with \Edwards\ \ref\Var{S. R. S. Varadhan, Appendix to
{\sl Euclidean Quantum Field Theory} by K.
Symanzik, in l{\sl Local Quantum Theory}, R. Jost Ed. (Varenna, 1968),
Academic Press, New York (1969) p. 285;
\hfill\break\noindent
J. Westwater, Commun.
Math. Phys. {\bf 72} (1980) 131; {\bf 79} (1981) 53;
\hfill\break\noindent
J. F. Le Gall,
Commun. Math. Phys. {\bf 104} (1986) 471; {\it ibid}. 505.};
the large distance behavior of intersection properties of independent
random walks at $d=4$ \ref\Law{G. F. Lawler, Commun. Math. Phys. {\bf 86}
(1982) 539,
\hfill\break\noindent
M. Aizenmann, Commun. Math. Phys. {\bf 97} (1985) 91,
\hfill\break\noindent
G. Felder and J. Fr\"ohlich, Commun. Math. Phys.
{\bf 97} (1985) 111.};
the large distance behavior of weakly self-avoiding polymers
at $d=4$ in constructive field theory \ref\AIagoM{D. Arnaudon, D. Iagolnitzer,
and J. Magnen, Phys. Lett. {\bf B 273} (1991) 268.}. These non-perturbative
studies always corroborate the results of the perturbative renormalization
group analysis.
\medskip
The existence of an underlying local field theory in the external
$d$-dimensional space, which is crucial to
ensure renormalizability and allows for predictions from the
perturbative expansion, is however directly related
to the 1-dimensional nature of the object.
When we now consider a {\it D-dimensional object} with $D\ne 1$,
embedded in $d$ dimensions, no
such equivalence with a $d$-dimensional local field theory exists.
Still, the approach ``\`a la des Cloizeaux" can be generalized, by
considering a $D$-dimensional field theory. For instance, the Edwards
Hamiltonian writes for a $D$-dimensional manifold with internal
coordinate $x$
\ref\RBall{R. C. Ball, (1981) unpublished.}
\ref\KN{M. Kardar and D. R. Nelson, Phys. Rev. Lett. {\bf 58}
(1987) 12.}
\ref\ArLub{J. A. Aronowitz and T. C. Lubensky, Europhys. Lett. {\bf 4}
(1987) 395.}
\nref\BDbis{B. Duplantier, Phys. Rev. Lett. {\bf 58} (1987) 2733;
and in
{\sl Statistical Mechanics of Membranes and
Surfaces}, Proceedings of the Fifth Jerusalem Winter School for Theoretical
Physics (1987), D. R. Nelson, T. Piran and S. Weinberg Eds., World Scientific,
Singapore (1989).}
\nref\KNerr{M. Kardar and D. R. Nelson, Phys. Rev. Lett. {\bf 58} (1987)
2280(E); Phys. Rev. {\bf A 38}
(1988) 966.}:
\eqn\EdwardsM{
{\cal H}={1\over 2} \int d^Dx\, \Big(\nabla_x \vec r
\cdot \nabla_x\vec r\Big) +\
{b\over 2}\int d^Dx\int d^Dx'\, \delta^{d}\big(\vec r(x)-\vec r(x')\big)
\ .}
This describes a polymerized or ``tethered" manifold with a fixed internal
metric (to be distinguished from the case of fluid membranes,
with a fluctuating metric). The self-avoidance interaction term
leads to a perturbative expansion in $b$, with poles in
$\epsilon = 4D-d(2-D)$. This method has been used to first order
in $\epsilon$ \KN \ArLub
, and leads to first order estimates of
critical exponents \KN \ArLub \BDbis \KNerr ,
{\it assuming that renormalizability holds}
and that a Renormalization Group equation can thus be used.

Two crucial questions remain however open, which show that new mathematical
deve\-lopments are required:
\item{(I)} A perturbative approach cannot be performed {\it directly}
at $D$ larger or equal to 2. Indeed, for $D\ge 2$ (and $d\ge 0$),
$\epsilon$ is never small ($\epsilon \ge 8$). The double
expansion in $b$ and $\epsilon$ requires to consider the case
of {\it real} non-integer $D$ (typically $1\le D< 2$).
The term of order $N$ in the perturbative
expansion being an integral over $2N$ (resp. $N$) interaction points
in case (i) (resp. case (ii)) in internal $D$-dimensional space,
the meaning of these integrations for non integer $D$ has to be
defined.
\item{(II)}
Since, as a $D$-dimensional field theory, the theory is either
non local (case (i)) or local (case (ii)) but with a singular potential
with explodes at the origin $\vec r =0$ (typically
$1/|\vec r |^\gamma$ or $\delta^d(\vec r)$), standard methods of local
field theory do not apply. Since furthermore, as mentioned above,
we cannot rely (as for $D=1$) on an equivalence with a $d$-dimensional
local field theory, the question arises of the actual renormalizability
of the theory, and in particular of the validity of the use of a
(for instance first order) Renormalization Group equation to predict
a scaling behavior.
\medskip
Beyond the one-loop calculations of \KN\ArLub\BDbis\KNerr\ for the model of
self-avoiding random manifold,
which assumes renormalizability, a next step in a general analysis of
the problem of renormalization for interacting extended object with
dimensionality $D\ne 1$ has been performed by one of the present authors in
\ref\BD{B. Duplantier, Phys. Rev. Lett. {\bf 62} (1989) 2337.}.
In \BD\ a model describing the simple avoidance interaction of a
$D$-dimensional fluctuating manifold with a fixed Euclidean
element was considered. The leading UV divergences of the model
were analyzed in perturbation theory and resummed, so that the consistency
of a renormalization group equation at one loop was established for this model.
A similar direct approach has been applied to the Edwards manifold model
\EdwardsM , and the one-loop renormalizability established
\ref\DHK{B. Duplantier, T. Hwa and M. Kardar, Phys. Rev. Lett. {\bf 64}
(1990) 2022.}.
\medskip
The purpose of this paper is to present a general, mathematically
rigorous, framework to study these questions, and to analyze the
renormalizability of models of interacting objects {\it to all
orders} in perturbation theory.
In this paper, we shall discuss the simple model of reference \BD , of a
$D$-dimensional fluctuating manifold interacting with a single
fixed point (or more generally a fixed
Euclidean element), defined by the following Hamiltonian
\eqn\HamM{
{\cal H}={1\over 2} \int d^Dx\, \Big(\nabla_x \vec r
\cdot \nabla_x\vec r\Big) +\
{b}\int d^Dx\, \delta^{d}\big(\vec r(x)\big)
\ .}
We prove perturbative renormalizability for
this model, to all orders in perturbation theory, from the
internal-space formulation of \BD .
For that purpose we rely on methods devised in perturbative field theory, in
particular by Berg\`ere and Lam, for renormalizing the Feynman amplitudes
in the so-called $\alpha$-parameter or Schwinger representation.
Indeed, our construction can be seen as a generalization of renormalization
theory in Schwinger representation to the case of a $D$-dimensional
$\alpha$-parameter space.

\medskip
This paper is organized as follows.

In section 2, we present the model of a $D$-dimensional manifold
interacting with a single fixed point, discuss its physical relevance
for the problem of entropic repulsion by an
impurity (case of repulsive interaction) and of delocalization transition
(case of attractive interaction), and discuss its formal
perturbative expansion. To each order $N$ of
the expansion corresponds a unique diagram, which is an integral
over the positions of $N$ points in the $D$-dimensional Euclidean internal
space.

In section 3, we give a mathematical prescription to define
the analytic continuation of those integrals for non-integer $D$.
The basic idea relies on the concept of ``distance geometry":
we use the Euclidean invariance in $\RR^D$ to replace
the integral over $N$ $D$-dimensional vectors
labeling the positions of the points,
$x_i=\{x^\mu_i ; \mu=1,\ldots,D\}$ ($i=1,\ldots, N$)
by an integral over the ${N(N-1)\over 2}$ mutual squared distances $a_{ij}=
(x_i-x_j)^2$, with possible constraints. The dimension $D$ appears then
only as a parameter for
the measure term of the $a_{ij}$'s (which by analytic continuation in
$D$ has in general to be considered as a distribution).

In section 4, we analyze the short-distance (ultraviolet)
divergences of these analytically continued integrals. We show that they lead
to poles in $\epsilon \equiv D-d(2-D)/2$. They correspond to an
upper critical dimension $d^\star = 2D/(2-D)$.
We also analyze the large distance
(infrared) divergences which occur when the internal dimensions of the
$D$-dimensional manifold go to infinity.
We show how to regularize these infrared divergences, simply by keeping a
finite size for the manifold, in order to concentrate on the
ultraviolet divergences.

The next four sections are devoted to the analysis and proof
of renormalizability of the theory. Our analysis relies in fact
heavily on concepts and mathematical tools developed in the
$70$'s for the theory of perturbative renormalization of
``ordinary" local field theories
\ref\BPHZ{N. N. Bogoliubov and O. S. Parasiuk, Acta Math. {\bf 97} (1957) 227;
\hfill\break\noindent
K. Hepp, Commun. Math. Phys. {\bf 2} (1966) 301;
\hfill\break\noindent
W. Zimmermann, Commun. Math. Phys. {\bf 15} (1969) 208.
}.
Since these concepts have to be
strongly modified for our problem, and since they are not so well-known,
they will be introduced from the beginning in this paper,
which is therefore (hopefully) entirely self-contained.

In section 5, we show that a diagram of arbitrary order $N$ is
finite when $\epsilon>0$. For that purpose,
we introduce a ``sector decomposition" of the domain of integration
over distances in internal space, which is analogous to the Hepp sectors of
renormalization theory.

The next three sections are devoted to the proof of renormalizability
of the theory, that is the
possibility of absorbing the poles at $\epsilon =0 $ into a redefinition
of the coupling constant of the model, thus making the diagrams
finite at $\epsilon =0$ by appropriate counterterms.
Renormalized amplitudes are defined in section 6, by subtraction of
suitable counterterms. These counterterms are organized in families
of divergent ``subdiagrams", which correspond to the concepts
of ``forest" and of ``nest".

To prove finiteness, we need to reorganize in each ``sector" the
counterterms. Such a reorganization is presented
in section 7, and requires an elaborate ``equivalence classes
of nests" construction, inspired from
\ref\BergLam{M. C. Berg\`ere and Y.-M. P. Lam,
J. Math. Phys. {\bf 17} (1976) 1546.}.

Finally, in section 8, we show that the subtracted amplitudes are
finite at $\epsilon = 0$, as long as the integration over the squared
distances $a_{ij}$ is given by a measure,
while in section 9 we show that this remains true in the general case
where the measure term is a distribution. This ends the proof of the
renormalizability of the model. The rest of section 9 is devoted to
some physical consequences of this renormalizability property, such as
the existence of a Wilson-Fisher $\epsilon$-expansion and of
universal scaling behaviors.

In the concluding section 10, we summarize our work and discuss
various prospects, in particular for the problem of self-avoiding
random manifolds.

A lot of technical points are relegated into various appendices.
\medskip
The reader not interested in the details of the proof of renormalizability
may skip (at least in a first reading$\ldots$) sections 5, 7 and 8.

\newsec{The model}
\subsec{The action}
We first define the model that we shall study and the formal structure of
its perturbative expansion, without taking care of the possible infinities
which may arise from short and/or large distance divergences.
It is the purpose of next sections (in particular section 4)
to define proper regularization schemes.

We start with the manifold Hamiltonian \BD :
\eqn\Ham{{\cal H}=\int_{\cal V} d^Dx \left[{1\over 2}\rvec(x) (-\Delta)^{k\over
2}
\rvec(x) + b\ \delta^d(\rvec(x))\right]\ ,}
where $x$ labels the internal position in the $D$-dimensional
manifold with volume ${\cal V}$
and $\rvec(x)$ is
the corresponding position in the $d$-dimensional Euclidean space.
For the physical case $k=2$, the first term in \Ham corresponds to
the elastic energy of the Gaussian manifold (the internal tension is set
to unity).
For reasons of mathematical convenience, which will be clear in the following,
we shall consider in full
generality the more general class of elastic Hamiltonian with $k\ge 2$. This
allows in particular to define in a proper way a consistent analytic
continuation in the internal dimension $D$.
The case $k=4$ corresponds to a manifold with vanishing tension but with
bending rigidity.
The absence from Eq. \Ham\ of a two-point self-avoidance interaction term
(as compared to Eq. \EdwardsM )
means that we are dealing with  a ``phantom" manifold which can
intersect itself freely.
The second term in \Ham\ corresponds to the interaction of the manifold with
a fixed impurity, that is a
single point in the external $d$-dimensional space, here fixed at the
origin $\rvec={\vec 0}$.
The coupling constant $b$ may be either positive (repulsive interaction)
or negative (attractive interaction).

As mentioned in the Introduction, this model is interesting
as a toy model for the more complex problem of self-avoiding manifolds.
In both cases the interaction term is a singular $\delta$-function, and
similar mathematical techniques can be used to write perturbative expansions
and to study their properties.
In the present case the interaction is much simpler,
since it corresponds to a 1-body
interaction, instead of a 2-body interaction in the case of self-avoidance.
This model is also interesting in its own, since the Hamiltonian \Ham\ can also
be used to describe the (attractive or repulsive) interaction of a fluctuating
$D$-dimensional manifold with a fixed $D'$-dimensional Euclidean subspace in
a $d'$-dimensional Euclidean space \BD , with $d'=d+D'$. In this case $\rvec$
describes the $d$ coordinates of the fluctuating manifold {\it orthogonal}
to the fixed Euclidean $D'$-subspace.
The case $D=1$, corresponding to a polymer interacting with some fixed object,
has been already considered by several authors
\DWFreed\ \ref\Rubin{R. J. Rubin, J. Math. Phys. {\bf 8} (1967) 576, and
references therein.}
\ref\Joanny{ J.-F. Joanny, J. Phys. France {\bf 49} (1988) 1981.} .
The case $D=1$, $d=2$ corresponds for instance to a polymer interacting with a
rigid rod in 3-dimensional space.
If $D=D'$ this model can also be used to describe a ``directed manifold"
(parallel to a flat Euclidean subspace). In this case the coordinates of
the point with internal coordinate $x$ in the external $d'$-dimensional space
are $(x,\rvec(x))$, and the first
$D=D'$ longitudinal degrees of freedom are fixed.
For instance the case
$D=D'=1$ describes a ``directed polymer" interacting with a parallel rod in
$d'$-dimensional space, the case $D=D'=2$, $d'=3$ ($d=1$) describes a SOS-like
fluctuating interface interacting with a parallel plane, ...

\midinsert
\figinsert{8.truecm}{\hsize}{\figcap\dessin{
A $D$-dimensional fluctuating manifold (here $D=2$) interacting:
(a) with a point at the origin in $\RR^d$ (here $d=3$),
(b) with a fixed $D'$-dimensional Euclidean subspace of $\RR^{d'}$
(here $D'=2$, $d=1$, $d'=d+D'=3$).
(c) A ``directed" manifold interacting
with a ``parallel" flat subspace of same dimension $D$ in $\RR^{d'}$.
}}
\endinsert

The ``engineering" dimensions of the position field $\rvec$ and of the
coupling constant $b$ are respectively
\eqn\scaldim{\eqalign{
[\rvec]=\left[x^\nu\right] \qquad&\qquad\nu={k-D\over 2}\ ,\cr
[b]=\left[x^{-\epsilon}\right]\qquad&\qquad\epsilon=D-\nu d\ .\cr}
}
Therefore the interaction is expected to be relevant
(that is to change the large distance properties of the manifold)
if $\epsilon>0$,
that is if $D>D^\star$,
where $D^\star$ is the critical internal dimension, given by
\eqn\Dcrit{
D^\star =\ k\,{d \over d+2}\ ,
}
or equivalently if $d<d^\star$, where $d^\star$ is the
critical embedding dimension
\eqn\dcrit{
d^\star\ =\ {2D \over k-D}\ ,
}
simply equal to the fractal dimension of the manifold.
In particular, this model possesses an upper critical
dimension $0<d^\star<\infty$ for a ``membrane" dimension $0<D<k$.
For the standard interface model ($k=2$), we recover the conditions
$0<D<2$ \KN\ \ArLub \BD .
The exponent $\nu$ plays the role of the size exponent of the elastic manifold.
For fluctuating interfaces, that is ``directed manifolds", it is also
called in the literature the wandering exponent, and denoted by $\zeta$
\ref\Fisher{M. E. Fisher, in {\sl Statistical Mechanics of Membranes and
Surfaces}, Proceedings of the Fifth Jerusalem Winter School for Theoretical
Physics (1987), D. R. Nelson, T. Piran and S. Weinberg Eds., World Scientific,
Singapore (1989).}.
$\nu$ has its natural range between $0$ (collapsed manifold) and $1$
(stretched manifold).  This corresponds exactly to
\eqn\domdef{k-2\le D\le k\ ,}
or equivalently to the ``physical" conditions $D\le d^\star \le \infty$.

\medskip
In \BD , a {\it dimensionless effective coupling constant} $g$ was introduced,
which measures the effective strength of the interaction as a function
of the length scale $X$ measuring the linear internal extent of the manifold,
defined by $\CV= X^D$.
In the vicinity of the critical dimension ($\epsilon \simeq 0$), and for the
physical case $k=2$, a one-loop calculation \BD\ shows that this effective
coupling constant obeys a renormalization group (RG) flow equation, which
writes
\eqn\RGW{
X{\partial g \over \partial X}\,=\,W(g)\,=
\,\epsilon\,g-{1\over 2}S_D\,g^2
\,+\,{\cal O}(g^3)}
with $S_D\,=\,2\,\pi^{D\over 2}/\Gamma({D\over 2})$
the volume of the unit sphere in $\RR^D$. Apart from the trivial
$g=0$ solution, this flow equation has a fixed point solution at the
non-trivial
zero of the Wilson function $W(g)$
\eqn\fixedpoint{g^\star={2\epsilon\over S_D}+\CO (\epsilon^2 )}
At large negative $g$, $W(g)$ behaves like
\eqn\TopleaseBD{W(g)\simeq D\,g \log (-g).}

\midinsert
\figinsert{7.5truecm}{\hsize}{
\figcap\RGflows{The Wilson $W$ function and the Renormalization Group (IR)
flow (for increasing manifold size $X$)
for the dimensionless coupling constant $g$:
(a) in the case $\epsilon>0$, (b) in the case $\epsilon<0$, (c) in the case
$\epsilon=0$.}
}
\endinsert
The physical consequences of these equations are the following:
\item{(I)}{
 $\epsilon >0$:
This corresponds to $D>D^\star$ or $d<d^\star$. The RG flow has an infrared
(IR) stable
fixed point at $g^\star>0$ and an IR unstable (ultraviolet (UV)
stable) fixed point at $g=0$,
as depicted on \RGflows .
For arbitrarily small negative $b$ (attractive interaction), $g$ is negative
and flows to $(-\infty)$ at large length scale $X$; the manifold
is localized (or pinned) at the origin $\rvec={\vec 0}$, and its
average distance
to the origin stays finite.
For arbitrarily small positive
$b$ (repulsive interaction), $g$ is positive and flows to $g^\star$ at large
$X$; the manifold is delocalized, and furthermore repelled from the origin.
The UV Gaussian fixed point at $g=0$ thus describes a
{\it delocalization transition}, whose critical properties
are given by mean field theory
\foot{This transition occurs at vanishing $b$, which corresponds to infinite
temperature. Thus it cannot be induced by a simple change in the temperature
but requires a qualitative change from attractive to repulsive interaction.}.
The nontrivial IR fixed point at $g=g^\star$ describes the
{\it universal large distance properties}
of the delocalized state \BD ,\Joanny , and of the long range repulsive
force away from the origin generated by the fluctuations of the manifold.
}
\item{(II)}{$\epsilon <0$:
This corresponds to $D<D^\star$ or $d>d^\star$. The RG flow has now an
IR unstable (UV stable) fixed point at $g^\star <0$ and an IR stable
fixed point at $g=0$ .
The delocalization transition now occurs for some $b=b^\star<0$, {\it i.e.}
for a non-zero, large enough attractive interaction.
For $b<b^\star$, $g$ is negative and flows to $(-\infty)$ at large $X$;
the manifold is pinned at the origin.
The UV non-trivial fixed point at $b=b^\star$ describes the
delocalization transition.
At this point $g(b^\star)=g^\star$ for any value of the size $X$.
The critical properties of the transition are now anomalous, {\it i.e.}
no longer given by mean field theory.
For smaller attractive interaction ($b^\star<b<0$),
$g$ is negative but now flows to $0$ at large $X$. For repulsive interactions
($b>0$), $g$ is positive and flows to $0$ at large $X$. In these latter two
cases ($b>b^\star$),
the manifold is delocalized, and no longer feels at large distance
the existence of the singular interaction at the origin, since the IR behavior
is now governed by the trivial Gaussian fixed point at $b=0$.
}
\item{(III)}{$\epsilon=0$:
Finally, at the critical dimension, we are in the marginal situation where
the localization transition occurs at $g=0$ ($b=0$),
but where calculable logarithmic corrections
to scaling occur \BD .}

As discussed previously, this picture is valid provided that the
renormalization
group calculations which lead to \RGflows\ make sense. This point has been
discussed at one loop by one of us in \BD . For the case of a one-dimensional
manifold ($D=1$), exact solutions corroborate this picture.
Finally, let us mention the exact treatment of the renormalization group flow
for small $b$ ($b\simeq 0$) for the problem of interface pinning of
\ref\DMRR{F. Dunlop, J. Magnen, V. Rivasseau and P. Roche, J. Stat. Phys.
{\bf 66} (1992) 71.}.
This corresponds to the case $D=2$, $d=1$ and $k=2$ ($\epsilon =2$).

\subsec{The partition function}
The partition function $\CZ$ for the model is defined by
\eqn\partfunc{\CZ=\int {\cal D}[\rvec]\exp(-{\cal H})\ .}
Its perturbative expansion in the coupling constant $b$ is
\eqn\expansion{\CZ=\sum_{N=0}^{\infty}{(-b)^N\over N! }\,\CZ_N\ ,}
where
\eqn\zn{\CZ_N={\big \langle} \int_{\cal V} \prod_{i=1}^{N} d^Dx_i\,
\delta^d(\rvec(x_i))
{\big \rangle}_0}
and $\langle \ldots \rangle_0$ is the average with respect to
the Gaussian measure
$\exp\left[ -\int_{\cal V}d^D x {1\over 2} \rvec
(-\Delta)^{k\over 2} \rvec\right]$.
The evaluation of $\CZ_N$ is best performed in Fourier space by
introducing the vertex function
\eqn\vertex{V(x,\kvec )=\exp (i\,\kvec \cdot \rvec(x))\ ,}
with $\kvec $ a $d$-dimensional vector, and by writing $\CZ_N$ as
\eqn\znbis{\CZ_N={\big \langle} \prod_{i=1}^{N} \int_{\cal V} d^Dx_i
\int {d^d \kvec_i\over (2\pi)^d} V(x_i,\kvec_i)
{\big \rangle}_0\ .}
We compute the above functional average by taking care of the
overall displacement of the manifold ( zero-mode):
\eqn\zeromod{\rvec_G={1\over {\cal V}}\int_{\cal V} d^Dx\ \rvec(x)\ .}
We have explicitly:
\eqn\aver{{\big\langle}\prod_{i=1}^{N}V(x_i,\kvec_i){\big\rangle}_0=
\int \kern -2.pt d^d\rvec_0\int\kern -2.pt{\cal
D}[\rvec(x)]\delta^d(\rvec_G-\rvec_0)
\exp\left[{-\int_{\cal V}d^Dx{1\over 2}\rvec(-\Delta)^{k\over 2}
\rvec}+{i\sum_{i=1}^{N}\kvec_i\cdot\rvec(x_i)}\right]}
Performing the shift $\rvec = \rvec_G +\rvect$, we get:
\eqn\averbis{
\int d^d\rvec_0 \int {\cal D}[\rvect(x)] \delta^d(\rvect_G)
\exp \left[{-\int_{\cal V} d^Dx{1\over 2} \rvect (-\Delta)^{k\over 2}
\rvect } + {i\sum_{i=1}^{N} \kvec_i \cdot (\rvect(x_i)+\rvec_0})\right]\ .}
Integrating over the displacement $\rvec_0$, and performing the Gaussian
average, with normalization
\eqn\norm{
\int {\cal D}[\rvect(x)] \delta^d(\rvect_G)
\exp \left[{-{1\over 2}\int_{\cal V} d^Dx\, \rvect (-\Delta)^{k\over 2}
\rvect } \right]\ =\ 1\ ,}
we finally get
\eqn\znter{\CZ_N=\int \prod_{i=1}^{N} {d^Dx_i\,d^d\kvec_i \over (2\pi)^d}
\, (2\pi)^d\,\delta^d(\sum_{i=1}^{N}\kvec_i)\,\exp\left[
-{1\over 2}\sum_{i,j=1}^{N} \kvec_i \cdot \kvec_j \,G(x_i,x_j)\right]}
where $G(x,y)$ is the propagator, solution (in infinite flat $D$-dimensional
space) of
\eqn\eqprop{(-\Delta_x)^{{k\over 2}} G(x,y)=\delta^D (x-y)\ ,}
namely:
\eqn\prop{
G(x,y)\,=\,{1 \over 2^k\,\pi^{D\over2}}\,{\Gamma({D-k\over2})\over
\Gamma({k\over 2})}\,|x-y|^{k-D}
\ .}
This propagator, which is a Coulomb-like potential, will play a fundamental
role in what follows. In the range of parameters \domdef\ , it vanishes
at $|x-y|=0$.

\noindent
The first term of the expansion of $\CZ$ ($N=0$) is simply the (infinite)
volume
of external space
\eqn\Zzero{
\CZ_0\,=\,(2\pi)^d\delta^d(\kvec ={\vec 0})\,=\,\int d^d\rvec_G\equiv V_{\RR^d}
\ .}
But the next terms are finite. Indeed, for $N> 0$
we can deal with the $\delta^d$ constraint in \znter\ by setting
$\kvec_1=-\sum\limits_{i=2}^{N}\kvec_i$.
The integration over $\kvec$ becomes Gaussian and leads for $N=1$ to
\eqn\Zone{
\CZ_1\ =\ \int d^D x_1\ =\ {\cal V}
}
and for $N>1$ to the basic formula \BD :
\eqn\ZN{\encadre{\hbox{$\displaystyle
\CZ_N\ =\ (2\pi)^{-{d\over2}(N-1)}\,\int\prod_{i=1}^N d^Dx_i
\,\left(\det\left[ \Pi_{ij}\right]_{\scriptscriptstyle 2\le i,j\le N}
\right)^{-{d\over 2}}
$}}\ ,}
where $\Pi_{ij}$ ($2\le i,j \le N$) is the $(N-1)\times (N-1)$ matrix
\eqn\Pimatrix{
\Pi_{ij}\,=\,G(x_i,x_j)-G(x_1,x_j)-G(x_i,x_1)+G(x_1,x_1)
\ .}
Notice that $\Pi_{ij}$ is function of the point $x_1$ which acts as a
reference point, and that
$G(x_1,x_1)$ is actually equal to zero.
\subsec{Correlation functions}
Similarly, all expectation values of observables can be obtained from
the partition functions with inserted vertex operators \vertex
\eqn\zvertex{
\CZ^{(M)}(X_a,\kvec_a)\,=\,\CZ\cdot \langle \prod_{a=1}^M V(X_a,\kvec_a)
\rangle\,=\,\int{\cal D}[\rvec]\exp(-{\cal H}+\sum_{a=1}^M i\,\kvec_a
\cdot \rvec(X_a) )
\ .}
Each term of their perturbative expansion
\eqn\zvertexn{
\CZ^{(M)}(X_a,\kvec_a)\,=\,\sum_{N=0}^\infty {(-b)^N
\over N!}\CZ_N^{(M)}(X_a,\kvec_a)
}
can be computed by the same techniques. The final result is for $N>1$
\eqn\ZvertexN{
\CZ_N^{(M)}(X_a,\kvec_a)=
(2\pi)^{-{d\over2}(N-1)}\int\prod_{i=1}^N d^Dx_i
\,\left(\det\left[ \Pi_{ij}\right]_{\scriptscriptstyle 2\le i,j\le N}
\right)^{-{d\over 2}}
\exp \left( -{1\over2}\sum_{a,b=1}^{M}\kvec_a \cdot \kvec_b\, \Delta_{ab}
\right)
}
where $\Delta_{ab}$ is a ratio of determinants:
\eqn\Deltamatr{
\Delta_{ab}\,=\,{\det\nolimits_N
\left(\matrix{\Pi_{ab}&\Pi_{aj}\cr\Pi_{ib}&\Pi_{ij}\cr}\right)
\over
\det\nolimits_{N-1} (\Pi_{ij})
}\ ,}
with an obvious extension of the definition of the $\Pi$ matrix \Pimatrix\
to include external points (in particular $\Pi_{ab}=G(X_a,X_b)-G(x_1,X_b)
-G(X_a,x_1)+G(x_1,x_1)$).
The cases $N=0$ and $N=1$ require a specific analysis.
For $N=0$ we get simply
\eqn\Zvertexzero{
\CZ_0^{(M)}(X_a,\kvec_a)\,
=\,(2\pi)^d\delta^d(\sum_{a=1}^{M} \kvec_a )\,\exp ( -{1\over 2}
\sum_{a,b=1}^{M}\,\kvec_a \cdot \kvec_b\,G(X_a,X_b) )
\ ,}
and for $N=1$
\eqn\Zvertexone{
\CZ_1^{(M)}(X_a,\kvec_a)\ =\ \int d^D x_1\,
\exp ( -{1\over 2}
\sum_{a,b=1}^{M}\,\kvec_a \cdot \kvec_b\,\Pi_{ab} )
}
(Notice in this last equation that $\Pi_{ab}$ actually depends on $x_1$).
\subsec{Mean squared distances}
 From \Zvertexzero\ one can in particular derive
the mean squared distance between any two points $x$ and $y$
for the free model ($b=0$):
\eqn\msqext{
{1\over 2d}\,{\langle (\rvec(x)-\rvec(y))^2\rangle}_0\ =\ -\,G(x,y)
\ =\ {1 \over 4^\nu (4\pi)^{D/2}}{\Gamma(1-\nu)\over\nu\Gamma(
\nu+{D \over 2 })}|x-y|^{2\nu}
\ ,}
which is IR- and UV-finite and positive for $0<\nu<1$ ($k-2<D<k$).

\newsec{Analytic continuation in the internal dimension ${\tit D}$}
\subsec{Independent set of parameters: ${\tit D}$, {\tentit\char'027} and
{\tentit\char'017}}

We now want to give a meaning to the above expressions for
arbitrary real $D$, $d$ and $k$, so as to have a continuous
approach to the ``physical" elastic membrane problem $D=2$
and $k=2$.
As is clear from \ZvertexN , the general observables of the form
\zvertex\ depend on the external dimension $d$ only through:
{\it (i)} the external invariants $\kvec_a\cdot\kvec_b$,
{\it (ii)} the exponent $-d/2$ in \ZvertexN .
We can therefore, as usual in field theory, consider $d$ as a continuous
parameter. The same is true for the exponent $k$ associated with the
internal Laplacian, which appears only as a parameter in the
propagator \prop\ .
Since we shall be interested in the range $d$ close to $d^\star$,
it is natural to substitute to
the continuous parameters
$d$ and $k$ the set of continuous parameters $\epsilon$ and $\nu$.
Their relevant range is $\epsilon \simeq 0$ (where we expect a non-trivial
universal fixed point) and $0<\nu<1$ (where the manifold is crumpled, that
is neither collapsed nor stretched).

The analytic continuation in the internal dimension $D$ is a new feature
of this model and requires a separate analysis, namely that of the
signification of the measure $\prod\limits_i d^Dx_i$ for non-integer $D$.
We now discuss equivalent geometric definitions of this measure,
which have a natural extension to non-integer $D$.

\subsec{Distance geometry in $\tit D$ dimensions}

\midinsert
\figinsert{12.5truecm}{6.5 truecm}{%\hsize}{
\figcap\repf{
Equivalent representations of the positions of a given set of $N$
interaction points (here $N=6$). The points are described (a) by their
position $x_i$ in $\RR^D$ or $\RR^{N-1}$ or (b) by the set of their
mutual squared distances $a_{ij}=(x_i-x_j)^2$ or (c) by their relative vector
$y_i=x_{i+1}-x_1$ in $\RR^D$ or $\RR^{N-1}$ (relative to the point $x_1$)
or (d) by the line vectors (labeled by $\alpha$) of an arbitrary
spanning tree joining these points.}
}
\endinsert
We are looking at generalized integrals of the type
$\int d^Dx_1\ldots d^Dx_N\,f(x_1,\ldots,x_N)$ where $f$ is invariant
by rotation in $D$-dimensional space and thus depends only on the invariant
scalar products
\eqn\scalprod{u_{ij}=x_i\cdot x_j\ }
which form a symmetric matrix $[u_{ij}]$.
For $D\ge N$ we can reduce the integration over the $x_i$'s to an integral
over the $u_{ij}$'s of the form (see Appendix A)
\eqn\intu{
\int \prod_{i=1}^N d^Dx_i\, f(u_{ij})\ =\ \int_{{\cal U}_N} \prod_{i \le j}
du_{ij}
\,\sigma_N^{(D)}([u_{ij}])\, f([u_{ij}])
\ ,}
where
\eqn\measuij{
\sigma_N^{(D)}([u_{ij}])\ =\ {S_{D}\over 2}{S_{D-1}\over
2}\ldots{S_{D-N+1}\over 2}\,
\left( \det\nolimits_N [u_{ij}]\right)^{D-N-1\over 2}
\ .}
$S_D$ is the volume of the unit sphere in $\RR^D$, $S_D=2\ \pi^{D\over 2} /
\Gamma ({D\over 2})$.
The domain of integration ${\cal U}_N$ for $u_{ij}$ is such that $u_{ij}$ is
the actual
scalar product of vectors in Euclidean space, {\it i.e.}
$[u_{ij}]$ is a positive matrix.

If moreover the integrand is translationally invariant in $D$-dimensional
space, we can go to relative vectors $y_{i}=x_{i+1}-x_1$
($1\le i\le N-1$) and reduce by one unit
the number of points, {\it i.e} use $\sigma^{(D)}_{N-1}([y_i\cdot y_j ])$ .
\eqn\measuone{
\prod_{i=1}^N d^Dx_i\ =\ d^Dx_1\prod_{1\le i \le j\le N-1} d(y_{i}\cdot y_{j})
\ {S_{D}\over 2}{S_{D-1}\over 2}\ldots {S_{D-N+2}\over 2}\,
\left( \det\nolimits_{N-1} [y_{i}\cdot y_{j}]\right)^{D-N\over 2}
\ .}
This is equivalent to a measure expressed uniquely in terms of
the complete set of \break $N(N-1)/2$ squared distances
\eqn\aij{
a_{ij}=(x_i-x_j)^2
}
by simply rewriting $y_i\cdot y_j$ as
\eqn\xtoa{\eqalign{y_{i-1}\cdot y_{j-1}&=D_{ij}(a)\cr
D_{ij}(a)&\equiv {1\over 2}(a_{i1}+a_{j1}-a_{ij})
\qquad 2\le i,j \le N\cr}}
Finally, after the simple change of variables \xtoa\ %(see Appendix A)
we arrive, for a translationally and rotationally invariant integrand,
at an integral over distances
\eqn\inta{
\int_{\RR^D} \prod_{i=1}^N d^Dx_i\, f(a_{ij})\ =\ {\cal V}\ \int_{{\cal A}_N}
\prod_{1\le i < j\le N} da_{ij}
\,\mu_N^{(D)}([a_{ij}])\, f([a_{ij}])
\ .}
where
\eqn\measaij{
\mu_N^{(D)}([a_{ij}])\ =\ 2^{-{(N-1)(N-2)\over 2}}\,
{S_{D}\over 2}{S_{D-1}\over 2}\ldots{S_{D-N+2}\over 2}\,
\left( \det\nolimits_{N-1} [D_{ij}(a)]_{\scriptscriptstyle 2\le i,j \le N}
\right)^{D-N\over 2}
\ .}
This last formula is valid for $D\ge N-1$. Indeed, $D=N-1$ is the smallest
dimension for which $N$ linearly independent points can be embedded in
Euclidean
space. The domain of integration
${\cal A}_N$ for $a_{ij}$ is then simply the set for which
$[D_{ij}(a)]$ is a positive
matrix.

\nref\Blum{L. M. Blumenthal, {\sl Theory and Applications of Distance
Geometry},
Clarendon Press, Oxford (1953).}
\nref\Cox{H. M. S. Coxeter, Aeq. Math. {\bf 1} (1968) 104.}
In \measaij\ appears the important quantity
\eqn\pna{P_N(a)\equiv \det\nolimits_{N-1}[D_{ij}(a)]
=\det\nolimits_{N-1}[y_i\cdot y_j
]}
which is a homogeneous polynomial of degree $N-1$ in the $a_{ij}$.
$P_N(a)$ is actually fully symmetric under permutations of the indices $i$ or
$j$ in $[a_{ij}]$, as can be seen from its expression as a Cayley-Menger
determinant
\foot{This determinant appears, in a different disguise, in a letter by
Descartes to the Princess Elisabeth of Bohemia (1643), as quoted by
Coxeter in \Cox .}
well-known in distance geometry
\Blum
\eqn\CayMen{P_N(a)\ =\ {(-1)^N \over 2^{N-1}}\,
\left| \matrix{0&1&1&\ldots&1\cr 1&0&a_{12}&\ldots&a_{1N}\cr
1&a_{21}&0&\ldots&a_{2N}\cr \vdots&\vdots&\vdots&\ddots&\vdots\cr 1&a_{N1}&
a_{N2}&\ldots&0\cr } \right|
\ .}
We have for instance for $N=2$ and $3$ points
\eqn\Ptwo{P_2(a)\,=\,a_{12}\ ,\quad
P_3(a)\,=\,{1 \over 4}(2 a_{12}a_{23}+2a_{23}a_{31}+2a_{31}a_{12}-a_{12}^2
-a_{23}^2-a_{31}^2)}
The matrix $[D_{ij}]$ will be positive iff any bordered principal minor
$P_K(a)$ is $\ge 0$ for any $K\le N$:
\eqn\princmin{P_K(a)\ =\ {(-1)^K \over 2^{K-1}}\,
\left| \matrix{0&1&1&\ldots&1\cr 1&0&a_{12}&\ldots&a_{1K}\cr
1&a_{21}&0&\ldots&a_{2K}\cr \vdots&\vdots&\vdots&\ddots&\vdots\cr 1&a_{K1}&
a_{K2}&\ldots&0\cr } \right|\,\ge\,0
\ .}
For $K=2$, this is simply the positivity condition $a_{12}\ge 0$.
For $K=3$, one recovers the familiar triangular inequality
\eqn\trineq{
(a_{12}-a_{13}-a_{23})^2\le 4 a_{13} a_{23}\quad\Leftrightarrow\quad
|a_{13}^{1\over 2}-a_{23}^{1\over 2}|\le a_{12}^{1\over 2}
\le a_{13}^{1\over 2}+a_{23}^{1\over 2}
\ .}
For $K>3$ one gets more general inequalities which are the necessary and
sufficient conditions for the $a_{ij}$ to be realized as squared
distances between
$N$ points of the Euclidean space $\RR^{N-1}$.
The volume ${\cal V}(x_1,\ldots ,x_K)$
of the
(possibly degenerate) parallelotope \ref\EDM{{\sl Encyclopedic Dictionary of
Mathematics}, Second Edition, edited by Kiyosi It\^ o, The MIT Press,
Cambridge,
London (1987).} ($(K-1)$-dimensional parallelepiped)
with vertices $x_1,x_2, \ldots ,x_K$ is given by
\eqn\volsim{{\cal V}^2(x_1,\ldots ,x_K)\ =\ P_K(a)
\ .}
Thus $P_K(a)=0$ indicates that the first $K$ points are linearly
dependent, {\it i.e.} can be embedded in  $\RR^{K-2}$.

For $D\le N-2$, the expression \measaij\ becomes singular due to the appearance
of zeros in the sphere volumes $S_{D-K+2}$ for $D+2\le K\le N$ on the one hand,
and due to divergences of the term $(P_N(a))^{{D-N\over 2}}$, which occur
when $P_N(a)$ vanishes, that is on the boundary of the domain ${\cal A}_N$,
on the other hand. Nevertheless $\mu^{(D)}_N(a)$ can
now be considered as a distribution with a support in
submanifolds of ${\cal A}_N$ of dimension $D(N-{D+1\over 2})$,
which correspond to $D$-dimensional
Euclidean subspaces of $\RR^{N-1}$. One therefore
still reproduces the natural Euclidean measure in $\RR^D$,
as can be shown by analytic continuation,
which we now describe.

\subsec{Analytic continuations in $\tit D$}
\noindent{\it $\diamond $ 1- Distance geometry for non-integer $D$}

The first way to define integrals of the form
$\int d^D x_1 \ldots d^Dx_N\,f(x_1,\ldots,x_N)$ for non integer $D$ is to
start from \intu\ and \measuij\ or equivalently from \inta \ and \measaij .
The measures \measuij\ and \measaij\ now involve $D$
as a simple parameter and
therefore provide a natural basis for analytic continuation.
For {\it real} $D > N-2$, $\mu^{(D)}_N(a)$ remains a positive measure
density on ${\cal A}_N$.
Therefore it can be considered as a distribution, over the space $\RR^{N(N-1)
\over 2}$ of all squared distances $a_{ij}$, with support ${\cal A}_N$
(i.e. by definition it vanishes outside ${\cal A}_N$).
As a distribution it can be extended to $0\le D\le N-2$ by analytic
continuation.
This amounts to treat by a finite part prescription all the divergences
which occur at the boundaries of ${\cal A}_N$
(see below the spherical coordinate representation for more details).
As a distribution, it is not singular for positive {\it integer} $D\le N-2$,
but
becomes a measure density concentrated on the submanifold such as
the principal minors $P_K(a)$ vanish for all $K$ such that $D+1 < K \le N$.

As an example let us consider the case of two points.
For $N=2$ we have the distribution (denoting $\chi({\cal A})$
the characteristic function of support ${\cal A}$)
\eqn\disttwo{
\mu_2^{(D)}(a)\ \chi({\cal A}_2)\ =\ {\pi^{D\over 2}\over \Gamma({D\over
2})}\left|
a_{12}\right|^{{D \over 2}-1}
\theta (a_{12})
}
When $D\to 0$ the r.h.s. of \disttwo\ tends to
\eqn\distto{
{\pi^{D\over 2}\over \Gamma({D\over 2})}\left|
a_{12}\right|^{{D \over 2}-1}
\theta (a_{12})\ { \buildrel {\scriptscriptstyle D\to 0} \over \longrightarrow}
\ \delta(a_{12})
}
Thus the support of the distribution becomes restricted to
the zero-dimensional subspace
(where all points coincide).

Similarly for $N=3$ we have
\eqn\distthree{
\eqalign{
\mu_3^{(D)}(a)\ \chi({\cal A}_3)\ &=\
{1\over 2}{\pi^{D\over 2}\over \Gamma({D\over 2})}
{\pi^{D-1\over 2}\over \Gamma({D-1\over 2})}
\left|\det\nolimits_2D\right|^{{D-1 \over 2}-1}
\theta(\det\nolimits_2 D)\  \theta (a_{12})\theta (a_{13})
\theta (a_{23})\cr
& { \buildrel {\scriptscriptstyle D\to 1} \over \longrightarrow}
\ {1\over 2} \delta(\det\nolimits_2D)\ \theta(a_{12}) \theta(a_{13})
 \theta(a_{23})\cr
}
}
where $\det\nolimits_2 D\equiv P_3(a)$ reads:
\eqn\dettwoD{
\det\nolimits_2 D={1\over 4}
( a_{12}^{1\over 2}+a_{13}^{1\over 2}+a_{23}^{1\over 2})
(a_{12}^{1\over 2}+a_{13}^{1\over 2}-a_{23}^{1\over 2})
(a_{13}^{1\over 2}+a_{23}^{1\over 2}-a_{12}^{1\over 2})
(a_{12}^{1\over 2}+a_{23}^{1\over 2}-a_{13}^{1\over 2})
\ .}
Separating three different boundary sectors of ${\cal A}_3$, we get
\eqn\distter{
\mu^{(1)}_3(a)\chi ({\cal A}_3)da_{12}da_{13}da_{23}\ =\
2\delta(a_{12}^{1\over 2}+a_{23}^{1\over 2}-a_{13}^{1\over 2})
\ \theta(a_{12}) \theta(a_{23})\theta (a_{13})
da_{12}^{1\over 2}\ da_{23}^{1\over 2}\ da_{13}^{1\over 2}
\ +\ {\rm perm.}
}
which represents indeed all possible relative positions of three points
on an oriented line.

\medskip
\noindent{\it $\diamond$ 2- Cartesian coordinates in $\RR^{N-1}$}

Realizing that $N-1$ is the minimal dimension of Euclidean space in which one
can embed $N$ points with given squared distances $a_{ij}$ (in ${\cal A}_N$),
we can use \measuone\ back to reexpress the measure over the scalar products
$d(y_i\cdot y_j)$ as a measure over $N$ points in $\RR^{N-1}$
\eqn\measubis{
\prod_{i=1}^{N-1} d^{N-1}y_i\ =\ \prod_{1\le i \le j\le N-1} d(y_{i}\cdot
y_{j})
\ {S_{N-1}\over 2}{S_{N-2}\over 2}\ldots {S_{1}\over 2}\,
\left( \det\nolimits_{N-1} [y_{i}\cdot y_{j}]\right)^{-{1\over 2}}
\ .}
Thus we can implement the analytic continuation in $D$ by modifying the
Euclidean measure in $\RR^{N-1}$ by
a suitable analytic measure term:
\eqn\DtoNbis{
\prod_{i=1}^{N-1} d^Dy_i\ \equiv \ \prod_{i=1}^{N-1} d^{N-1}y_i
{S_DS_{D-1}\ldots S_{D-N+2} \over S_{N-1}S_{N-2}\ldots S_1}
\bigg[
\det [y_i \cdot y_j]_{1\le i,j \le N-1}
\bigg]^{{D-N+1\over 2}}
\ .}
Analytic continuation can thus be summarized in the following compact formula,
which is a formal rewriting of \DtoNbis :
\eqn\Compac{\encadre {\vbox{\hbox{$\displaystyle \prod_{i=1}^{N-1}d^Dy_i=
\prod_{i=1}^{N-1}d^{N-1}y_i\ \Omega (D,N)
\, \left({\cal V}\left(0,y_1,
\ldots,y_{N-1}\right)\right)^{D-N+1}$}
\hbox{$\qquad \displaystyle \Omega (D,N)\ =\
{{\rm Vol}({\rm SO}(D))\over {\rm Vol}({\rm SO}
(D-N+1)){\rm Vol}({\rm SO}(N-1))}$}}}\ ,  }
where ${\rm Vol}({\rm SO}(D))$ is the volume of the special orthogonal
group in $D$ dimensions:
\eqn\volsod{{\rm Vol}({\rm SO}(D))={S_D\over 2}{S_{D-1}\over 2}\ldots
{S_1\over 2}\ .}
When $M$ external points $X_a$ are present (that is points over which
we do not integrate), Eq. \Compac\ has to be replaced
by the more general formula:
\eqn\Compacext{\vbox{\hbox{$\displaystyle \prod_{i=1}^{N}d^Dx_i=
\prod_{i=1}^{N}d^{M+N-1}x_i\ \Omega (D,M,N)
\, \left({{\cal V}\left(x_1,x_2,
\ldots,x_N,X_1,\ldots,X_{M}\right)\over
{\cal V}\left(X_1,\ldots,X_M\right)}\right)^{D-M-N+1}$}
\hbox{$\qquad \displaystyle \Omega (D,M,N)\ =\
{{\rm Vol}({\rm SO}(D-M+1))\over {\rm Vol}({\rm SO}
(D-M-N+1)){\rm Vol}({\rm SO}(N))}$}}\ .}

\medskip
\noindent{\it $\diamond$ 3- Spherical coordinates}

A third (equivalent) way to perform an analytic continuation in $D$
is the use of spherical coordinates.
We first consider again the case of $N$ points in $\RR^D$ with $D$ integer
and $D \ge N-1$.
We take $x_1$ as the center of the spherical coordinates, and describe the
$N-1$ other points by their relative coordinate, as before
\eqn\yi{y_i=x_{i+1}-x_1\qquad i=1,\ldots ,N-1\ .}
Introducing generalized spherical coordinates for the $y_i$, we write
\eqn\spheric{
\eqalign{
y_{i,1}\  &=|y_i|\cos \theta_{i,1} \cr
y_{i,2}\  &=|y_i|\sin \theta_{i,1} \cos \theta_{i,2}\cr
&\vdots \cr
y_{i,D-1} &=|y_i|\sin \theta_{i,1} \sin \theta_{i,2} \ldots \sin \theta_{i,D-2}
\cos \theta_{i,D-1} \cr
y_{i,D}\  &=|y_i|\sin \theta_{i,1} \sin \theta_{i,2} \ldots \sin \theta_{i,D-2}
\sin \theta_{i,D-1} \cr
}
}
where $\theta_{i,n} \in [0,\pi ]$ for $1\le n \le D-2$ and
$\theta_{i,D-1}\in [0,2\pi [$.
The corresponding measure is given by
\eqn\Jac{d^Dy_i=|y_i|^{D-1}d|y_i|\ \prod_{n = 1}^{D-1} (\sin
\theta_{i,n})^{D-1-n} d\theta_{i,n}
\ .}
For rotationally invariant integrands, we can furthermore fix successively
\eqn\thetzero{\theta_{i,n}=0 \qquad n \ge i}
Taking care of the successive rotational symmetries, we arrive at
\eqn\meastheta{\encadre{\hbox{$\displaystyle{
\prod_{i=1}^{N-1} d^Dy_i\ =\ S_DS_{D-1}\ldots S_{D-N+2}\ \prod_{i=1}^{N-1}
|y_i|^{D-1}d|y_i|\prod_{i=2}^{N-1}\prod_{n =1}^{i-1}
(\sin \theta_{i,n})^{D-1-
n} d\theta_{i,n}
}$}}}
with all the $\theta_{i,n}$ now integrated from $0$ to $\pi$.
In \meastheta\ , $D$ again appears only as a parameter. This therefore
provides another natural path to analytic continuation in $D$.
Indeed, possible singularities at integer $D$ arise from the negative
powers of the $\sin\theta_{i,n}$'s, which diverge at $\theta_{i,n}=0$ or $\pi$.
It is clear from the spherical coordinates representation \spheric\ that
when some of the $\theta$'s are equal to $0$ or $\pi$ the vectors
$y_i$ are not linearly independent and the $N$ points $x_i$ are on an
Euclidean subspace with dimension smaller than $N-1$.
Away from integer values of $D$ (with $0<D<N-1$), these divergences can be
treated by the standard finite part prescription (independently for
each $\theta_{i,n}$).
To prove that for integer $D$, \meastheta\ remains a distribution and
can be rewritten as a finite measure localized on some subspace
(corresponding to spherical coordinates in some $D$ dimensional submanifold)
requires a more elaborate discussion, not presented here.

This analytic continuation \meastheta\ is totally equivalent to the
analytic continuation \Compac , as can be seen by
going back as before to coordinates in $\RR^{N-1}$. Using \meastheta ,
we have formally
\eqn\DtoN{
\prod_{i=1}^{N-1} d^Dy_i\ =\ \prod_{i=1}^{N-1} d^{N-1}y_i
{S_DS_{D-1}\ldots S_{D-N+2} \over S_{N-1}S_{N-2}\ldots S_1}
\bigg[ \prod_{i=1}^{N-1}|y_i|\times  \prod_{i=2}^{N-1} \prod_{n
=1}^{i-1} \sin \theta_{i, n}
\bigg]^{D-N+1}
}
where the $\theta_{i,n}$ are spherical angles in $\RR^{N-1}$.
We read on this equation
the angular representation of the squared parallelotope volume \Blum\
\eqn\DetSin{
\eqalign{
P_N(a)\ =\ \det [y_i\cdot y_j]_{1\le i,j\le N-1} \ &=\
\prod_{i=1}^{N-1}|y_i|^2 \times \prod_{i=2}^{N-1} \prod_{n
=1 }^{i-1} \sin^2 \theta_{i,n }\cr &={\cal V}^2
(0,y_1,\ldots ,y_{N-1})\ .\cr
}
}
and \DtoN\ is therefore identical to \Compac .

\noindent Finally, when $M$ external points are present, \meastheta\ has to be
replaced by
\eqn\measthetaext{
\prod_{i=1}^{N} d^Dx_i=S_{D-M+1}S_{D-M}\ldots S_{D-M-N+2}\
\prod_{i=1}^{N}
|x_i|^{D-1}d|x_i|\prod_{i=1}^{N}\prod_{n =1}^{M+i-2}
(\sin \theta_{i,n})^{D-1-
n} d\theta_{i,n}
}
where the $\theta_{i,n}$ are the $M+i-2$
successive relative spherical angles for $x_i$ necessary to
assign position to the vector $x_i-X_1$ with respect to the $M-1$ external
vectors $X_2-X_1,\ldots,X_M-X_1$
and to the $i-1$ internal vectors $x_j-X_1$ for $j<i$,
in a reference frame where $X_1$ is at the origin.

\subsec{Factorization}

Of course, for integer $D$, the measure $\prod\limits_id^Dy_i$ is naturally
factorized, when applied to a product of functions of independent variables:
\eqn\factoriz{
\int \prod_{k=1}^{P+Q}d^Dy_k\,f(\{y_{k; k=1,P}\})
\,g(\{y_{k; k=P+1,P+Q}\})
\ =\ \int \prod_{i=1}^P d^Dy_i\,f(y_i)\cdot\int \prod_{j=P+1}^{P+Q} d^Dy_j\,
g(y_j)
\ .}

This important factorization property becomes non trivial when extended
to arbitrary $D$, as can be seen from \DtoNbis\ .
Still, if we consider the scalar product matrix
$[u_{ij}]_{1\le i,j \le P+Q}$ and denote by $[u]_P$ (respectively
$[u]_Q$ ) the submatrix $[u_{ij}]_{1\le i,j \le P}$ (respectively
$[u_{ij}]_{P+1\le i,j \le P+Q}$), one has (see Appendix B)
\eqn\factorid{\eqalign{
\int_{{\cal U}_{P+Q}} d[u]\,\sigma^{(D)}_{P+Q}([u])
f([u]_P) g([u]_Q)
\ &=\ \int_{{\cal U}_P} d[u]_P\,\sigma^{(D)}_P
([u]_P)\, f([u]_P)\cr
&\qquad \times
\int_{{\cal U}_Q} d[u]_Q\,\sigma^{(D)}_Q ([u]_Q)\, g([u]_Q) \cr
}}
which means that the integration over the mixed scalar products
$u_{ij}$, $1\le i\le P < j\le P+Q$ can be performed and amounts to factorize
$\sigma^{(D)}_{P+Q}$ into $\sigma^{(D)}_P \sigma^{(D)}_Q$.
{\it The factorization property of the measure is thus preserved under analytic
continuation in $D$.}

\subsec{The interaction as a Cayley-Menger determinant}
The $N$ point interaction term
$\left( \det\nolimits_{N-1} \left[ \Pi_{ij}\right]\right)^{-{d \over 2}}$
depends explicitly on $D$ through the occurrence of the Green function
\msqext\ and is readily analytically continued to non-integer $D$.
Let us recall that
we consider $D$, $\nu$ and $\epsilon$ as the three independent
parameters of the model, so that $d$ itself is a function of $D$ given by
$d=(D-\epsilon)/\nu$.
 From a distinct, geometrical point of view, it is particularly interesting to
notice that the interaction term
also involves a determinant of the Cayley-Menger type
with $a_{ij}$ replaced by its power $(a_{ij})^{\nu}$
\eqn\CayMengnu{P_N(a^{\nu})\ \equiv \ {(-1)^N \over 2^{N-1}}\,
\left| \matrix{0&1&1&\ldots&1\cr 1&0&a_{12}^{\nu}&\ldots&a_{1N}^{\nu}\cr
1&a_{21}^{\nu}&0&\ldots&a_{2N}^{\nu}\cr \vdots&\vdots&\vdots&\ddots&\vdots\cr
1&a_{N1}^{\nu}&
a_{N2}^{\nu}&\ldots&0\cr } \right|
\ .}
Indeed, from definition \Pimatrix\ and from \msqext\ , we have
\eqn\PiD{
\Pi_{ij}\ =\ A_D(\nu)\
D_{ij}(a^\nu)
\ ,}
with
\eqn\Dijanu{D_{ij}(a^\nu)\ =\ {1\over 2}(a_{i1}^\nu +a_{j1}^\nu
-a_{ij}^\nu) }
and the factor
\eqn\ADnu{A_D(\nu)\ =\ {2\over 4^\nu (4\pi)^{D\over 2}}
{\Gamma (1-\nu) \over
\nu \Gamma(\nu +{D\over 2})}
\ ,}
and therefore
\eqn\detPiD{
\det\nolimits_{N-1} [\Pi_{ij}]\ =\ [A_D(\nu)]^{N-1}\ P_N(a^\nu)
\ .}

Finally we have the compact formula, analytic in $D$, $\epsilon$ and $\nu$,
for the term of order $N$ of the partition
function \expansion
\eqn\ZNa{\encadre{\hbox{$\displaystyle
\CZ_N={\cal V}\left( 2\pi A_D(\nu)\right)^{-{d\over 2}(N-1)}\prod_{K=2}^{N}
\big( {S_{D-K+2}
\over 2^{K-1}}\big)\int_{{\cal A}_N} \prod_{1\le i < j \le N} da_{ij}
[P_N(a)]^{{D-N\over 2}}[P_N(a^\nu)]^{-{d\over 2}}$}}
}
with again $d=(D-\epsilon)/\nu$.

\subsec{Analytic expression of $\tit \CZ_N$ in Cartesian coordinates}
An immediate corollary of the above formalism is the following
alternative formula for $\CZ_N$, now in Cartesian coordinates in $\RR^{N-1}$,
which provides an equivalent definition of the analytic continuation of
$\CZ_N$:
\eqn\ZNcart{\eqalign{
\CZ_N\ = (2\pi)^{-{d\over 2}(N-1)}\,{\cal V}\,
\int \prod_{i=1}^{N-1} d^{N-1}y_i
{S_D\ldots S_{D-N+2} \over S_{N-1}\ldots S_1}
&\left(
\det\left[y_i \cdot y_j\right]_{1\le i,j \le N-1}
\right)^{{D-N+1\over 2}}\cr & \times
\left(\det\left[ \Pi_{ij}\right]_{\scriptscriptstyle
2\le i ,j \le N}
\right)^{-{d\over 2}}\ .\cr }
}

\subsec{Determinant attached to trees}
In the following, we shall find useful to express  both
the measure and the interaction contributions in terms of more
general variables
$\lambda_\alpha$ obtained from the positions $x_i$ and
attached to arbitrary oriented trees.
A spanning tree is a connected graph whose vertices are the previous $N$
points $x_i$, and without loops. This graph therefore has $N-1$ internal
lines labeled by $\alpha =1,\ldots ,N-1$ for which one also specifies
an orientation.
An oriented tree is characterized by its incidence $N\times (N-1)$
matrix $[\epsilon_{i\alpha}]$ defined by
$\epsilon_{i\alpha}=1$ if the line $\alpha$ is incident to $i$
and points toward $i$, $\epsilon_{i\alpha}=-1$ if $\alpha$ is
incident to $i$ and points outward $i$, $\epsilon_{i\alpha}=0$
otherwise. One has
\eqn\propeps{
\sum_{i=1}^{N}\epsilon_{i\alpha }=0\ .}
For each line $\alpha$ of the tree
we define the {\it line vector} (or {\it edge vector})
$\lambda_\alpha$ in $\RR^{N-1}$
by
\eqn\lamb{\lambda_\alpha = \sum_{i=1}^N \epsilon_{i\alpha }\ x_i\ =\
\sum_{i=1}^{N-1} \epsilon_{i+1 \alpha} \ y_i\ ,}
where the $y_i$'s have been defined in \yi .

\noindent{\it $\diamond$ Expression for the measure}

Since the Jacobian of the linear transformation \lamb\ from the $y_i$'s
to the $\lambda_\alpha$'s is
$|\det'[\epsilon ]|=|\det[\epsilon_{i\alpha}]_{\buildrel {{\scriptscriptstyle
2\le i\le N\ \ }}
\over {{\scriptscriptstyle 1\le \alpha\le N-1}} }  |=1$ and
$\det[\lambda_\alpha \cdot \lambda_\beta]=
(\det'[\epsilon ])^2 \det [y_i \cdot y_j]=\det [y_i \cdot y_j]$.
one has directly from \DtoNbis
\eqn\DtoNlamb{\eqalign{
\prod_{i=1}^{N-1} d^Dy_i\ &=  \prod_{\alpha =1}^{N-1} d^D\lambda_\alpha \cr
&\equiv \ \prod_{\alpha =1}^{N-1} d^{N-1}\lambda_\alpha
{S_DS_{D-1}\ldots S_{D-N+2} \over S_{N-1}S_{N-2}\ldots S_1}
\bigg[
\det [\lambda_\alpha \cdot \lambda_\beta]_{1\le \alpha,\beta \le N-1}
\bigg]^{{D-N+1\over 2}} \ .\cr
}
}
This also means that one can replace in \measuone\ the integration
over the matrix
elements $u_{ij}=y_i\cdot y_j$ by an integration over matrix elements
$u_{\alpha\beta}=\lambda_\alpha\cdot\lambda_\beta$ associated with an arbitrary
tree.

\noindent{\it $\diamond$ Expression for the interaction}

We now derive the expression of the determinant $P_N(a^\nu)$
which enters the interaction factor in terms of the $\lambda_\alpha$'s.
Equation \ZN\ was actually a particular representation of the
interaction, associated with a particular choice of a tree, namely
the star centered at $x_1$ and lines pointing toward the other points.
This can be seen in our choice $\kvec_1=-\sum_{i=2}^{N}\kvec_i$
to account for the $\delta^d(\sum\limits_i \kvec_i )$ constraint in the
momentum integral
\znter\ .
We can generalize this construction to an arbitrary oriented tree ${\bf T}$
by writing $\kvec_i$ as
\eqn\qalpha{
\kvec_i = -\sum_{\alpha =1}^{N-1} \epsilon_{i\alpha }\ \qvec_\alpha
\ .}
These vectors $\qvec_\alpha $ can be seen as flowing along the lines of
the tree while the vectors $\kvec_i$ can be thought of as being injected
at the nodes of the tree.
Equation \qalpha\ expresses the momentum conservation at the nodes and
moreover, together with \propeps , ensures $\sum\limits_i \kvec_i ={\vec 0}$
for any set of $\qvec_\alpha$'s.
Using then
\eqn\ktoq{
\prod_{i=1}^{N} d^d\kvec_i\, \delta^d(\sum_{i=1}^N \kvec_i)\ =\
\prod_{\alpha =1}^{N-1}d^d\qvec_\alpha
\ ,}
we get for the interaction term \znter\
\eqn\znq{\eqalign{
\CZ_N& ={\cal V}\int \prod_{\alpha =1}^{N-1} {d^D\lambda_\alpha
d^d\qvec_\alpha \over (2\pi)^d}
\exp\left[ -{1\over 2}\sum_{\alpha ,\beta =1}^{N-1}
 \qvec_\alpha \cdot \qvec_\beta \,
\Pi^{{\bf T}}_{\alpha
\beta }\right]\cr
& =\ (2\pi)^{-{d\over2}(N-1)}{\cal V}\int\prod_{\alpha =1}^{N-1} d^D
\lambda_\alpha
\,\left(\det\left[ \Pi^{{\bf T}}_{\alpha \beta }\right]_{\scriptscriptstyle
1\le \alpha
,\beta \le N-1}
\right)^{-{d\over 2}}\ ,\cr
}}
where we take advantage of \DtoNlamb\ and define a new matrix $\Pi^{{\bf T}}$
attached
to the tree ${\bf T}$:
\eqn\PiT{
\Pi^{{\bf T}}_{\alpha \beta}= \ \sum_{i,j =1}^{N} \epsilon_{i\alpha }
G(x_i,x_j) \epsilon_{j\beta }
\ .}
Indeed $\det [ \Pi_{\alpha\beta}^{{\bf T}} ]$ is independent of the choice of
the tree
${\bf T}$.

In terms of pairs of oriented lines $\alpha, \beta $ of the tree, with
extremities $(i_\alpha ,{i'}_\alpha )$ and $(i_\beta ,{i'}_\beta )$
respectively,
the matrix element $\Pi^{{\bf T}}_{\alpha \beta }$ is associated with the
quadrilateral $(i_\alpha , {i'}_\alpha ; i_\beta ,{i'}_\beta)$
\eqn\Piquad{
\Pi^{{\bf T}}_{\alpha \beta }= G(x_{i_\alpha }, x_{i_\beta } )
+ G(x_{{i'}_\alpha }, x_{{i'}_\beta }) - G(x_{i_\alpha }, x_{{i'}_\beta })
- G(x_{{i'}_\alpha }, x_{i_\beta })\ .}
It can be viewed as an interaction potential between two dipoles
$\lambda_\alpha $ and $\lambda_\beta $ and has the following pictorial
representation:\par
\vbox{
\vskip 1.5truecm
%\vbox{\special{psfile=quadrilatere.ps}}
\eqn\Pifig{
\Pi^{{\bf T}}_{\alpha \beta}\ =\hskip 6.truecm \
}
\vskip 1.5truecm
}
\noindent {\it $\diamond$ Expression for correlation functions}

For correlation functions $\CZ^{(M)}(X_a,\kvec_a)$ \zvertex\
one can generalize the above construction simply
{\it (i)} by considering
the spanning star tree ${\bf T}_{\rm ex}$ with line vectors $\Lambda_a=X_a
-X_1$ ($a>1$) for the external
points, {\it (ii)} by choosing an arbitrary tree ${\bf T}_{\rm in}$
with line vectors
$\lambda_\alpha$ for the internal points, and {\it (iii)}
by attaching these two trees
by a line vector $\Lambda_1$ joining the external point $X_1$
to an arbitrary internal point.
In this way, we obtain a larger tree ${\bf T}$
to which we can associate a generalized form of \ZvertexN\ :

\eqn\ZvertexT{\eqalign{
\CZ_N^{(M)}(X_a,\kvec_a)\ =
\ (2\pi)^{-{d\over2}(N-1)}\,\int d^D\Lambda_1\prod_{\alpha =1}^{N-1}
d^D\lambda_\alpha
&\,\left(\det\left[ \Pi^{{\bf T}}_{\alpha \beta }\right]_{\scriptscriptstyle
1\le
\alpha ,\beta \le N-1}
\right)^{-{d\over 2}}\cr
&\quad\times\exp \left( -{1\over2}\sum_{a,b=1}^{M}\,\kvec_a \cdot \kvec_b\,
\Delta_{ab} \right)
\cr}}
\eqn\DeltamaT{
\Delta_{ab}\,=\,{\det\nolimits_N
\left(\matrix{\Pi^{{\bf T}}_{ab}&\Pi^{{\bf T}}_{a\beta }\cr\Pi^{{\bf
T}}_{\alpha b}&
\Pi^{{\bf T}}_{\alpha \beta }\cr}\right)
\over
\det\nolimits_{N-1} (\Pi^{{\bf T}}_{\alpha \beta })
}\ .}
As discussed above the determinants in \DeltamaT\ are independent of the
tree ${\bf T}$ chosen.
In \ZvertexT ,
the integral over the $\lambda_\alpha$'s and $\Lambda_1$ has to be understood,
for real $D$, as
\eqn\dlambter{\eqalign{
\int d^D\Lambda_1\prod_{\alpha=1}^{N-1} d^D\lambda_\alpha\ &=\
\int d^{M+N-1}\Lambda_1\prod_{\alpha=1}^{N-1} d^{M+N-1}\lambda_\alpha\
\  {S_{D-M+1}\ldots S_{D-M-N+2}\over S_{N}\ldots S_1}\,
\cr &\qquad \times \left\{ {
 \left[ \det_{N+M-1}
  \left(\matrix{\Lambda_a\!\cdot\!\Lambda_b&
 \Lambda_a\!\cdot\!\lambda_\beta\cr
 \lambda_\alpha\!\cdot\!\Lambda_b&\lambda_\alpha\!\cdot\!\lambda_\beta\cr}
 \right)
 \right]
\over
 \left[ \det_{M-1}\left(\Lambda_a\!\cdot\!\Lambda_b\right)
 _{2\le a,b\le M}\right]
}\right\}^{D-M-N+1 \over 2}
\cr}}
and \ZvertexT\ is a function of the invariants $a_{ab}=(X_a-X_b)^2$, which
are quadratic forms in terms of the line vectors $\Lambda_a$.
\subsec{
The limit ${\tit D=}{\bf 1}$ and the Schwinger representation
}
As an example, for a manifold with internal dimension $D=1$, one can recover
the standard Schwinger representation\foot{In the context of polymers,
it is also known as the Fixman representation \Fixman\ .} of an interacting
field theory
with interaction term $({\bf\Phi})^2 (\vec 0)$
(see subsection 6.1 for further details),
here in direct correspondence with the continuous Edwards-like model for a
polymer interacting with a single fixed point at the origin.
Choosing $D=1$ and $k=2$ in \Ham\ corresponding to the Gaussian weight of
a Brownian chain, one has $\nu={1\over 2}$ and the propagator along the chain
\eqn\propone{G(x,y)=-{1\over 2}|x-y|\ .}
Furthermore, for the perturbative order $N$, the measure term \measaij\
reconstructs in the limit $D=1$ (like in \distter\ ) the measure over
all relative distances
of $N$ ordered points along the chain, as well as all their permutations.
For a given permutation $x_{i_1}\le \ldots \le x_{i_N}$, the measure term is
simply
\eqn\measuuu{\prod_{\alpha =1}^{N-1}da^{1\over 2}_{i_\alpha i_{\alpha+1}}
\ .}
Choosing as a particular tree ${\bf T}$ the successive oriented links
$(i_\alpha,{i'}_{\alpha })=
(i_\alpha,i_{\alpha+1})$ the matrix $\Pi^{{\bf T}}$ \Piquad\ is diagonal
\eqn\PiPol{\Pi_{\alpha\beta}^{{\bf T}}\ =\ s_\alpha\,\delta_{\alpha\beta}
\quad{\rm with}\quad s_\alpha\,=\, a_{i_\alpha i_{\alpha+1}}^{1\over 2}
\,=\,x_{i_{\alpha+1}}-x_{i_\alpha}
\ .}
The $s_\alpha$ are nothing but the usual Schwinger parameters (proper
time) for the propagator lines $\alpha$, or in polymer theory the lengths
of the successive polymer segments.
The interaction gives for the partition function a term of the form
\eqn\ZnPol{
\CZ_N\ = \ \int \prod_{\alpha =1}^{N-1} ds_\alpha\,
(\ldots ) \Big(\prod_{\alpha=1}^{N-1} s_\alpha \Big)^{-{d\over 2}}
}
which is nothing but the Schwinger representation for the ``daisy" diagram
in $d$ dimensions.
\midinsert
\figinsert{5.truecm}{\hsize}{
\figcap\daisyf{The daisy diagram corresponding to the term
\ZnPol\ .}
}
\endinsert

\newsec{Ultraviolet and infrared properties of the integrand}
\subsec{Existence and positiveness of the integrand}
The rules that we have proposed above for defining the perturbative
expansion of the model in non-integer dimension $D$ remain formal.
Indeed, we have not shown yet that the integrands do exist and that
the integrals are convergent (for $D$ large enough), and define an analytic
function in $D$.
Let us concentrate on the $N$-th term for the partition function, $\CZ_N$,
which is explicited by the integral \ZNa\ in terms of distance
variables $a_{ij}$,
by the integral \ZNcart\ in terms of Cartesian coordinates in $\RR^{N-1}$
or by the integral \znq\ in terms of tree-variables $\lambda_\alpha$.
We shall furthermore assume in the following sections that $D\ge N-1$,
that is $D$ large enough for $\mu^{(D)}_N$ to be a measure density
(similarly, for $\CZ_N^{(M)}$, we shall assume $D\ge N+M-1$).
We shall discuss in section 9 how our results can be extended to smaller $D$.

\smallskip
\noindent{\it $\diamond$ Schoenberg's theorem}
\smallskip

First, in view of the formula \ZNa , the positiveness of the Cayley-Menger
determinant $P_N(a^\nu)$ \CayMengnu\ has to be ensured inside the domain of
integration ${\cal A}_N$ for the variables $a_{ij}$.
For $0<\nu \le 1$, this actually is just a consequence
of a remarkable theorem in distance geometry due to I.J. Schoenberg
\ref\Schoen{I. J. Schoenberg, Ann. Math. {\bf 38} (1937) 787.}.
\vskip .5 true cm
\noindent{THEOREM I}
{\it
If we change the metric of the Euclidean space $\RR^m$ from the Euclidean
distance $d(x,y)=|x-y|$ to the new distance}
\eqn\dtodnu{
{\tilde d}(x,y)=(d(x,y))^\nu \qquad \qquad
0<\nu \le 1
}
{\it
the new metric space $\RR_{(\nu )}^m$ thus arising may be embedded
isometrically in the Hilbert space $\RR^{\infty }$ with the $L^2$-norm.}

\noindent A practical (equivalent) statement is that any set of $N$ distinct
points of $\RR_{(\nu )}^m$ can be embedded in the Euclidean space $\RR^{N-1}$.
In our language, this means that, if the $a_{ij}$ are actual squared
distances of $N$ points in $\RR^{N-1}$, then $a_{ij}^{\nu }$ with $0<\nu \le 1$
can also be realized as actual squared distances between $N$ transformed
points in $\RR^{N-1}$. An immediate consequence is that $P_N(a^\nu )\ge 0 $,
as well as all the lower rank polynomials $P_K(a^\nu )\ge 0$.
\vskip .5 true cm
We moreover have the useful refined result for $0<\nu < 1$ \Schoen :
\vskip .5 true cm
\noindent{THEOREM II}
{\it
If $x_1,\ldots , x_N$ are $N$ distinct points in $\RR^{m}$, and $[a_{ij}]$
the corresponding squared distance matrix, the matrix
$D_{ij}(a^\nu )={1\over 2}(a_{i1}^\nu + a_{j1}^\nu - a_{ij}^\nu )$,
($0<\nu <1$), is positive definite.}
\vskip .5 true cm
\noindent The positiveness is a consequence of the previous theorem.
The novelty here concerns the definiteness and
states that the determinant $P_N(a^\nu )$ vanishes if and only if
two points at least coincide, that is $a_{ij}=0$ for some $i\neq j$.
Notice that this property does not hold for the case $\nu =1$ for which we
already know that $P_N(a)$ vanishes as soon the $a_{ij}$ can be realized
as distances between $N$ points in $\RR^{K}$ for $K\le N-2$, which can
be obtained with
none of the $a_{ij}$ ($i\neq j$) vanishing.

\subsec{Short distance divergences}
The above result ensures that for $0<\nu<1$ the only possible divergences
in \ZNa\ occur when some distances $a_{ij}$ go to $0$ (UV divergences)
or $\infty$ (IR divergences).
Let us first discuss the UV behavior.

If one scales the distances by a global factor $\rho$:
\eqn\arho{a_{ij}\longrightarrow \rho^2 a_{ij}\ ,}
the measure term in \ZNa\ is scaled according to
\eqn\measrho{
\prod_{1\le i<j\le N}da_{ij}[P_N(a)]^{{D-N\over 2}} \longrightarrow
\rho^{D(N-1)}\,\prod_{1\le i<j\le N}da_{ij}[P_N(a)]^{{D-N\over 2}}
}
while the interaction term scales as
\eqn\interho{
[P_N(a^\nu)]^{-{d\over 2}}\longrightarrow \rho^{-(N-1)\nu d}
[P_N(a^\nu)]^{-{d\over 2}}
\ .}
We therefore obtain a global scaling factor $\rho^{(N-1)(D-\nu d)}
=\rho^{(N-1)\epsilon}$.
This means that the contribution to $\CZ_N$ of the region of ${\cal A}_N$
such that all distances $a_{ij}\le \rho^2$ is of order $\rho^{(N-1)\epsilon}$,
indicating that
$\CZ_N$ is superficially UV convergent
for $\epsilon >0$, but divergent for $\epsilon\le 0$.

Similarly, we expect that when the distances between some subset of $P$
points are $\le\rho^2$, we get a contribution of order $\rho^{(P-1)\epsilon}$
to $\CZ_N$.
This is indeed what occurs,
due to the following crucial
factorization property of the interaction term.

\medskip
\noindent{THEOREM }{\it Short distance factorization of the interaction term}

Consider the subset ${\cal P}$ of (for instance)
the first $P$ interacting points (considered
as embedded in $\RR^{N-1}$)
$x_1,\ldots ,x_P$ and let us contract it toward one of its points,
which we choose to be $x_1$. We set
\eqn\xrho{
x_k(\rho )\,=\,\left\{\matrix{x_1+\rho (x_k-x_1)\hfil\quad &{\rm if}\ 1\le k\le
P \cr
x_k\hfil\quad &{\rm if}\ P<k\le N \ .\cr}\right.
 }
\midinsert
\figinsert{11.truecm}{7.truecm}{%\hsize}{
\figcap\contf{Schematic picture of the short-distance factorization
of the interaction term relative to some set $\CG$ of $N$ interaction points
(here $N=10$).
When the points of a subset $\CP$ of $\CG$ are
contracted toward one of its point $x_1$, the interaction term factorizes
into the product of the interaction term relative to $\CP$ and the interaction
term relative to ${\overline{\CP}}=(\CG\setminus\CP)\cup\{x_1\}$.}
}
\endinsert
This defines a mapping in distance variables
\eqn\aijrho{
a_{ij}(\rho)\,=\,\left\{\matrix{
\rho^2 a_{ij}\hfill\quad&{\rm if}\ 1\le i\le j\le P\hfill\cr
a_{1j}-\rho(a_{1i}+a_{1j}-a_{ij})+\rho^2(a_{1i})\hfil\quad&{\rm if}\ 1\le i
\le P<j\le N\cr
a_{ij}\hfill\quad&{\rm if}\ P<i\le j\le N\hfill\ .\cr}\right.
}
Then, in the limit $\rho\to 0$, the determinant of the matrix
$D_{ij}(a^\nu)$ \Dijanu\ factorizes as
\eqn\factdet{
\eqalign{
\det\nolimits_{N-1}[D_{ij}\left( a^\nu(\rho) \right) ]\ &=\
\rho^{2\nu (P-1)}\,
\det\nolimits_{P-1}[D_{ij}(a^\nu)]_{2\le i,j \le P}\,\cr &\quad\times
\det\nolimits_{N-P}[D_{ij}(a^\nu)]_{P+1\le i,j \le N}
\left\{1+{\cal O}(\rho^{2\delta})\right\}\cr
}}
with
\eqn\defdelt{\delta=\min(\nu, 1-\nu)>0 \ .}

\medskip\noindent{\it Proof:}\medskip
The matrix $D_{ij}$ transforms under a contraction according to
\eqn\Dijrho{
D_{ij}(a^\nu(\rho))\,=\,\left\{\matrix{
\rho^{2\nu}  D_{ij}(a^\nu) \hfill\quad&{\rm if}\ 1\le i\le j\le P\hfill\cr
&\cr
{1\over 2}\Big\{\rho^{2\nu} a_{1i}^\nu + a_{1j}^\nu\hfill &\cr
\hfill \ -[a_{1j}-\rho(a_{1i}+a_{1j}-a_{ij})+\rho^2(a_{1i})]^\nu \big\}
\hfill\quad&{\rm if}\ 1\le i
\le P<j\le N\hfill\cr
&\cr
D_{ij}(a^\nu)\hfill\quad&{\rm if}\ P<i\le j\le N\hfill\ .\cr}\right.
}
For small $\rho$, the mixed term $D_{ij}$ $i\le P < j$ has the expansion
\eqn\expDij{
\eqalign{
D_{ij}(a^\nu (\rho))
&=\rho^{2\nu }a_{1i}^\nu + \rho\, \nu
a_{1j}^{\nu -1}(a_{1i}+a_{1j}-a_{ij})+{\cal O}(\rho^2)
\cr
&=\rho^\nu{\cal O}(\rho^\delta)
\cr}
}
since the leading term is $\propto \rho^{2\nu }$ or $\propto \rho $, depending
on whether $\nu $ is greater or less than $1/2$.
Thus we can write the matrix $D_{ij}(a^\nu(\rho))$ in blocks associated
respectively with the subset ${\cal P}$ and
${\cal \overline{P}}=\{x_1\}\cup \{x_{P+1}, \ldots, x_N\}$
\eqn\factmat{
D(a^\nu(\rho))\ =\ \bordermatrix{&{\scriptscriptstyle 2,\ldots\ldots, P}&
{\scriptscriptstyle P+1, \dots, N}\cr
&\rho^{2\nu}D_{\cal P}(a^\nu)&\rho^\nu{\cal O}(\rho^\delta)\cr
&\rho^\nu{\cal O}(\rho^\delta)&D_{\cal \overline{P}}(a^\nu)\cr}
\ .}
Hence
\foot{This follows for instance from
$\det\left(\matrix{A&B\cr B^t&C}\right)=\det(A)\det(B)\det(1-A^{-1}BC^{-1}B^t)$
for invertible matrices $A$ and $C$.}
$\det (D(a^\nu(\rho)))=\rho^{2\nu(P-1)}\left[\det(D_{\cal P}(a^\nu))
\det(D_{\cal\overline{P}}(a^\nu))+{\cal O}(\rho^{2\delta})\right]$.
Furthermore, from Schoenberg's theorem, if $\det(D_{\cal P}(a^\nu))$ or
$\det(D_{\cal\overline P}(a^\nu))$ vanishes, some subset of points $x_k(\rho)$
coincide for any $\rho$ and therefore $\det(D(a^\nu(\rho)))$ also vanishes.
The equivalence in Eq. \factdet\
and the theorem follow.

\medskip
The consequences of this theorem are twofold.
First, as expected, when a subset ${\cal P}$
of $P$ points coalesce to a single point
$p$, this gives a divergence in $\CZ_N$, as well as in any correlation function
$\CZ_N^{(M)}$, since from \ZvertexN\ the same interaction determinant
$(\det(\Pi))^{-{d\over 2}}$ is present.
Second, this divergence is formally equal to the global
divergence of the partition function
amplitude $\CZ_P$ for the $P$ contracted points times the amplitude obtained
by replacing those points by the single contraction point $p$,
$\CZ_{N-P+1}$.
This is a key point for ensuring renormalizability, since this shows
that short-distance divergences can be absorbed into an effective interaction
term, thanks to a short-distance operator product expansion for ``interaction
operators"
\eqn\OPE{
\prod_{i\in {\cal P}}\delta^d(\rvec(x_i))\ \buildrel {\forall i,\ x_i\to x_p}
\over
\sim  \ |{\rm size}({\cal P})|^{-d\nu (P-1)}
\,\delta^d(\rvec(x_p))
\ ,}
where ${\rm size}({\cal P})$ is a ``typical distance" between the
points $x_i$ of ${\cal P}$ in $D$-dimensional space (which depends on the
precise way the limit $x_i\to x_p$ is taken).

It is the purpose of the next sections to give a precise meaning to these
assertions, to provide rigorous arguments, and to discuss their consequences
for the physics of the model.
\medskip
One can regularize those short-distance divergences and make the integrals
\ZNa , \ZvertexT\ UV-finite by changing the short-distance behavior of the
propagator $G(x,y)$. However, it is both convenient and natural to use
dimensional regularization, that is to consider the amplitudes as analytic
functions of the parameters $D$ (the dimension of internal space), $\nu$
(the scaling dimension of the field $\rvec$), and $\epsilon$ (the scaling
dimension of the interaction).
As we shall argue below, for fixed $D$ and $0<\nu<1$, the amplitudes are
expected to be UV-finite, and therefore analytic functions of $\epsilon$
in the half-plane ${\rm Re}(\epsilon)>0$.
Because of the short-distance behavior of its integrand, $\CZ_N$ will
exhibit  poles at $\epsilon=0$.
For instance, the singular contribution to the integral
\ZNa\ arising from the integration over the global dilation parameter
of the $N$-interaction point set gives a single pole $\propto 1/\epsilon$.
More generally, we expect that
multiple poles in $1/\epsilon^{k}$ ($1\le k \le N-1$) will
occur at $\epsilon=0$, corresponding to
the dominant singularities appearing when $k$ successive
subsets of interaction points coalesce \BD .
Apart from these poles at $\epsilon=0$,
subdominant divergences will be shown to give poles in the
$\epsilon$ plane for ${\rm Re}(\epsilon)\le -\delta/(N-1)$.
In field theory,
the factorization property of the integrand under partial
contractions of subdiagrams determines
the pole structure of the resulting Feynman amplitude and is the
key point that ensures renormalizability.
Here, although the interacting manifold model is not mapped
onto a standard field theory,
a similar pole structure of $\CZ_N$ will be found,
due to the
factorization property of the interaction term that we just discussed.

\subsec{IR regularization}
By similar power counting arguments ({\it i.e.} dimensional analysis),
it is expected that the integrals
will diverge for large distances $a_{ij}\to\infty$ (when ${\rm Re}(\epsilon )
\ge 0$).
As usual in field theory, we shall deal with this problem by introducing
an infra-red regulator, and by showing that such a regulator does not change
the short-distance properties and the renormalization of the model.

The simplest kind of regulator is to work in a finite $D$-dimensional space,
{\it i.e.} to consider a ``membrane" of finite size.
This is in fact what is usually done for the continuous polymer Edwards model.
Indeed, the polymer is taken to have a finite total ``length" $S$, which
amounts to constrain the length variables $s_\alpha$ in \ZnPol\
by a measure term \break
$(S-\sum\limits_\alpha s_\alpha)\, \theta(S-\sum\limits_\alpha s_\alpha)$.

In our case, our formulation of the model in non-integer dimension relies on
the invariance of the observables under Euclidean motions in $\RR^D$.
A simple way to keep a similar symmetry over a finite manifold is to start
from the $D$-dimensional hypersphere ${\cal S}_D$ with radius $R$ and volume
${\cal V}_{{\cal S}_D}= S_{D+1}R^D$, so that the group of
invariance is now ${\rm SO}(D+1)$.
One can easily generalize the concept of distance geometry on ${\cal S}_D$,
and its analytic continuation for non-integer $D$.
Indeed, we can embed the sphere into $\RR^{D+1}$, and write
the integral of a ${\rm SO}(D+1)$ invariant function of $N$ variables
as an integral over scalar products
$u_{ij}=x_i\cdot x_j$:
\eqn\intuS{
\int \prod_{i=1}^N d^{D+1}x_i \,\delta(|x_i|-R) f(u_{ij})\ =\
\int_{{\cal U}_N(R)}\prod_{i<j}du_{ij}\, \sigma_N^{(D)}([u_{ij}],R)\,
f([u_{ij}])
\ ,}
with $u_{ij}=R^2$ if $i=j$, and the measure
\eqn\measuS{
\sigma_N^{(D)}([u_{ij}],R)\ =\ S_{D+1}\ldots S_{D-N+2}\, R^N\,
\left(\det\nolimits_N [u_{ij}
]\right)^{D-N\over 2}
\ ,}
${\cal U}_N(R)$ being the domain of $u_{ij}$ ($i<j$)
where the matrix $[u_{ij}]$ is positive with all the $u_{ii}$ set equal
to $R^2$.
Equivalently we can express the integral \intuS\ in terms of squared distances
$a_{ij}=2(R^2-u_{ij})$ {\it in $D+1$-dimensional space} (this defines
the so-called cord distance on ${\cal S}_D$ which differs from the
geodesic distance):
\eqn\intaS{
\int \prod_{i=1}^N d^{D+1}x_i \,\delta(|x_i|-R) f(a_{ij})\ =\
{\cal V}_{{\cal S}_D}
\int_{{\cal A}_N(R)}\prod_{i<j} da_{ij}\,\mu_N^{(D)}([a_{ij}],R)\,
f([a_{ij}])
\ ,}
with the measure
\eqn\measaS{
\mu_N^{(D)}([a_{ij}],R)=2^{-{N(N-1)\over 2}}\, S_{D}\ldots S_{D-N+2}\,
\left({1\over R^2}\det\nolimits_N [R^2-{1\over 2}a_{ij}]\right)^{{D-N}\over 2}
}
and ${\cal A}_N(R)$ the domain of $a_{ij}$ where the matrix $[R^2-{1\over 2}
a_{ij}]$
is positive. In particular, the positiveness of the $2\times 2$ minors
ensures for any two points the diameter inequality  $a_{ij}\le 4R^2$.
Hence, ${\cal A}_N(R)$ is a bounded subset of $\RR^{N(N-1)\over 2}$.
\medskip
One can check the identity:
\eqn\detSR{\det\nolimits_N([R^2-{1\over 2}a_{ij}])= R^2\det\nolimits_{N-1}
([D_{ij}(a)])+\det\nolimits_N ([-{1\over 2}a_{ij}])}
where $D_{ij}(a)$ is defined in \xtoa\ (indeed the $N-2$ highest degree
terms in the polynomial expansion in $R^2$ of the l.h.s of \detSR\
vanish identically !).
This implies that in the thermodynamic limit $R\to\infty$ one recovers the
measure \measaij\ in Euclidean (infinite flat) space.
Conversely, for a finite $R$,
formula \detSR\ shows that, at short-distances, the measure is dominated
by the first term of the r.h.s, {\it i.e.} the Euclidean one, while the
second term, which is one degree higher in $a_{ij}$,
becomes relevant for distances of order $R$ only, hence providing
an IR regulator.

It remains to write the expression for the interaction term.
In fact, the latter is the same as in \ZN , with the
matrix $\Pi_{ij}$ \Pimatrix ,
or more generally the tree matrix $\Pi^{{\bf T}}_{\alpha\beta}$ \PiT ,
involving the massless propagator
$G(x,y)=\big[(-\Delta)^{k\over 2}\big]^{-1}(x,y)$
now on ${\cal S}_D$.
There is however no general simple analytic expression for
$G(x_i,x_j)$ as a function of
the distance variable $a_{ij}$ defined above for general $D$ and $k$.
For definiteness, another simple possibility then consists in keeping a
propagator on the sphere
of the form
\msqext
\eqn\propS{
-\,G(x_i,x_j)
\ =\ {1 \over 4^\nu (4\pi)^{D/2}}{\Gamma(1-\nu)\over\nu\Gamma(
\nu+{D \over 2 })}\,|a_{ij}|^{\nu}
\ .}
This amounts to modify the ``elastic" term of the Hamiltonian \Ham\ by
finite volume corrections
\eqn\fvcorr{
\rvec (x) (-\Delta)^{k\over 2}\rvec(x)\,\to\,
\rvec (x) \left[ (-\Delta)^{k\over 2}+{\rm cst} R^{-2}
(-\Delta)^{k-2\over 2})+{\rm cst} R^{-4}
(-\Delta)^{k-4\over 2})+\ldots\right]\rvec(x)
}
which change its large distance behavior (IR regulator), but not its
short-distance behavior. In particular, Schoenberg's theorem II, which
is readily satisfied by the propagator $G$ given by \propS , is expected
to remain valid for the exact massless propagator on the sphere.
The corrections in \fvcorr\ vanish in the limit $R\to\infty$.
In the following, we will keep in mind that the model is defined with
the measure \measaS\ and the propagator \propS\ . However, since we shall
be concerned with the UV renormalization of the model, we shall use
formally the simpler Euclidean ($R\to \infty$) limit \measaij\ of \measaS\ .
As discussed above, they actually share the same short-distance properties.

\newsec{Absolute convergence for {\tentit\char'017}$\ {\tit =D\ -}$
{\tentit\char'027} ${\tit d\, >}\ {\bf 0}$}

In this section, we want to prove that:
\vskip .5 true cm
\noindent{THEOREM I}
{\it
For $\epsilon >0$ ({\it i.e.} $d<d^\star$),
the integrals $\CZ_N$ and $\CZ_N^{(M)}$ are absolutely (UV) convergent.}
\vskip .5true cm
As in field theory, this actually is
a consequence of (i) the superficial convergence of $\CZ_P$ for any $P\le N$
and (ii) the basic factorization property \factdet , and generalizations
thereof. Since the formalism developed above can be thought of as a natural
generalization of the Schwinger representation of Feynman integrals, the
proof of absolute convergence will be inspired by the standard method
based on decomposition into Hepp sectors \ref\IZ{C. Itzykson and J.B. Zuber,
{\sl Quantum Field Theory}, McGraw-Hill, New York (1985).}.
As discussed just above, we shall always assume the (implicit) presence
of an IR regulator.

\subsec{Generalized Hepp sectors}
\midinsert
\figinsert{14.truecm}{\hsize}{
\figcap\heppfig{(a) An example of construction of the ordered tree ${\bf T}=
(\lambda_1, \lambda_2,\lambda_3,\lambda_4)$ for a set of interaction
points with $|\lambda_1 |\le |\lambda_2 |\le |\lambda_3 |\le |\lambda_4 |$.
This tree defines the generalized Hepp sector $\CH^{\bf T}$
to which this set of points belongs. (b) Moving the point $x_2$ toward
the point $x_1$ results in a change of generalized Hepp sector.}
}
\endinsert
We start with formula \ZNcart\ and partition the domain of integration
for the $y_i$'s into generalized Hepp sectors as follows.
Let us consider
the $N$ points in $\RR^{N-1}$ with Cartesian coordinates
$0,y_1,\ldots,y_{N-1}$.
We first singularize the pair of points having the minimum mutual
distance, and define $\lambda_1$ as the vector in $\RR^{N-1}$ joining
these two points, with an arbitrary orientation. We define $\lambda_2$
in a similar way, as the vector associated with the minimal distance
among all the remaining mutual distances. $\lambda_2$ can (i) either
share one of its extremities with $\lambda_1$, or (ii) be disjoint. At the
next step, we define $\lambda_3$ as the vector associated with the
minimal distance among all the remaining ones {\it and} such that
$(\lambda_1,\lambda_2,\lambda_3)$ do not form a closed loop (this may
occur only in case (i)). We iterate this construction, by requiring
at each step that no loop ever appears, up to the emergence of the last vector
$\lambda_{N-1}$. We thus have constructed an oriented {\it ordered}
tree ${\bf T}$ with line vectors $(\lambda_1,
\ldots,\lambda_{N-1})$, which spans the $N$ points and is such that
\eqn\ineqmin{|\lambda_1|\le |\lambda_2|\le \ldots \le |\lambda_{N-1}|\ .}
We shall denote ${\bf T}=(\lambda_1,\ldots,\lambda_{N-1})$ although
the tree ${\bf T}$ is not strictly speaking characterized by the line
vectors $\lambda_\alpha$ but only by the incidence matrix $\epsilon_{i\alpha}$
of the linear transformation from the $x_i$ (or $y_i$) to the $\lambda_\alpha$.
With any ordered tree ${\bf T}$, we can therefore associate the Hepp sector
${\cal H}^{{\bf T}}$ defined as the domain of the $y_i$'s in $\RR^{N-1}$
leading after this construction to this ordered tree ${\bf T}$, regardless of
its orientation. It is clear that $\RR^{N-1}=\bigcup\limits_{{\bf T}}\CH^{{\bf
T}}$.
\medskip
In a given sector ${\cal H}^{{\bf T}}$, we make a change of variables from the
$y_i$'s to the $\lambda_\alpha$'s associated with the ordered tree ${\bf T}$
(with
an arbitrary choice of orientation)
and, in particular, use $\Pi^{{\bf T}}_{\alpha \beta}$ to evaluate the
interaction
term. We parametrize the $\lambda_\alpha $ by their spherical coordinates in
$\RR^{N-1}$, namely by their modules $|\lambda_\alpha |$ and relative angles
$\theta_{\alpha,1},\ldots \theta_{\alpha,\alpha-1}$ as in \spheric\ and
\thetzero\ .
The variables $|\lambda_\alpha|$ will play the role of the Schwinger parameters
$s_\alpha$ in field theory. Since $|\lambda_1|\le |\lambda_2|\le \ldots
\le |\lambda_{N-1}|$, it is natural to rewrite the $|\lambda |$'s as
\eqn\lamtobeta{\eqalign{
|\lambda_1|\ &=\beta_1\beta_2\ldots \beta_{N-1}\cr
|\lambda_2|\ &=\beta_2\ldots \beta_{N-1}\cr
&\vdots \cr
|\lambda_{N-1}|&=\beta_{N-1}\cr
}}
with $0\le \beta_\alpha \le 1$ for $1\le \alpha \le N-2$ and $0\le
\beta_{N-1}<\infty$ (in the Euclidean version of the problem, thus
without IR regulator). The domain of integration ${\cal D}^{{\bf T}}$ for the
$\beta$ and $\theta$
variables which reconstructs the domain ${\cal H}^{{\bf T}}$ for the $y_i$'s in
$\RR^{N-1}$,
depends on the topology of the ordered tree. For instance, the value
$\beta_\alpha =1$ can in general be reached inside the sector only for
some domain of the angle $\theta$ between $\lambda_\alpha$ and
$\lambda_{\alpha+1}$. Still, the domain ${\cal D}^{{\bf T}}$
has the following general structure:
\eqn\domainT{\matrix{
0 \le \theta_{\alpha,n} \le \pi \hfill \quad & 1\le n < \alpha \le N-1
\hfill \cr
&&\cr
\beta_\alpha^{\rm min}({\bf T}\ ;\beta_{\gamma : \gamma < \alpha} \,
;\theta\hbox{ 's})
\le \beta_\alpha \le
\beta_\alpha^{\rm max}({\bf T}\ ;\beta_{\gamma : \gamma < \alpha} \,
;\theta\hbox{ 's})
\hfill \quad & 1\le \alpha \le N-2 \hfill \cr
&&\cr
0 \le \beta_{N-1} \hfill \quad & \hfill \cr
}}
where
$\beta_\alpha^{\rm min}({\bf T}\ ;\beta\hbox{ 's},\theta\hbox{ 's})$
and
$\beta_\alpha^{\rm max}({\bf T}\ ;\beta\hbox{ 's},\theta\hbox{ 's})$
are
(positive and possibly vanishing) functions of the $\theta$'s and of
the $\beta_\gamma$'s for $\gamma < \alpha$. The inequality $\beta_\alpha^{\rm
min
} > \beta_\alpha^{\rm max}$ for some $\theta$'s and
$\beta_{\gamma:\gamma<\alpha}$ would indicate that such a partial configuration
of $\theta$'s and
$\beta_\gamma$ always lies outside
the given sector.
The only important properties of ${\cal D}^{{\bf T}}$ that we shall use are:

\noindent (i) ${\cal D}^{{\bf T}}$ is by definition bounded, if one excepts the
variable
$\beta_{N-1}$, since $\beta_\alpha^{\rm max}({\bf T};\beta{\rm 's};\theta{\rm
's})\le 1$ by construction. The variable $\beta_{N-1}$ itself stays bounded
due to the implicit presence of an IR regulator.

\noindent (ii) $\det([\Pi^{{\bf T}}_{\alpha, \beta }])$, when expressed in
terms
of the $\beta$'s and the $\theta$'s, is a continuous function
of these variables  and vanishes in ${\cal D}^{{\bf T}}$ if and only
if one at least of the $\beta$'s vanishes. Indeed, from Schoenberg's theorem,
$\det([\Pi^{{\bf T}}_{\alpha,\beta}])=0$ iff two points coincide, that is if
their
mutual distance is zero. Since this distance is by construction larger than
or equal to $|\lambda_1|$ in the sector, this implies $|\lambda_1|=0$,
or equivalently
$\beta_1\beta_2\ldots \beta_{N-1}=0$.

\subsec{Absolute convergence}

It is enough to prove the absolute convergence in each Hepp sector ${\cal
H}^{{\bf T}}$.
Omitting global factors in \ZNcart\ we consider the integral:
\eqn\conve{\eqalign{
\int_{{\cal H}^{{\bf T}}} \prod_{i=1}^{N-1} &d^{N-1}y_i
\left(
\det\left[y_i \cdot y_j\right]
\right)^{{D-N+1\over 2}}
\left(\det\left[ \Pi_{ij}\right]
\right)^{-{d\over 2}} \cr
=&\int_{{\cal D}^{{\bf T}}} \prod_{\alpha =1}^{N-1}(\beta_\alpha)^{\alpha D-1}
d\beta_\alpha
\,\prod_{\alpha=2}^{N-1}\prod_{n =1}^{\alpha -1}
\left(\sin\left( \theta_{\alpha,n}\right)\right)^{D-1-n}
d\theta_{\alpha,n}\left(
\det\left[\Pi^{{\bf T}}_{\alpha \beta}(\beta\hbox{'s},\theta\hbox{'s})
\right]\right)^{-{d\over 2}}\ .\cr}
}
As already mentioned, we shall limit ourselves to the case
$D\ge N-1$. We shall discuss in section 9 how our results can then be
extended to $D< N-1$. The product of sinuses in \conve\ is thus a
bounded function on ${\cal D}^{{\bf T}}$. Possible ultraviolet divergences
may only arise
from the vanishing of $\det[\Pi_{\alpha\beta}]$, that is when some
$\beta$'s vanish. For $\epsilon > 0$ ($d<d^\star=D/\nu$),
it is sufficient to show that, on ${\cal D}^{{\bf T}}$,
\eqn\toshow{\prod_{\alpha =1}^{N-1}(\beta_{\alpha})^{\alpha D}\left(
\det \left[\Pi_{\alpha \beta}\right]\right)^{-{d\over 2}}=
{\cal O}(\prod_{\alpha =1}^{N-1}(\beta_\alpha)^{\alpha \epsilon})
\ .}
As is clear from its definition, $\Pi^{{\bf T}}_{\alpha\beta}$ vanishes when
$\lambda_\alpha$ and/or $\lambda_\beta$ vanish. The key point is
that while $\Pi^{{\bf T}}_{\alpha\alpha}=
A_D(\nu)|\lambda_\alpha|^{2\nu}$,
 $\Pi^{{\bf T}}_{\alpha\beta }$
vanishes more rapidly than $|\lambda_\alpha|^\nu|\lambda_\beta|^\nu$
if $\alpha\ne \beta$ (see Appendix C). This property is best expressed
by introducing the ``normalized" matrix
\eqn\Upsil{\Avril^{{\bf T}}_{\alpha\beta}\equiv{1\over A_D(\nu)}{\Pi^{{\bf
T}}_{\alpha\beta}
\over |\lambda_\alpha|^\nu|\lambda_\beta|^\nu}}
(such that
$\Avril^{{\bf T}}_{\alpha\alpha}=1$).
\medskip
\noindent In term of the $\beta$'s, we can write:
\eqn\Pibeta{
\matrix{
\Pi^{{\bf T}}_{\alpha\alpha}&=A_D(\nu)\,
\beta_\alpha^{2\nu}\beta_{\alpha +1}^{2\nu}\ldots
\beta_{N-1}^{2\nu}\Avril^{{\bf T}}_{\alpha\alpha}\hfill& \cr
&&\cr
\Pi^{{\bf T}}_{\alpha\beta}&= A_D(\nu)\,
\beta_\alpha^\nu \ldots \beta_{\beta-1}^\nu \beta_\beta^{2\nu}\ldots
\beta_{N-1}^{2\nu}\Avril^{{\bf T}}_{\alpha \beta }(\beta\hbox{ 's},\theta\hbox{
's})
\hfill & \quad (\alpha<\beta )\ ,\cr
}}
leading to the identity
\eqn\boundDet{
\det\nolimits_{N-1} \left(\left[\Pi^{{\bf T}}_{\alpha\beta}\right]\right)\,=\,
(A_D(\nu))^{N-1}\,\beta_1^{2\nu}\ldots\beta_{N-1}^{2\nu(N-1)}\,
\det\nolimits_{N-1} \left(\left[\Avril^{{\bf T}}_{\alpha\beta}\right]\right)
\ .}
This amounts to factorize out the maximal powers of $\beta$'s.
In particular, $\det (\Avril^{{\bf T}} )$ is independent of
$\beta_{N-1}$.
In order to obtain \toshow, one has to show that on ${\cal D}^{{\bf T}}$
the positive quantity
$\det (\Avril^{{\bf T}} )$
in \boundDet\ {\it cannot vanish} and is actually bounded from below by
a strictly positive number. This property is proven in Appendix C.
Indeed, if $\det(\Avril^{{\bf T}})$ were to  vanish, $\det(\Pi^{{\bf T}})$
would
also vanish
and, from Schoenberg's theorem, some
subset of the $\beta$'s must vanish.
This corresponds to contract successively some subsets of points (by a
contracting scale factor $\beta$) to single points.
A generalization of the factorization property
\factdet\ (see Appendix C) shows that, in such a limit, the determinant
$\det(\Pi^{{\bf T}})$ factorizes
into a product of similar determinants associated with subtrees of ${\bf T}$.
The normalized determinant $\det (\Avril^{{\bf T}})$ then
becomes  exactly equal to a product of normalized subdeterminants, each of them
corresponding to
a subtree of ${\bf T}$. In the sector, these subtrees have no coinciding (with
vanishing distance) points
and therefore their determinants do not vanish. Thus,
$\det(\Avril^{{\bf T}})$ does not vanish even in this limit where
 some $\beta$'s tend to zero.
\medskip
 From the above results, the quantity $\det(\Avril^{{\bf T}} )$ in \boundDet\ ,
seen as a function of
$\beta_\gamma$ ($1\le \gamma \le N-2$) and of the $\theta$'s, is
a continuous positive non-vanishing function
on the {\it compact} restriction of ${\cal D}^{{\bf T}}$ obtained by
omitting the (here dummy) variable $\beta_{N-1}$.
Therefore it admits a strictly positive lower bound on
${\cal D}^{{\bf T}}$ and thus (since $d>0$)
\eqn\bDetd{
(\det\nolimits_{N-1} [\Pi^{{\bf T}}_{\alpha\beta}])^{-{d\over 2}}\,
<\, {\rm cst}\cdot
\beta_1^{-d\nu}\ldots\beta_{N-1}^{-(N-1)d\nu}
\ ,}
which is equivalent to \toshow .
The convergence of the integral \conve\ in the Hepp sector ${\cal H}^{{\bf T}}$
for $\epsilon=D-\nu d >0$ follows.
\medskip
We thus have proven the convergence of the generic perturbative term $\CZ_N$
of the partition function $\CZ$ (for $D\ge N-1$). Similarly, the perturbative
terms
$\CZ_N^{(M)}$ (Eq. \ZvertexN )
of the vertex operators $\CZ^{(M)}$ (Eq. \zvertex\ )
can be shown to be UV convergent for $\epsilon >0$ and $D$ large enough
($D\ge N+M-1$).
This follows from
the same decomposition into Hepp sectors and the use of \ZvertexT\ .
The proof is then exactly the same up to the following modifications:
\item{(I)} The measure term in \conve\ is replaced by a measure
similar to \measthetaext\ for tree variables. The difference between this
measure and that of \conve\ concerns only angular terms, which are bounded
functions on $\CD^{\bf T}$ (provided now that $D\ge N+M-1$).
\item{(II)} The exponential term,
depending of the external momenta, has for argument a negative quadratic
form $\displaystyle -{1\over 2}\sum_{a,b}\kvec_a\cdot\kvec_b\, \Delta_{ab}$,
and
is therefore bounded between $0$ and $1$.
\par
\noindent The above proof therefore carries over to this generalized case.

\newsec{The subtraction operation {\titlermsb R}}

\subsec{Renormalization: introductory remarks}

The purpose of renormalization is to show that the short-distance divergences
that occur at $\epsilon=0$ can be absorbed into a redefinition of the
coupling constants of the model. If true, this property allows us (i)
to give a meaning to the theory at $\epsilon =0$, and (ii) to write a
Renormalization Group equation and deduce the scaling behavior of the model for
$\epsilon \lg 0$.
 From the analysis of divergences performed
in sections 4 and 5, we expect that the correlation functions can be made
finite by a simple renormalization of the bare
coupling constant $b$ in the action
\Ham \BD
\eqn\brenorm{
b\ =\ \mu^{\epsilon}\,{\hat b}_R\,Z({\hat b}_R,\epsilon)
}
where $\mu$ is an (internal) momentum scale and ${\hat b}_R$ a finite
dimensionless renormalized coupling constant.
In the case of a finite manifold with volume $\CV_{\CS_D}$,
a convenient and natural choice of momentum scale is
$\mu=R^{-1}\propto (\CV_{\CS_D})^{-1/D}$.
The renormalization factor $Z({\hat b}_R,\epsilon)$ will be an implicit
function of
the parameters $D$ (internal dimension of the manifold) and $\nu$ (scaling
dimension of the $\rvec$ field).
It will be defined in perturbation theory as
\eqn\counter{
Z({\hat b}_R,\epsilon)\ =\ 1\,+\,{\hat b}_R\, a_1(\epsilon)\,+\,
{\hat b}_R^2\, a_2(\epsilon)\,
+\,\ldots}
where the coefficients $a_n$ diverge as $\epsilon^{-n}$ when $\epsilon\to 0$.

If it is possible to construct, at least in perturbation theory, a
function $Z$ such that the partition function $\CZ(b)$ \partfunc\ and the
correlation functions $\CZ^{(M)}(X_a,\kvec_a;b)$ \zvertex\ are UV-finite in the
limit $\epsilon\to 0$, ${\hat b}_R$ and $\mu$ finite, then the model will be
perturbatively renormalizable.
The validity of the approach initiated in \KN ,\ArLub\ and \BD\ will then
be ensured,
since the standard techniques of renormalization group theory can
be applied to the model, and can (in principle) be
extended to all orders in perturbation theory.

\medskip
It is interesting to compare our construction
with what is usually
done for a ``standard" local field theory, such as the
${\rm O}(n)$-symmetric $\Phi^4_{\scriptscriptstyle D}$ theory, whose
action in $D$-dimensions is written as:
\eqn\phifouract{
\CH\,=\,\int_\CV d^Dx\,\left[ {1\over 4}(\partial{\bf \Phi})^2+{m^2\over 2}
{\bf \Phi}^2
+{b \over 2} ({\bf \Phi}^2)^2\right]
}
where ${\bf \Phi}=\{\Phi^i,\, i=1,\ldots n\}$ is an $n$-component field.
There are basically two kinds of approaches to prove renormalizability of this
theory at the critical dimension $D=4$.

The first approach (\`a la Wilson) consists in introducing explicitly
a short-distance cut-off, in integrating over the high momenta modes, and
in showing that the UV divergent terms in the effective action which arise from
this integration can be absorbed into a redefinition of the physical coupling
constants of the theory, so that a finite continuum limit can be reached
by letting the cut-off go to zero and the bare coupling constants flow along
RG trajectories
\ref\WilKog{K. Wilson and J. Kogut, Phys. Rep. {\bf 12C} (1974) 75.}
\ref\Polch{J. Polchinski, Nucl. Phys. {\bf B231} (1984) 269.}.
This approach is physically transparent, appropriate for
the applications of renormalization
group to critical phenomena in statistical mechanics
\DombGreen ,
and has in some cases gained a
rigorous status at the non-perturbative level
\ref\RivBook{
V. Rivasseau, {\sl From Perturbative to Constructive Renormalization},
Princeton University Press, Princeton (1991); and references therein.}.
However, it requires a formulation of the theory through a
lattice regularization, or a phase space formulation, which is
possible for integer space dimension $D$ only. It does not seem possible (up
to now)
to apply these methods in the framework of an analytic continuation in
non integer space dimension so as for instance to justify
the $\epsilon$-expansion used in the description of critical phenomena by a
$\Phi^4_{4-\epsilon}$ theory.

The second, perturbative approach \`a la
Bo\-go\-liu\-bov--Pa\-ra\-siuk--Hepp--Zim\-mer\-man
(BPHZ)
\BPHZ\
consists in working in
perturbation theory and in constructing, directly or by a recursive process,
a subtraction operation on the Feynman amplitudes of the theory, which makes
all the terms of perturbation theory finite and well defined through convergent
integrals; then one shows that this operation corresponds, in the
field theory language, to a
renormalization of the action by local counterterms, and that it  preserves the
equation of motions of the theory and the Ward identities associated with
its symmetries. From the statistical mechanics point of view, this amounts
to a change of variables from microscopic to effective coupling constants.
Renormalization group equations and scaling behaviors are then derived from the
renormalized theory.
This BPHZ formulation of renormalization has a simple and general perturbative
formulation for theories in non integer dimensions $D$, since there are now
well defined recipes
of "dimensional regularization" which allows to construct Feynman amplitudes
for non-integer $D$, and to study their properties, either in the real space
representation, or in the momentum space representation, or in the so-called
Schwinger parametric $\alpha$-representation.
The BPHZ subtraction operation can then easily be extended to the case
of non-integer space dimensions, at least in momentum space or in the
$\alpha$-representation.
\medskip
For our model the action \Ham\ can be seen also as a local field theory
in $D$-dimensional space for a scalar $d$-component $\rvec$ field
\eqn\Hambis{{\cal H}=\int_{\cal V} d^Dx \left[{1\over 2}\rvec(x)
(-\Delta)^{k\over 2}
\rvec(x) + b\ \delta^d(\rvec(x))\right]\ ,}
but the interaction $\delta$-term is singular and non-polynomial, which
makes the perturbative expansion very different from that of
the ordinary case, since
it does not involve usual Feynman diagrams. Furthermore, the dimension
of the interaction term depends explicitly on the number of components of
the field, here $d$.

In principle, nothing prevents the application of a renormalization program
\`a la Wilson in the physical case of objects described by \Hambis\ with
integer dimension ($D=1,\ 2$).
Some preliminary rigorous results have indeed been obtained (for the case
$D=2$, $d=1$, $k=2$) in \DMRR .
However, it is probably impossible to study by such methods the
renormalizability of the model at (or near) its critical dimension $D^\star$
(Eq. \Dcrit ),
since the latter is in general
non-integer (even for integer $d$), and between 0 and 2 (for the elastic
membrane $k=2$ case).
The so-called ``functional renormalization", which
is an approximate renormalization
group scheme, has also been applied to the study of the specific case $d=1$ in
\ref\GroLip{
S. Grotehans and R. Lipowsky, Phys. Rev. {\bf A 45} (1992) 8644.}.
Such schemes are well defined by analytic continuations at non integer $D$ but
are only approximate and have no rigorous status.

On the other hand, in section 3, we constructed a perturbation
theory for the model in non integer dimension $D$, via distance geometry,
which correspond to a dimensional regularization scheme
in (internal) {\it real--} or {\it position--space}.
In sections 4 and 5, we have shown that the
structure of the UV divergences of the amplitude (poles in $\epsilon$)
is quite similar to that of Feynman amplitudes of ordinary local field
theories.
It is the purpose of the rest of this article to show that it is possible to
develop a BPHZ-like formalism to prove renormalizability of this model.
In this Section we shall propose a subtraction operation, which will appear to
be a generalization of the BPHZ subtraction operation
for ordinary Feynman integrals, with a similar structure in term of the
so-called ``Zimmerman forests".
This subtraction operator, which in our case acts directly on the integrands
of interaction terms like \ZNa\ and \znq , involving positions $x_i$ or squared
distances $a_{ij}$, will be shown to make the
integrals UV-finite (for $\epsilon=0$), and to correspond to
a renormalization of the coupling constant $b$.
This will ensure (in perturbation theory) the renormalizability of the model,
the validity of renormalization group equations, and of an $\epsilon$-expansion
about the critical dimension.

\medskip
Another fundamental structure underlies our approach, since the position
variables $x_i$, (or the $a_{ij}$ in distance geometry) can be thought of as
a $D$-dimensional generalization of the Feynman $\alpha$-parameters in
the Schwinger representation. In field theory,
this representation consists in writing the propagators in terms of an
auxiliary
$\alpha$ parameter via a
Laplace transform of the free field propagator (in momentum space)
\eqn\alpharep{
{1 \over p^2+m^2}\ =\ \int_0^\infty d\alpha\,{\rm e}^{-\alpha (p^2+m^2)}
}
and in writing all the Feynman amplitudes as multiple integrals over these
$\alpha$-variables.
As we have seen for our model \Hambis\ in section 3,
the integrals giving the perturbative
terms \ZNa\ of the partition function have a form generalizing that of
a Feynman amplitude in $\alpha$-representation.
Indeed, the subtraction operation and the mathematical
techniques that we shall use to prove renormalizability are in fact extensions
of techniques developed by Berg\`ere and Lam in
\BergLam\
to study renormalization of local field theory precisely
in the $\alpha$-representation.

This analogy of the internal position $D$-space representation of a statistical
mechanics model with the $\alpha$-representation of a local field theory
is not surprising. Indeed, for $D=1$,
it is well-known that the Edwards model for self-avoiding polymer \Edwards\
embedded in $d$ dimensions, can be formulated as a local $\Phi^4$ theory in
$d$-dimensional space,
with Hamiltonian \phifouract\ (with $D$ now formally replaced by $d$),
in the limit where the number of components of the field ${\bf \Phi}$, $n$,
goes to zero (this is the well-known de Gennes equivalence). The length
$S$ of the polymer
is conjugate, via a Laplace transform, to the squared-mass $m^2$ of the
corresponding $n\to 0$ field theory.
Similarly, for our model \Hambis\ (and for $k=2$),
in the case $D=1$ (polymer interacting with
an impurity), the same mapping allows to write it as a $n\to 0$ field theory
in the {\it external} $d$-dimensional space $\RR^d$, with Hamiltonian
\eqn\HamEquiv{
\CH[{\bf \Phi}]\ =\ \int_{\RR^d} d^d \rvec\,\left( {1\over 4}
(\partial{\bf \Phi})^2\,+\,{m^2 \over 2} {\bf \Phi}^2 \right)
\ +\ b\,{\bf \Phi}^2(\rvec={\vec 0}) .
}
The interaction between the polymer and the impurity located at the origin is
represented by the last term in the r.h.s. of \HamEquiv , which is a
singular mass term located at the origin.
Here also, the length of the polymer $S$ (which corresponds to the radius
$R$ of the manifold in the case $D=1$) is conjugate to the squared mass
$m^2$ of the field ${\bf \Phi}$ in \HamEquiv . The diagrams associated with
\HamEquiv\ are ``daisy diagrams" identical to those of \daisyf ,
with the $\alpha$-parameters for the propagators
identified with the internal relative distances in \ZnPol\
$s_\beta=|x_{i_{\beta+1}} -x_{i_\beta}|$ between successive vertices $i_\beta$
and $i_{\beta+1}$ in the {\it internal} 1-dimensional
manifold, i.e. the polymer itself (see Eq. \PiPol ).

Thus, it will appear that our BPHZ renormalization scheme in position space
for the theory \Ham\ defined in $\RR^D$ is a generalization to continuous
values of $D$ of the
ordinary BPHZ renormalization in $\alpha$-representation of the theory
\HamEquiv\ defined in $\RR^d$, with $\alpha$ viewed as a $D=1$ relative
position
\foot{
For this theory, the renormalization is quite trivial, since the only divergent
diagram is the tadpole, i.e. the petal of the daisy.}.
Finally let us stress that this remarkable mathematical analogy makes
us hope that in a
similar way, it will be possible to develop
renormalization techniques in position space for the non-local theory
\EdwardsM\
(which describes self-avoiding $D$-dimensional manifold), which would
reduce for $D=1$ to the ordinary renormalization theory for the Edwards model
(formulated either as a direct renormalization \`a la
des Cloizeaux for the Edwards model, or equivalently as a
BPHZ renormalization for the
$n=0$ $\Phi^4_d$ field theory in the $\alpha$-representation).

\subsec{The subtraction operation}
We first give a heuristic presentation of
the recursive subtraction process that we shall use to
prove renormalizability.
As we have seen, the term of order $N$ of a $M$-point correlation function,
$\CZ_N^{(M)}(X_a,\kvec_a)$,
is given by an integral over the positions of $N$ internal points
\ZvertexN , that we write schematically
(omitting the external momenta $\kvec_a$,
and the parameters $D$, $\nu$ and $\epsilon$), and denoting by $\CG$ the set of
these $N$ internal points
\eqn\ZverSch{
\CZ_N^{(M)}(X_a)\ =\ \int \prod_{i\in\CG}d^Dx_i\, I_\CG(x_i,X_a).}

To subtract the ``superficial UV divergences" which occur in the integral
\ZverSch\ when some subset $\CP$ of points collapses toward a single point,
we can use the factorization theorem of section 4.2 ( and Appendix C), which
implies that when the points of the subset $\CP$ tend altogether
toward some (arbitrary)
point $p$ in $\CP$, the integrand in \ZverSch\ behaves as
\eqn\IntScal{
I_\CG(x_i,X_a)\ \sim \ I_{\CG\setminusp\CP}(x_i,X_a)\cdot I_\CP(x_i)}
where $I_{\CG\setminusp\CP}$ denotes the integrand of the ``reduced" set
$\CG\setminusp\CP$
with the $M$ external points (with positions
$X_a$), and the $N-{\rm Card}(\CP)+1$ internal points obtained from $\CG$
by removing all the points of $\CP$ but $p$, and $I_\CP$ is the integrand
for the subset $\CP$ with no external points and ${\rm Card}(\CP)$ internal
points.
Therefore, we expect that by subtracting the divergences associated with all
families
of {\it mutually disjoint subsets} $\{\CP_k\}$ in $\CG$, we deal with all
{\it superficial} short-distance divergences.
This can be performed by changing in \ZverSch\
the integrand into
\eqn\IntSubOne{
I_\CG(x_i,X_a)\ \to\ I_\CG^{\bullet}(x_i,X_a)\equiv
I_\CG(x_i,X_a)\,+\,\sum_{\{\CP_k\}}
I_{\CG\setminuspk\{\CP_k\}}(x_i,X_a)
\prod_{k}\big( - I_{\CP_k}(x_i)\big) ,}
where the reduced set $\CG\setminuspk\{\CP_k\}$ is obtained by replacing each
subset
$\CP_k$ by one of its points $p_k$ (chosen arbitrarily).
To this subtracted amplitude, we associate the subtracted partition function
term:
\eqn\ZOO{\CZ_N^{\bullet (M)} (X_a) \equiv \int \prod_{i\in \CG}d^Dx_i
I_\CG^{\bullet}(x_i,X_a).}
Thanks to the factorization property of the measure (section 3.4), we can
integrate separately
$I_{\CG\setminuspk\{\CP_k\}}$ over the positions of the internal points
of ${\CG\setminuspk\{\CP_k\}}$, and
each counterterm $I_{\CP_k}$ over the positions of all the points of
$\CP_k$ but one, $p_k$, thus obtaining for each of these counterterms
one term of the expansion of
the partition function $\CZ$.
Evaluating all subset integrals in \IntSubOne\ leads to the explicit formula:
\eqn\ZOZ{\CZ^{\bullet (M)}_N (X_a)=\sum_{N'=1}^N {1\over (N')!}
\CZ^{(M)}_{N'}(X_a)
\sum_{\{N_k,\, k=1,\ldots ,
N'\}\atop N_k\ge 1,\,\sum\limits_k N_k=N} {N!\over \prod\limits_k (N_k)!}
\prod_{k:\atop \,N_k > 1}
(-{\CZ_{N_k}\over \CV_{\CS_D}})}
where $\CV_{\CS_D}$ is the internal volume of the manifold.
One can check that
this subtraction operation on integrands corresponds to a perturbative
expansion
of the partition functions $\CZ^{(M)}$
with respect to a ``renormalized" coupling
constant $b_\bullet$ such that:
\eqn\Zbul{\eqalign{\CZ^{(M)} & = \CZ^{(M)}(b)
= \sum_N (-b)^N {\CZ^{(M)}_N \over N!} \cr
& = \CZ^{\bullet (M)} (b_\bullet) \equiv \sum_{N'} (-b_\bullet)^{N'}
{\CZ^{\bullet (M)}_{N'}\over N'!}\cr } }
with $b_\bullet$ given implicitly by the equation:
\eqn\shiftb{
b\ =\  b_\bullet \,+\,{1\over \CV_{\CS_D}}\sum_{N=2}^\infty {(-b_\bullet)^N
\over N!}\CZ_N
}
However, this subtraction is not sufficient to make the $\CZ^{(M)}$ finite,
since it does not deal with sub-divergences inside the subsets
$\CP$.
As in standard renormalization theory, one deals with that problem by
repeating this subtraction operation inside these subsets,
that is by subtracting to each $I_{\CP_k}$ the divergent parts associated
to families of mutually disjoint subsets in $\CP_k$, and iterating the
process.
One thus obtains at a given order $N$
a subtraction operation expressed in terms of the sets $\CF=\{\CP_k\}$
of {\it mutually disjoint or strictly included}
subsets $\CP_k$ of $\CG$. In analogy with renormalization theory
in field theory, such a set $\CF$ will
be called a {\it forest}\foot{
In renormalization theory, a forest is a family of diagrams $\CP_k$ such that
for any $k\ne l$ one has either $\CP_k\subset\CP_l$, or $\CP_l\subset\CP_k$, or
$\CP_k\cap\CP_l={\hbox {\O}}$.}
of $\CG$.

In addition, for a given forest $\CF$, at each subtraction step, that is for
each subset $\CP_k$ of $\CF$, we have to specify a {\it root} $p_k$ of $\CP_k$,
toward which we contract $\CP_k$, in order to calculate the associated
counterterm. It is quite clear that, after integration over the position
variables, the result of the subtraction operation does not
depend on the specific choice of roots. However,
it is natural to choose for each forest a set of roots in a way which
is consistent with the geometrical picture of the
subtraction operation as successive contractions of subsets toward their
root.
This leads to the notion of a {\it compatibly rooted} forest, which will be
discussed below.

\medskip
After these somehow heuristic considerations, let us give the precise
definition of the subtraction operation that we shall use.

\par \noindent Let us consider a set ${\cal G}$ of $N$ abstract points,
that we call vertices.

\medskip\noindent{\it Definition 6.1}
\par A {\it rooted subset} of $\CG$ is a couple ($\CP,p$) of a subset $\CP$ of
$\CG$ and of a vertex $p$ which belongs to $\CP$, that we call the {\it root}
of $\CP$.

\medskip\noindent{\it Definition 6.2}
\par A {\it forest} $\CF$ of $\CG$ is a set of subsets $\CP_i$ of $\CG$ such
that:
\item{-} two elements of $\CF$ are disjoint or strictly included into
one another, {\it i.e.}
$$\eqalign{
&\CP_i\neq\CP_j\qquad {\rm if}\ i\neq j\cr
{\rm and}\qquad & \CP_i\cap\CP_j=\CP_i\ {\rm , or}\ \CP_j\ {\rm , or}\
\hbox{\O}
, \qquad \forall i,j.\cr}$$
\item{-} all elements of $\CF$ have at least two elements, i.e.
$${\rm Card}(\CP_i)=|\CP_i| > 1 .$$
\par\noindent Let us note that, by convention, the empty
set $\hbox{\O}$ is a forest.

\medskip\noindent{\it Definition 6.3}
\par A {\it rooted forest} $\CF_\oplus$ is a set of rooted subsets
$(\CP_i,p_i)$
of $\CG$ such that $\{\CP_i\}$ is a forest.

\medskip\noindent{\it Definition 6.4}
\par A {\it compatibly rooted forest} is a rooted forest such that, if,
for some $i$ $j$, $\CP_i\supset\CP_j$ and $p_i\in\CP_j$, then $p_i=p_j$.

\medskip\noindent{\it Definition 6.5}
\par Finally to any rooted forest $\CF_\oplus$ we associate
its compatibly rooted
forest ${}^c\CF_\oplus$ by simply changing its roots according
to the following recursion:
\item{-} First, replace the root $p_i$ of each $\CP_i$ of the forest
by the root $p_j$ of the {\it smallest} subset $\CP_j$ of the forest such that
$p_i\in\CP_j$ ($p_j$ may coincide with $p_i$). One thus obtain a new rooted
forest.
\item{-} Then, repeat this process for the new forest. One can easily show
that after a finite number of iterations ($\le {\rm Card}(\CF)$), this process
will leave the roots unchanged, so that one obtains a compatibly rooted forest
${}^c\CF_\oplus$.
\par \noindent Of course, a forest $\CF_\oplus$ is compatibly rooted iff
${}^c\CF_\oplus=\CF_\oplus$.
\medskip\noindent{\it $\diamond$ Dilation operation}
\par For a rooted subset $(\CP,p)$, we define the dilation operation
$\CD^\rho_{(\CP,p)}$ as the transformation acting on the positions of
the vertices according to (as in \xrho ):
\eqn\xrhobis{
\CD^\rho_{(\CP,p)}\quad:\quad x_i\ \to\ x_i(\rho )\,=\,\left\{\matrix{
x_p+\rho \  (x_i-x_p)\hfil\quad &{\rm if}\ i\in \CP\cr
x_i\hfil\quad &{\rm if}\ i\not\in \CP\cr}\right.
}
or equivalently in distance space, according to:
\eqn\aijrhobis{
\CD^\rho_{(\CP,p)}\quad:\quad a_{ij}\ \to\ a_{ij}(\rho)\,=\,\left\{\matrix{
\rho^2 a_{ij}\hfill\quad&{\rm if}\ i\in\CP,\ j\in\CP\hfill\cr
a_{pj}-\rho \ (a_{pi}+a_{pj}-a_{ij})+\rho^2 \ a_{pi}\hfil\quad&{\rm if}\
i\in\CP,\ j\not\in\CP\cr
a_{ij}\hfill\quad&{\rm if}\ i\not\in\CP,\ j\not\in\CP\hfill\ .\cr}\right.
}
More generally, for a function $I$, expressed as a function of the positions
$x_i$ or the distances $a_{ij}$, we denote by $\CD^\rho_{(\CP,p)}I$ the value
of this function at the positions (or distances) modified according to
\xrhobis\ (or \aijrhobis ).

\medskip\noindent{\it $\diamond$ Taylor operator}
\par We then define the ``Taylor" operator $\Tay_{(\CP,p)}$ acting
on functions $I$ by:
\eqn\taylor{\Tay_{(\CP,p)}I=\lim_{\rho \to 0} \rho^{d\nu(|\CP|-1)}
\CD^\rho_{(\CP,p)}I .}
The functions that we shall consider are
the integrands in \ZvertexN\ and \ZN ,
which are of the form
\eqn\Ampl{\eqalign{
I_{\CG}(x_i,X_a)\hfill &=\left(\det\left[\Pi\left(x_{i,\,i\in
\CG}\right)\right]
\right)^{ -{d\over 2}}\ \exp \left(-{1\over 2}\sum_{a,b}\kvec_a\cdot\kvec_b\,
\Delta_{ab}(x_{i,\,i\in \CG},X_a)\right)\hfill \cr
I_{\CG}(x_i)\hfill &=\left(\det\left[\Pi\left(x_{i,\,i\in \CG}\right)\right]
\right)^{ -{d\over 2}}\hfill \cr}
}
where the $\Pi$ and $\Delta$ matrices,
defined in \Pimatrix , \Piquad\ and \Deltamatr , \DeltamaT ,
are functions
of the positions of the internal vertices $i$ in $\CG$
and external vertices $a$.
On such functions, the effect of $\Tay_{(\CP,p)}$ is to keep the most
singular term in $\rho$ when performing the dilation $\CD^\rho_{(\CP,p)}$.
For instance one operator $\Tay_{(\CP,p)}$ factorizes $I_\CG$ into
\eqn\TayI{\Tay_{(\CP,p)}I_\CG(x_i,X_a)=I_{\CP}(x_i)I_{\CG\setminusp
\CP}(x_i,X_a)}
where
\eqn\BDeq{\CG\setminusp\CP\equiv\CG\setminus(\CP\setminus\{p\})}
is the reduced set obtained by
contracting $\CP$ into $p$ ( $\setminus$ is the usual subtraction of sets).
This operation can be repeated for rooted subsets which form a
{\it compatibly rooted} forest,
and the result does not depend on the order of the $\Tay$
operators in this case (commutativity).
The result is a product of integrands $I(x_i)$ of reduced internal subsets,
times the integrand $I(x_i,X_a)$ of the set $\CG$ reduced by all elements of
the forest.

\medskip\noindent{\it $\diamond$ The subtraction operator}
\par With those notations, we define the subtraction operation {\bf R} as a sum
of subtractions for all forests.
For a given forest $\CF$,
subtractions associated with different roots give different results
on the integrand. We shall sum over the subtractions for all compatibly
rooted forests $\CF_{\oplus^c}$, with some weight factor $W(\CF_{\oplus^c})$
associated with the (compatible) rooting of $\CF$.
In order to ensure the finiteness of the subtracted integrals,
the weights $W(\CF_{\oplus^c})$ must be such that the sum of the $W$ for all
rooted forests which correspond to the same unrooted forest $\CF$ gives $1$.
A convenient choice
of weight factor $W(\CF_{\oplus^c})$ for $\CF_{\oplus^c}$
is to make it
proportional to the number of different (not necessarily compatibly)-rooted
forests $\CF_\oplus$ which gives $\CF_{\oplus^c}$ by
the compatibilization
operation $^c$ (i.e. ${}^c\CF_\oplus=\CF_{\oplus^c}$).
Our final definition for {\bf R} is therefore expressed as a sum over all
rooted forests, or equivalently as a sum over all compatibly rooted forests.
It reads\foot{In this equation,
$\CF_{\oplus^c}$ denotes an arbitrary compatibly
rooted forest, while ${}^c\CF_\oplus$ denotes the compatibly rooted
forest obtained from the (non necessarily compatibly) rooted forest
$\CF_\oplus$ by the compatibilization procedure of Definition 6.5.}:
\eqn\Roper{\encadre{\hbox{$\displaystyle \eqalign{
 {\bf R}\ &=\ \sum_{\CF_\oplus}
\Bigg[ \prod_{(\CP,p)\in {}^c\CF_\oplus}{1\over |\CP |}
\big( -\Tay_{(\CP,p)} \big)\Bigg]
\cr
&=\ \sum_{\CF_{\oplus^c}}\,W(\CF_{\oplus^c})\,
\Bigg[ \prod_{(\CP,p)\in \CF_{\oplus^c}}
\!\big( -\Tay_{(\CP,p)} \big)\Bigg]
\cr}$}}}
The  weight factors are given explicitly by a product
over all {\it different roots} $p$ of $\CF_{\oplus^c}$
\eqn\Wexplic{
W(\CF_{\oplus^c})\,=\,
\prod_{
       {p\ {\rm root}}\atop
       {{\rm of}\,\CF_{\oplus^c}}
      }
{1\over |\CP_p|} ,
}
where  $\CP_p$ is the {\it largest} subset of the forest whose root is $p$.

\medskip \noindent{\it $\diamond$ Subtracted amplitudes and
renormalization}\par
We now restrict ourselves to the case of amplitudes defined in a finite volume,
by using the IR regulator introduced in section 4.3 ($D$-dimensional sphere),
that is by defining the integration over the positions of the vertices by
\intaS\ and \measaS .
The subtracted correlation functions at order $N$ are simply defined by
applying
the subtraction operator {\bf R} to the integrand of \ZverSch
\eqn\ZMRen{
{\CZ^{\bf R}}_N^{(M)}(X_a)\ \equiv \
\int \prod_{i\in\CG}d^D x_i\,{\bf R}\,[I_\CG(x_i,X_a)] .}

Let us note that, since the integrand for the partition function is homogeneous
under global rescaling, one has
\eqn\RoperonI{ {\bf R}\left[I_{\CG}\left(x_i\right)\right]=0 }
(as soon as $|\CG|\ge 2$, of course).
This means that with our choice of subtraction,
for $N\ge 2$, in the absence of external correlation points,
\eqn\ZNeqO{\CZ_N^{\bf R}=0\ ,\quad N\ge 2.}
\medskip
The purpose of the next sections is to  prove that this subtraction operation
makes all correlation functions finite, as summarized in the theorem:
\smallskip\noindent {THEOREM }
{\it For $0<\nu<1$, the renormalized
integral \ZMRen\ is convergent for $\epsilon=0$ and
defines a finite function
${\CZ^{\bf R}}_N^{(M)}(X_a)$ for $D\ge N+M-1$.}

\medskip

The renormalized correlation functions are defined by their perturbative
expansion in powers of a
renormalized coupling constant $b_R$
\eqn\RenExp{
{\CZ^{\bf R}}^{(M)}(X_a;b_R)\ =\ \sum_{N=0}^\infty \,{(-b_R)^N \over N!}\,
{\CZ^{\bf R}}_N^{(M)}(X_a).}
As discussed above, the forest structure of the subtraction operation {\bf R}
ensures that for $\epsilon>0$, there exists a renormalized coupling
constant $b_R(b)$ such that the renormalized correlation functions
${\CZ^{\bf R}}^{(M)}(X_a;b_R)$
are equal to the original ``bare" correlation functions
$\CZ^{(M)}(X_a;b)$ for the model \Ham .

The relation between $b$ and $b_R$ can be obtained directly from the identity
of the partition functions
\eqn\PartEqu{
\CZ(b)\ =\ \CZ^{\bf R}(b_R) .}
 From \ZNeqO\ we have
$\CZ^{\bf R}(b_R)\ \equiv \ (2\pi)^d\delta^d(\kvec = {\vec 0})\,
-\,b_R\,\CV_{\CS_D}\equiv V_{\RR^d}\,-\,b_R\,\CV_{\CS_D}$
and therefore, equating to $\CZ(b)$, we get
\eqn\brendir{b_R=-{1\over \CV_{\CS_D}}\left(\CZ-
V_{\RR^d}\right) ,}
or the explicit series expansion in $b$:
\eqn\bren{
b_R\ =\ b\,-\,{1\over \CV_{\CS_D}}\,\sum_{N\ge 2}{(-b)^N\over N!} \CZ_N .}
Notice that the fully renormalized coupling constant $b_R$ satisfies the
indentity
$b\ =\ b_R \,+\,{1\over \CV_{\CS_D}}\,\sum_{N\ge 2}{(-b)^N\over N!} \CZ_N$,
while the former partially renormalized coupling constant $b_\bullet$ (built
so as to absorb the superficial divergences) satisfies the {\it truncated}
equation \shiftb ,
$b\ =\ b_\bullet \,+\,{1\over \CV_{\CS_D}}\,\sum_{N\ge 2}{(-b_\bullet)^N
\over N!} \CZ_N$,
obtained from the equation for $b_R$ mentioned just above by replacing
$(-b)$ by $(-b_\bullet)$ in the r.h.s.

Equation \brendir\ shows that, in this scheme, renormalization
simply amounts to a change of variable from the microscopic $b$ to an
effective coupling constant $b_R$, directly proportional to the {\it connected}
partition function of the manifold interacting with a point.
This scheme is precisely that used in \BD , and generalizes that of the
``direct renormalization method" \desCloiz\ for the polymer
Edwards model.

Let us stress that $b_R$ as defined above is not dimensionless.
The corresponding dimensionless coupling constant can be conveniently
chosen as
\eqn\defofg{g=\Big( 2\pi \,A_D(\nu) \Big)^{-{d/ 2}}b_R
\CV_{\CS_D}^{1-\nu d/D}}
for which the Wilson function \RGW\ has been calculated explicitly
at one loop \BD .
In this subtraction scheme, the subtraction scale $\mu$ of the general
equation \brenorm\ is fixed
by the $D$-dimensional volume (which fixes the IR cut-off)
$\mu\sim (\CV_{\CS_D})^{- 1/D}$. In these notations, this
precisely corresponds to
${\hat b}_R=b_R\, (\CV_{\CS_D})^{\epsilon / D}$ and
$Z^{-1}({\hat b}_R,\epsilon)\equiv b_R / b = 1+
{1\over \CV_{\CS_D}}\sum_{N\ge 2}
(-b)^{N-1} {\CZ_N\over N!}$, where $b$ is an implicit function of $b_R$,
thus ${\hat b}_R$.
Of course, other subtraction schemes can be chosen where the subtraction scale
$\mu$ is not related to the volume of internal $D$-dimensional space.
They are needed in order to define the theory ({\it e.g.} the
normalized correlation functions) in the infinite volume limit.

\newsec{Reorganization of the counterterms}

\subsec{Formulation of the subtraction operation in terms of nests}

As we shall see later, it will be more convenient in the proof of the
finiteness of the renormalized amplitudes to express the subtraction
operation {\bf R} in term of {\it nested} subdiagrams.
In the formalism of BPHZ renormalization in the Schwinger representation
in field theory, a subdiagram is a set of lines (propagators) of a Feynman
graph (and has in general many connected components).
A {\it nest} is then a family of
subdiagrams $P_k$ which are nested, that is included into one another
(for any $k\ne l$, $P_k\subset P_l$ or $P_l\subset P_k$).

In our case we shall introduce a different notion of diagram, now in
terms of vertices, rather than lines. Indeed, we have seen that the natural
generalization of Schwinger parameters $s_\alpha$ is given by the larger
set of all  mutual distances $a_{ij}$ between points on the manifold.
In terms of links, we thus would have to deal with the large number of
interdependent mutual distances, which are constrained by
triangular inequalities. Therefore, we prefer to define diagrams in terms of
vertices.
Denoting again by $\CG$ a set of $N$ vertices, a diagram of $\CG$ will now
be a collection of disjoint vertex-subsets of $\CG$. Each of these subsets
of vertices can be thought of as a connected set (which stands for the
the complete set of its pairwise mutual distances in the link representation).
These ideas will be embodied in the following definitions.
\medskip
We recall that a partition $P$ of a set ${\cal S}$
is a set of mutually disjoint non empty subsets $\CS _i$
of ${\cal S}$, whose union is ${\cal S}$ itself.
\midinsert
\figinsert{6.truecm}{9.truecm}{%\hsize}{
\figcap\subdiagf{A subdiagram.}
}
\endinsert
\midinsert
\figinsert{6.truecm}{8.truecm}{%\hsize}{
\figcap\diagf{A complete diagram, with connected components $\CP_i$.}
}
\endinsert
\medskip\noindent{\it Definition 7.1} (see fig. \xfig\subdiagf , \xfig\diagf )

We shall call a {\it subdiagram} (respectively {\it complete diagram}) of $\CG$
any partition $P$ of some
subset $\CS$ of $\CG$ (respectively of $\CG$ itself).
The generic word {\it diagram} will be used in both cases.

The elements of this partition $P$ are called the {\it connected components}
of the diagram $P$.

\midinsert
\figinsert{6.truecm}{8.5truecm}{%\hsize}{
\figcap\contdiag{The complete diagram with connected components $\CP_i$
(dashed line)
is contained in the complete diagram with connected components $\CQ_j$
(full line).}
}
\endinsert
\medskip \noindent{\it Definition 7.2} (see \contdiag )

A diagram $P$ is {\it contained} in a diagram $Q$ if any connected component of
$P$ is included in one of the connected components of $Q$.
This will be denoted  $P\prec Q$.
\foot{Let us stress that $P\prec Q$ does not mean that $P$, considered
as a set (whose elements are subsets of $\CG$), is included in $Q$. Still
if $P\subset Q$, then $P\prec Q$.}
This defines a partial ordering among the diagrams of $\CG$.
\midinsert
\figinsert{6.truecm}{8.5truecm}{%\hsize}{
\figcap\intdiagf{The intersection diagram (dark-grey diagram) of two diagrams (
grey and white dia\-grams).}
}
\endinsert
\medskip\noindent{\it Definition 7.3} (see \intdiagf ) \par
We define the {\it intersection} of two diagrams $P$ and $Q$ as the maximal
diagram which is contained in both $P$ and $Q$ (it is unique),
and denote it by $P\wedge Q$. Its connected components are nothing but the
(non-empty) intersections of a connected component of $P$ and one of $Q$.
\midinsert
\figinsert{6.truecm}{8.5truecm}{%\hsize}{
\figcap\unidiagf{The union diagram (dark-grey diagram) of two diagrams (grey
and white diagrams).}
}
\endinsert
\medskip\noindent{\it Definition 7.4} (see \unidiagf ) \par
We define the {\it union} of two diagrams $P$ and $Q$ as the minimal
diagram which contains both $P$ and $Q$ (it is also unique), and denote it
by $P \vee Q$.
Let us note that the connected components of $P \vee Q$ are unions of connected
components of $P$ and $Q$, but in general not simply the
union of one connected component of $P$ and of one of $Q$.
\midinsert
\figinsert{6.truecm}{8.5truecm}{%\hsize}{
\figcap\godotdiagf{The minimal complete diagram $G_\odot$.}
}
\endinsert
Notice that the union and the intersection of complete diagrams of $\CG$
are complete. The maximal complete diagram of $\CG$ is $G=\{\CG\}$.
We shall denote by $G_\odot$
the (unique) minimal complete diagram of $\CG$. Its connected components
are the $N$ single vertex subsets of $\CG$ (see \godotdiagf ).
For any complete diagram $P$,
we have $G_\odot \prec P \prec G$.
\midinsert
\figinsert{6.truecm}{8.5truecm}{%\hsize}{
\figcap\susdiagf{The subtraction diagram (dark-grey diagram) of a diagram (grey
diagram) from another diagram (white diagram).}
}
\endinsert
\medskip\noindent{\it Definition 7.5} (see \susdiagf ) \par
We define the {\it subtraction}
of a diagram $P$ from a diagram $Q$ as the (unique)
maximal diagram contained in $Q$ and whose intersection with $P$ is empty,
and denote it by $Q\setminus P$.

\medskip
The usual properties of commutativity and
associativity are satisfied by $\wedge$ and $\vee$.
However these operations are not distributive with respect to one another.
They still satisfy the weaker relations
\eqn\quasidistr{\eqalign{
P\wedge (Q\vee R)&\succ (P\wedge Q)\vee (P\wedge R)\cr
P\vee (Q\wedge R)&\prec (P\vee Q)\wedge (P\vee R)\cr
}}
\midinsert
\figinsert{6.truecm}{8.5truecm}{%\hsize}{
\figcap\rootsubdiagf{A rooted subdiagram. The roots are specified by
squares.}
}
\endinsert
\medskip\noindent {\it Definition 7.6} (see \rootsubdiagf ) \par

A {\it rooted diagram}
$P_\oplus$ is a family $\{(\CP_1,p_1),\ldots,(\CP_k,p_k)\}$ of
rooted subsets
$(\CP_i , p_i)$ of $\CG$ such that
 $P=\{\CP_1,\ldots,\CP_k\}$ is a diagram of $\CG$.

\noindent
We call
\eqn\Pcomp{P={\rm comp}(P_\oplus)=\{\CP_1,\ldots,\CP_k\}}
the component diagram of $P_\oplus$, and
\eqn\Proot{\wp={\rm root}(P_\oplus)=\{\{p_1\},\ldots,\{p_k\}\}}
the root diagram of $P_\oplus$.
We shall use for a rooted diagram the equivalent notations:
\eqn\rodino{P_\oplus\equiv \big({\rm comp}(P_\oplus),{\rm root}(P_\oplus)
\big)\equiv\big(P,\wp \big) .}
\midinsert
\figinsert{6.truecm}{8.5truecm}{%\hsize}{
\figcap\rootdiagf{A complete rooted diagram. Its elements are rooted
subsets $(\CP_i,p_i)$.}
}
\endinsert
\medskip\noindent{\it Definition 7.7} (see \rootdiagf ) \par
A complete rooted diagram is a rooted diagram such that its component diagram
is complete.
\vfill\eject
\medskip\noindent{\it Definition 7.8}\par
A {\it nest} $\CN$ is a set of ${\scriptstyle T}+1$ complete diagrams
$\{T_0,T_1,\ldots,T_\sT\}$ such
that
\eqn\Nest{\eqalign{T_0&=G_\odot \cr T_0&\prec T_1\prec T_2\prec\ldots
\prec T_\sT .\cr}}
\midinsert
\figinsert{6.truecm}{8.5truecm}{%\hsize}{
\figcap\rootnestf{Two successive complete rooted diagrams $T_{\sJ\oplus}$,
with connected components $\CT_{\sJ,j}$ (dashed lines)  and $T_{\sJ+1\oplus}$
with connected components $\CT_{\sJ+1,k}$ (full lines) of a rooted nest.
The roots of these two diagrams are {\it not compatible}.}
}
\endinsert
\medskip\noindent{\it Definition 7.9} (see \rootnestf ) \par
A {\it rooted nest} $\CN_\oplus$ is a set of complete rooted diagrams
$\{T_{0\oplus},T_{1\oplus},\ldots,T_{\sT\oplus}\}$ such
that the associated component diagrams form a nest
\eqn\CompNest{
{\rm comp}(T_{0\oplus})\prec{\rm comp}(T_{1\oplus})\prec\ldots\prec{\rm
comp}(T_{\sT\oplus}) .}
\midinsert
\figinsert{6.truecm}{8.5truecm}{%\hsize}{
\figcap\comprootnestf{The two successive diagrams
of \rootnestf , with {\it compatible} roots.
The roots $w_{\sJ+1,k}$ have been obtained from the roots of
\rootnestf\ by the construction of {\it Definition 7.11}.}
}
\endinsert
\vfill \eject
\medskip\noindent{\it Definition 7.10} (see \comprootnestf ) \par
A rooted nest is said to be {\it compatibly rooted} if we have moreover
\eqn\RootNest{{\rm root}(T_{0\oplus})\succ {\rm
root}(T_{1\oplus})\succ\ldots\succ{\rm root}(T_{\sT\oplus}) .}
(Notice that ${\rm root}(T_{0\oplus})=G_\odot$).\par\noindent
At level ${\scriptstyle J}$, the generic element of the rooted nest
$\CN_\oplus$
reads explicitly:
\eqn\explic{T_{\sJ\oplus}=\big\{(\CT_{\sJ,j},t_{\sJ,j}),\, j=1,\ldots ,
{\rm Card}(T_{\sJ}) \big\} .}
Eq. \RootNest\  means that when we consider two successive rooted
complete diagrams
of the rooted nest, $T_{\sJ\oplus}$ and $T_{\sJ+1\oplus}$, if we consider a
connected component $\CT_{\sJ+1,k}$ of $T_{\sJ+1}$ and its root $t_{\sJ+1,k}$,
this root must
coincide with
the root $t_{\sJ,j}$ of the connected component $\CT_{\sJ,j}$ of $T_{\sJ}$
to which $t_{\sJ+1,k}$ belongs (since $T_{\sJ}$ is complete, $t_{\sJ+1,k}$
belongs necessarily to some connected component of $T_\sJ$).
This property then implies by recursion that, at each level
${\scriptstyle L}\le {\scriptstyle J}$, $t_{\sJ +1,k}$ coincides with the root
$t_{{\scriptscriptstyle L},l}$ of the connected component
$\CT_{{\scriptscriptstyle L},l}$ of $T_{\scriptscriptstyle L}$ to which
it belongs.
\medskip\noindent{\it Definition 7.11}\par
To any rooted nest $\CN_\oplus$ with elements given by \explic ,
we associate the compatibly rooted
nest
\eqn\coroot{\eqalign{\cn_\oplus&=\big\{{}^cT_{\sJ\oplus}\big\}\cr
{}^cT_{\sJ\oplus}&=\big\{(\CT_{\sJ,j},w_{\sJ,j}),\, j=1,\ldots ,
{\rm Card}(T_{\sJ}) \big\} ,\cr}
}
with the same connected components $\CT_{\sJ,j}$ at each level
$\scriptstyle J$, and whose roots $w_{\sJ,j}$
are obtained from the roots $t_{\sJ,j}$ by the following recursion:
\item{-} at level $0$, the roots of $T_0$ are fixed since ${\rm
root}(T_{0\oplus})=G_\odot$;
\item{-} at level $1$, we identify $w_{1,j}$ with the original root $t_{1,j}$,
that is set $w_{1,j}=t_{1,j}$ for all
$j=1,\ldots ,{\rm Card}(T_{1})$;
\item{-} at level ${\scriptstyle J}+1$ and for each connected component
$\CT_{J+1,k}$, we look for the component $\CT_{\sJ,j(k)}$ of the complete
diagram $T_\sJ$ at the preceding level ${\scriptstyle J}$ to which the original
root $t_{\sJ+1,k}$ belongs. The root $w_{\sJ,j(k)}$ has
already been constructed at level $\scriptstyle J$ and we make the
roots compatible between level ${\scriptstyle J}$ and ${\scriptstyle J}+1$
by substituting  to
the original root $t_{\sJ+1,k}$ the root $w_{\sJ+1,k}=w_{\sJ,j(k)}$ (Notice
that, since $T_\sJ\prec T_{\sJ+1}$, $\CT_{\sJ,j(k)} \subset \CT_{\sJ+1,k}$
and therefore $w_{\sJ,j(k)}\in \CT_{\sJ+1,k}$) .

\noindent By construction, the rooted nest $\cn_\oplus$ is compatibly rooted.
Of course, a rooted nest $\CN_\oplus$ is compatibly rooted if and only if
$\cn_\oplus=\CN_\oplus$, and in this case, $w_{\sJ,j}\equiv t_{\sJ,j}$ for all
${\scriptstyle J}$ and $j$.

\medskip
To a rooted diagram $T_\oplus$, we associate the Taylor operator
$\Tay_{T_\oplus}$ defined simply as the product of the Taylor operators
$\Tay_{(\CT,t)}$ of its rooted connected components:
\eqn\Taydiag{\Tay_{T_{\oplus}}=
\prod_{(\CT_{j},t_{j})\in T_{\oplus}}
\Tay_{(\CT_{j},t_{j})} }
with the convention $\Tay_{(\CT,t)}=\II$ if $|\CT|=1$ ({\it i.e.} $\CT=\{t\}$),
which in particular implies that $\Tay_{T_{0\oplus}}=\II$ for $T_0=G_\odot$.
We denote by $\|T_\oplus\|$ the product of the cardinals of the
connected components $\CT_j$
of the diagram ${\rm comp}(T_\oplus)$
\eqn\diagcard{
\|T_\oplus\|\ =\ \prod_{\CT_j\in {\rm comp}(T_\oplus)} |\CT_j|
}
\medskip\noindent{PROPOSITION:}\par
The subtraction operator ${\bf R}$ \Roper\ can be rewritten
as a sum over rooted nests:
\eqn\Roperbis{\encadre{\hbox{$\displaystyle \eqalign{
{\bf R}\ &=-\sum_{\CN_\oplus}
\Bigg[ \prod_{{}^cT_{\sJ\oplus}\in \cn_\oplus}
\big( -{1\over \|{}^cT_{\sJ\oplus}\|}\Tay_{{}^cT_{\sJ\oplus}}\big)
\Bigg]\cr &=-\sum_{\CN_{\oplus^c}}\, W(\CN_{\oplus^c})
\Bigg[ \prod_{T_{\sJ\oplus}\in \CN_{\oplus^c}}
\big( -\Tay_{T_{\sJ\oplus}}\big) \Bigg]\cr }$}}}
where the second formula is a sum over compatibly rooted nests with
the appropriate weight factor:
\eqn\Wnest{
W(\CN_{\oplus^c})\,=\,
\prod_{ w }
{1\over |\CT_w|}
}
with $\CT_w$ being as before the largest connected component (among
all connected components of all diagrams of $\CN$) whose root is $w$.
In \Wnest\ the products is over all vertices of $\CG$ since any point
$w$ of $\CG$ is the root of at least one connected component in the nest,
namely the connected component $\{w\}$ of $T_0$.
\medskip\noindent{\it Proof:}\par
The global $(-1)$ factor in \Roperbis\ is introduced to reverse the
global $(-1)$ sign coming from the contribution
$\big( -{1\over \|T_{0\oplus}\|}\Tay_{T_{0\oplus}}\big)=-\II$ which is
present for each nest (compatible or not).
\par \noindent
To prove that \Roperbis\ coincides with \Roper\ one can proceed in two
steps, that we indicate below. The details are left to the
reader.

We start from \Roperbis\ as a sum over compatibly rooted nests
$\CN_{\oplus^c}$,

First, we notice that the family of
all distinct rooted components, excluding single vertex components,
of the rooted diagrams of some compatibly rooted nest $\CN_{\oplus^c}$
form a compatibly rooted forest.
Moreover, if two different compatibly rooted nests yield
the same compatibly rooted forest $\CF_{\oplus^c}$, the products of
$\Tay$ for these two different nests give the same result, which
is nothing but the product of $\Tay$ associated with the
compatibly rooted forest $\CF_{\oplus^c}$. This allows us to regroup
all compatibly
rooted nests which yield the same compatibly rooted forest.

Second, we have to check that the $(-1)$ factors and weights associated with
each diagram of this group of nests sum up in order to give the correct factor
$W(\CF_{\oplus^c})$
\Wexplic\ for this forest.
This can be seen in two steps:
First, the weights
$W(\CN_{\oplus^c})$ \Wnest\ are in fact equal to
$W(\CF_{\oplus^c})$ \Wexplic\ , for each $\CN_{\oplus^c}$ yielding
$\CF_{\oplus^c}$.
Therefore, at that stage, we
can forget about the roots and the weights $W$ and concentrate
on the $(-1)$ factors associated to the diagrams of the nests. It remains
to show that, when summing over all nests $\CN$ which yield a given
forest $\CF$, one has $(-1)\times\sum\limits_{\CN \to \CF}(-1)^{{\rm
Card}(\CN)}
= (-1)^{{\rm Card}(\CF)}$.
This relation can be easily checked for forests made out of two subsets,
which are either disjoint
or included into one another\foot{
If $\CF=\{S_1,S_2\}$, either $S_1\cap S_2={\hbox {\O}}$
and there are three nests
$\{G_\odot,G_\odot\vee\{S_1,S_2\}\}$,
$\{G_\odot,G_\odot\vee\{S_1\},G_\odot\vee\{S_1,S_2\}\}$ and
$\{G_\odot,G_\odot\vee\{S_2\},G_\odot\vee\{S_1,S_2\}\}$,
with respectively $2$, $3$ and $3$ diagrams,
or $S_1\subset S_2$ and there is only one nest
$\{G_\odot,G_\odot\vee\{S_1\},G_\odot\vee\{S_2\}\}$ with $3$ diagrams.
}
, and then extended by a recursion on the number
of elements of the forest.

\subsec{Sectors}

\medskip\noindent{\it Definition 7.12: Saturated nest}

A {\it saturated nest} $\CS$ of $\CG$ is a nest with $N={\rm Card} (\CG)$
(distinct) elements \foot{We use superscripts here in $R^\sI$ rather than
subscripts
as before in $T_\sJ$ for future convenience.}, which we call
$R^0,\ldots, R^{N-1}$.

\noindent The cardinal of a saturated nest is therefore maximal.
A saturated nest is actually constructed from $G_\odot $
(the complete diagram  made of $N$ single point connected components)
by fusing recursively at each level $R^\sI$ exactly two connected components of
the preceding level $R^{\sI-1}$ until $G=\{\CG\}$ is obtained.
A saturated nest is therefore characterized as follows:
\item{-} its minimal diagram is $R^0=G_\odot$,
\item{-} its maximal diagram is $R^{N-1}=G=\{\CG\}$,
\item{-} ${\rm Card}(R^{\sI+1})={\rm Card}(R^\sI)-1$ for all ${\scriptstyle{I}}
=0, \ldots, N-1$
\medskip\noindent{\it $\diamond$ Saturated nest associated with ordered
trees}\par
The notion of saturated nest occurs naturally when spanning integration points
by {\it trees}, as was done formally in section 3.7 .
Indeed,
let us consider a tree
${\bf T} =(\lambda_\alpha;\, \alpha=1, \ldots N-1)$,
considered as {\it ordered} by increasing values
of $\alpha$ (this order will actually correspond to increasing mutual
distances,
in a generalized sense to be made precise below).
Such an ordered tree ${\bf T}$
generates naturally a saturated nest $\CS({\bf T})$ as follows:
\item{-} $R^0=G_\odot$
\item{-} at level $\scriptstyle I$ ($1\le {\scriptstyle I}\le N-1$),
we consider the line $\alpha =
{\scriptstyle I}$ with end points $i_\alpha$, ${i'}_\alpha$ and set
$R^\sI=R^{\sI-1}\vee\{\{i_\alpha,{i'}_\alpha\}\}$, which corresponds to
the fusion of the connected component of $R^{\sI-1}$ containing $i_\alpha$
with that containing ${i'}_\alpha$.
\par \noindent Of course, different trees ${\bf T}$ can yield
the same $\CS({\bf T})$.
This allows us to classify trees into equivalence classes, by regrouping
all the trees ${\bf T}$ such that $\CS({\bf T})=\CS$ for any given
saturated nest $\CS$. If two ordered trees
${\bf T} =(\lambda_\alpha;\, \alpha=1, \ldots N-1)$ and
${\bf T'} =({\lambda'}_\alpha;\, \alpha=1, \ldots N-1)$ are
equivalent, then the transformation from $\lambda$ to $\lambda'$ is
such that:
\eqn\equtree{\lambda_\alpha=\pm {\lambda'}_\alpha+\sum_{\gamma<\alpha}
c^\gamma_\alpha {\lambda'}_\gamma\qquad\qquad c^\gamma_\alpha = 0,\pm 1}
where $c^\gamma_\alpha $ are coefficients equal to $0$ or $\pm 1$ (which are
in general further constrained so that ${\bf T}$ and ${\bf T'}$ actually
span the same
set of integration points).
\vfill\eject
\medskip\noindent{\it $\diamond$ Oriented ordered tree associated with a
compatibly
rooted saturated nest}

Conversely, if the saturated nest $\CS$ is {\it compatibly rooted},
there is a natural way to associate with $\CS_\oplus $ an oriented ordered
tree ${\bf T}(\CS_\oplus)$. Indeed,
by definition, a saturated nest $\CS=\{R^\sI\}$ is constructed
by fusing recursively at each level $R^\sI$ exactly two connected components
$\CR^{\sI-1,k}$ and $\CR^{\sI-1,k'}$ of
the preceding level $R^{\sI-1}$.
Denoting by $i_\sI$ and ${i'}_\sI$ their respective roots in
$R^{\sI-1}_\oplus$,
one of these roots, say $i_{\sI}$, is
the root of $\CR^{\sI-1,k}\cup\CR^{\sI-1,k'}$ in $R^{\sI}_{\oplus}$,
since the rooting is compatible. In this case
the other root ${i'}_\sI$ can no longer be the root of any
connected component of the diagrams $R^{\sI'}$ for ${\scriptstyle I}'
\ge {\scriptstyle I}$.
Therefore, if we define by $\lambda_{\sI}=x_{{i'}_\sI}-x_{i_\sI}$
the oriented line vector
joining the positions of the roots $i_\sI$ and ${i'}_\sI$,
the set of $\lambda_\sI$ for
${\scriptstyle I}=1,\ldots ,N-1$ defines an {\it oriented ordered}
(by ${\scriptstyle I}$) tree, which we denote by ${\bf T}(\CS_\oplus)$.
Of course, we have by construction $\CS({\bf T}(\CS_\oplus))=\CS$.
Moreover, one can easily check that the tree
${\bf T}(\CS_\oplus)$ has the following property: for any $\SI$ and $\SI'$,
the path on the tree joining the two origins $x_{\sI}$ and $x_{\sI'}$
of the vectors $\lambda_\sI$ and $\lambda_{\sI'}$ passes only through
vectors $\lambda_{\scriptscriptstyle K}$ for ${\scriptstyle K }
> \min (\SI,\SI')$.
\medskip\noindent
Although this construction does not play any role in the present section 7, it
will turn out to be useful in section 8.
\midinsert
\figinsert{10.5truecm}{\hsize}{
\figcap\treetonestf{(a) Saturated nest associated with an ordered tree
${\bf T}=(\lambda_1,\lambda_2,\lambda_3,\lambda_4)$. The nest is made
of four diagrams. Each diagram is represented by the contour of its
connected components with at least two vertices (the diagrams 1, 3 and 4 have
only one such connected component, the diagram 2 has two such connected
components). (b) Oriented ordered tree associated with a compatibly
rooted saturated nest. We have first assigned compatible roots to the
saturated nest
of (a) (here the diagrams 3 and 4, and the connected
component on the right of the diagram 2 have the same root) and then
constructed the oriented ordered tree from these roots.}
}
\endinsert

\medskip \noindent{\it Definition 7.13: Extended Hepp Sectors}\par
Now we want
to associate with an {\it unrooted} saturated nest $\CS$
an {\it extended} Hepp sector, defined from the
Hepp sectors attached to ordered trees constructed in section 5.1 .
\medskip \noindent
If we consider as in section 5.1 the $N$ points as being embedded
in $\RR^{N-1}$ with Cartesian
coordinates $0,y_1,\ldots,y_{N-1}$,
and denote as before ${\cal H}^{\bf T}$ the domain of the $y_i$'s defining
the Hepp sector attached to the ordered tree\foot{We recall
that the domain $\CH^{{\bf T}}$ corresponds to the domain where the
$\lambda_\alpha$'s obtained from the $y_i$'s by Eq. \lamb\ are actual
successive minimal distances, and in particular satisfy $|\lambda_1|
\le \ldots \le |\lambda_{N-1}|$.} ${\bf T}$,
we define the Hepp sector $\CH^\CS$ as the union of all Hepp sectors
attached to all ordered trees ${\bf T}$
such that $\CS({\bf T})=\CS$,
that is the domain of the
$y_i$'s given by:
\eqn\extHepp{\CH^\CS=\bigcup_{{\bf T}:\, \CS({\bf T})=\CS}
\,\CH^{{\bf T}} .}

\medskip
This extended Hepp sector is best described by the vectors
$\lambda_\alpha$ associated
with a given (arbitrary) tree ${\bf T}$ such that $\CS({\bf T})=\CS$. Let us
stress that now the $\lambda_\alpha$'s are no longer successive minimal
distances when the $y_i$'s move everywhere inside
$\CH^\CS$, but are so only for $y_i$'s inside the subset $\CH^{{\bf T}} $
of $\CH^\CS$.
In particular, the inequalities $|\lambda_\alpha | \le |\lambda_{\alpha +1}|$
of \ineqmin\ are not necessarily satisfied inside $\CH^\CS$.
Still, for $y_i$'s inside $\CH^\CS$, one can find a tree ${\bf T^0}$
such that
$\CS({\bf T^0})=\CS({\bf T})$ and $\{y_i\}\in \CH^{\bf T^0}$. The
$\lambda^0_\alpha$ associated
with ${\bf T^0}$ satisfy for this set of $y_i$'s the inequalities
$|\lambda^0_1|\le \ldots \le |\lambda^0_{N-1}|$. By construction, one has
inside $\CH^{{\bf T^0}}$ at each level $\alpha$:
$|\lambda^0_\alpha|\le |\lambda_\alpha|$ and, as in
\equtree , a relation between the $\lambda_\alpha$'s and
the $\lambda^0_\alpha$'s of the form $\lambda_\alpha=\pm
\lambda^0_\alpha +\sum_{\gamma<\alpha}c^\gamma_\alpha \lambda^0_\gamma$
with some coefficients $c^\gamma_\alpha$ equal to $0$ or $\pm 1$.
We can thus write:
\eqn\inlamb{\eqalign{|\lambda_\alpha |&=
|\pm \lambda^0_\alpha+\sum_{\gamma<\alpha}
c^\gamma_\alpha \lambda^0_\gamma|\cr &\le
|\lambda^0_\alpha |+\sum_{\gamma<\alpha}
|c^\gamma_\alpha | |\lambda^0_\gamma |\cr&\le
(1+\sum_{\gamma<\alpha}
|c^\gamma_\alpha |)|\lambda^0_\alpha |\cr&\le \alpha |\lambda^0_\alpha | .\cr}
}
We thus have the set of inequalities:
\eqn\inlambda{|\lambda^0_\alpha |\le |\lambda_\alpha |\le \alpha
|\lambda^0_\alpha |}
which, together with $|\lambda^0_\alpha |\le |\lambda^0_{\alpha +1}|$
implies
\eqn\ineqlamb{{|\lambda_\alpha |\over |\lambda_{\alpha +1}|}\le \alpha .}
This is an example of constraints satisfied by all tree variables compatible
with the nest $\CS$ in the extended sector $\CH^\CS$, which is a relaxed
extension of \ineqmin\ .
Another consequence of \inlambda\ is that if ${\bf T}$ and ${\bf T'}$
are two trees such that $\CS({\bf T})=\CS({\bf T'})=\CS$, then inside
$\CH^\CS$, the corresponding line vectors satisfy:
\eqn\raplamb{\eqalign{{1\over \alpha}&\le {|\lambda_\alpha |\over |{\lambda
'}_\alpha |
}\le \alpha\cr&{|\lambda_\alpha |\over
|{\lambda '}_{\alpha'} |}\le \alpha \qquad {\rm for}\quad \alpha'
> \alpha .\cr}}
These bounds will be useful in section 8.
\medskip
The corresponding extended Hepp sector $\CA_N^\CS$ in the space $\CA_N$ of
mutual squared distances $a_{ij}$ between vertices (see section 3.2)
can be described simply,
without reference to ordered trees.
Given a saturated nest $\CS=\{R^0,\ldots,R^{N-1}\}$, let us consider,
for a given diagram $R^\sI$, the smallest
squared distance between vertices which
belong to two {\it different} connected components of the diagram $R^\sI$
(minimal squared distance between connected components):
$${a_{\rm min}}(R^\sI)\ =
\,\min_{\CR^{\sI,k}\ne\CR^{\sI,l}\in R^\sI}\,
\left( \min_{i\in\CR^{\sI,k},j\in\CR^{\sI,l}}\, \left( a_{ij}\right)\right).$$
For the minimal diagram $R^0=G_\odot$ one has obviously
${a_{\rm min}}(G_\odot)=\min\limits_{i\ne j} (a_{ij})$, and by
convention for the maximal
diagram $G=\{\CG\}$ (which has only one connected component) we set
${a_{\rm min}}(G)=\infty$.
One can check that one has always, for any saturated nest,
${a_{\rm min}}(R^0)\le{a_{\rm min}}(R^1)\le\ldots\le{a_{\rm
min}}(R^{N-2})<{a_{\rm min}}(R^{N-1})$.

The extended Hepp sector $\CA_N^\CS$ associated with the saturated nest $\CS$
is the subset of $\CA_N$ such that
\eqn\ineqAS{
{a_{\rm min}}(R^0)<{a_{\rm min}}(R^1)<\ldots<{a_{\rm min}}(R^{N-2})<{a_{\rm
min}}(R^{N-1}).}
One can check that the sectors associated with two different saturated nests
are disjoints $\CA_N^\CS\cap\CA_N^{\CS'}=\hbox{\O}$, and that $\CA_N$ is
the union of the closure of sectors over all saturated nests
$\CA_N=\bigcup\limits_{\CS\ {\rm saturated }}{\overline {\CA_N^\CS}}$.

\subsec{Equivalence classes of nests: an example}

In order to prove the finiteness of subtracted correlation functions
${\CZ^{\bf R}}^{(M)}_N$ in \ZMRen\  when $\epsilon = 0$, we shall
proceed in a way similar to what was done in section 5, by decomposing
the domain of
integration over positions into {\it extended} Hepp sectors and prove that
the integration of ${\bf R}[I_\CG(x_i,X_a)]$ inside each extended Hepp
sector yields a finite result.
\par
We have seen that UV divergences arise generally when successive
subsets of points coalesce. Inside the Hepp sector $\CH^\CS$, these successions
must be compatible with the nested structure of $\CS$.
 From \Roperbis\ the subtracted integrand is a sum of contributions
associated with (rooted) nests $\CN_\oplus$, and many contributions (for
different nests) give the same divergences inside $\CH^\CS$.
The general strategy to prove that the subtracted integrand
${\bf R}[I_\CG(x_i,X_a)]$ is convergent inside the sector $\CH^\CS$ is
to regroup the nests giving the same UV divergences into equivalence classes,
and to show that all divergences cancel within each equivalence class.

\medskip
Let us first consider the simple example
of a sector associated with
a saturated nest $\CS$ such that, at some level ${\scriptstyle I}_0$,
the diagram $R\equiv R^{\sI_0}$ has one and only one
connected component $\CR$ with $|\CR |>1$ and let us focus on
the behavior of the subtracted integrand when the points of $\CR$ coalesce.
More precisely, let us consider the contribution in $\bf R$ of a rooted nest
$\CN_\oplus$ with one single rooted diagram $T_\oplus$ where $T_\oplus$
also has one and only one element $(\CT,w)$ with $|\CT|>1$ (notice
that the nest $\CN_\oplus$ is automatically compatible).
The corresponding contribution is (up to a factor ${-1\over |\CT |})$:
\eqn\exemp{\Tay_{(\CT,w)}I_\CG(x_i,X_a)=
I_{\CT}(x_i)I_{\CG\setminusw\CT}(x_i,X_a)}
where we used as before in \BDeq\ the short-hand notation
$\CG\setminusw \CT\equiv \CG\setminus (\CT \setminus \{w\})$
which simply
corresponds to replacing $\CT$ in $\CG$ by its single vertex $w$.
We now ask which are the nests whose contribution leads to the same
UV behavior when the points of $\CR$ coalesce, that is when
the positions $x_i$ for $i\in \CR$ tend altogether to an arbitrary position
$x_0$: we shall denote this limit by $\CR\to 0$.
In this limit, the first term $I_{\CT}(x_i)$ in the r.h.s. of Eq.
\exemp\ factorizes into $I_{\CR\cap\CT}(x_i)\, I_{\CT\setminuso(\CR\cap\CT)}
(x_i)$, where the notation $``\setminuso"$ means that the vertices of
$\CR\cap\CT$ have
been replaced by a single contraction vertex $0$ with position $x_0$.
The factorization
of the second term $I_{\CG\setminusw\CT}(x_i,X_a)$ depends on
whether or not the point $w$ belongs to $\CR$.
\vfill\eject
\medskip \noindent
- case (a): $w\in\CR$\par
If $w\in \CR$, then we get $I_{\CG\setminusw \CT}(x_i,X_a)
\to I_{\CR\setminuso(\CR\cap \CT)
}(x_i) \, I_{\CG\setminuso(\CR\cup\CT)}(x_i,X_a)$.
The contribution of $T_\oplus$ \exemp\ thus behaves as:
\eqn\exempa{\Tay_{(\CT,w)}I_\CG(x_i,X_a) \buildrel {\CR \to 0} \over \sim
I_{\CR\cap \CT}(x_i)\,
I_{\CT\setminuso(\CR\cap\CT)}(x_i)\, I_{(\CR\cup \CT)\setminuso\CT}(x_i)
\, I_{\CG\setminuso(\CR\cup\CT)}(x_i,X_a)},
where we used the fact that $\CR\setminuso(\CR\cap\CT)=(\CR\cup\CT)\setminuso
\CT$.
In view of \exempa ,
let us now consider the product of Taylor operators associated with
the larger rooted nest ${\tilde \CN}_\oplus$ defined as:
\eqn\newnesta{{\tilde \CN}_\oplus=\Big\{ \big\{(\CR\cap \CT,w)\big\},
\big\{(\CT,w)\big\},\big\{(\CR\cup\CT,
\bullet)\big\} \Big\} }
with $``\bullet"$ standing for an arbitrary compatible root\foot{
This root is either $w$ or some vertex of $\CR\setminus\CT$.}
\foot{We use here
the convention that a diagram is explicited
by keeping only each of its connected components having more that one element.
For instance, $\big\{(\CT,w)\big\}$ is a short-hand notation for
$(G_\odot\vee \{\CT\},\omega)$
which means that the diagram must be completed by the set of all remaining
isolated points not already in $\CT$, while $\omega$ consists of the root $w$
plus these isolated points. Similarly,
Eq. \newnesta\ is a short-hand notation for
${\tilde \CN}_\oplus=\Big\{ (G_\odot,G_\odot),(G_\odot\vee\big\{\CR\cap
\CT\big\},\omega),
(G_\odot\vee\big\{\CT\big\},\omega),(G_\odot\vee\big\{\CR\cup\CT
\big\},\bullet) \Big\}$ .}.
This new nest can be seen as resulting  from the superposition of the
two nests $\CN_\oplus$ and $\CS$ at level ${\scriptstyle I}_0$.
Applying the corresponding three $\Tay$ on $I_\CG$ one obtains
\eqn\exprima{\prod_{{\tilde T}_\oplus \in {{\tilde \CN}_\oplus}}\Tay_{{\tilde
T}_\oplus}\big[I_\CG(x_i,X_a)\big]\propto
I_{\CR\cap \CT}(x_i)\,
I_{\CT\setminusw(\CR\cap\CT)}(x_i)\, I_{(\CR\cup \CT)\setminusw\CT}(x_i)
\, I_{\CG\setminusbullet(\CR\cup\CT)}(x_i,X_a) .}
In the same limit when all
points in $\CR$ coalesce to the single point $0$,
$w$ and the compatible root $\bullet$ are replaced by $0$ since they
both belong to $\CR$, and \exprima\ is equal to the r.h.s. of \exempa .

\medskip \noindent
- case (b): $w\notin\CR$\par
If $w\notin \CR$, then we get $I_{\CG\setminusw \CT}(x_i,X_a) \to
I_{(\CR\setminus\CT)
}(x_i) \, I_{(\CG\setminusw\CT)\setminuso(\CR\setminus\CT)}(x_i,X_a)$ and
the contribution of $T_\oplus$ \exemp\ behaves as:
\eqn\exempb{\Tay_{(\CT,w)}I_\CG(x_i,X_a) \buildrel {\CR \to 0} \over \sim
I_{\CR\cap \CT}(x_i)\,
I_{\CT\setminuso(\CR\cap\CT)}(x_i)\, I_{\CR\setminus \CT}(x_i)
\, I_{(\CG\setminusw\CT)\setminuso(\CR\setminus\CT)}(x_i,X_a) .}
The larger rooted nest ${{\tilde \CN}_\oplus}$ which gives a similar
contribution when
$\CR\to 0$ is now defined as:
\eqn\newnestb{{{\tilde \CN}_\oplus}=\Big\{ \big\{(\CR\cap \CT,\bullet)\big\},
\big\{(\CT,w)\big\},\big\{(\CR
\setminus\CT, \bullet),(\CT,w)\big\} \Big\} .}
Notice that the largest element of ${{\tilde \CN}_\oplus}$ is now a diagram
with
{\it two} connected components $\CR\backslash\CT$ and $\CT$.
\midinsert
\figinsert{13.truecm}{6.truecm}{%\hsize}{
\figcap\largnestf{Schematic picture of the rooted nest ${\tilde \CN}_\oplus$
when the root $w$ of $\CT$ (a) belongs to $\CR$, or (b) does not
belong to $\CR$.}
}
\endinsert
The two cases (a) and (b) can be unified in a single formula. If we
denote $T_\oplus$ by $(T,\omega)$ where $\omega={\rm root} (T_\oplus)=
\big\{ \{ w\} \big\}$, the nest ${{\tilde \CN}_\oplus}$ can be written in both
cases as:
\eqn\newnestd{{{\tilde \CN}_\oplus}=\Big\{ (R\wedge T, \bullet ),
(T,\omega),(R\veew T ,\bullet) \Big\} }
where we introduce the union operation $\veew$ of an unrooted
diagram $R$ and a rooted diagram $(T,\omega)$
\eqn\unionr{R\veew T \equiv \Big[ R\setminus (T\setminus \omega ) \Big] \vee
T=\Big[R \setminus \big\{ {\rm comp} (T_\oplus)\setminus {\rm root } (T_\oplus)
\big\} \Big] \vee {\rm comp}(T_\oplus)}
where $``\setminus"$ is the subtraction operation acting on diagrams as
defined in definition 7.5 in section 7 .
The result of this operation is an unrooted diagram equal to $\{\CR\cup\CT\}$
if
the root $w$ of $T$ belongs to the connected component $\CR$ of $R$, and
equal to $\{(\CR\backslash\CT,\CT)\}$ if $w$ does not belong to $\CR$:
\eqn\veeweq{R\veew T = \left\{ \matrix{
\{\CR\cup \CT \}& \qquad {\rm if}
\qquad & w \in \CR \cr
\{ \CR\setminus \CT, \CT\}&\qquad {\rm if} \qquad & w\notin \CR . \cr } \right.
}

The operation $R \veew T$ thus consists in a {\it fusion} operation of
$\CR $ and $\CT$ into $\CR\cup \CT$,
followed by a {\it cutting out} of $\CT$ from $\CR\cup \CT$
if the root $w$ is not shared by $\CR$.
\midinsert
\figinsert{8.truecm}{\hsize}{
\figcap\runometf{The unrooted complete diagram $R\veew T$ (thick full lines)
obtained from the unrooted complete diagram $R$ (dashed lines) and
the complete rooted diagram $T$ (thin full lines). The diagram $R\veew T$
is obtained by fusing each connected component of $T$ to the
connected component of $R$ to which its root belongs, and cutting
it out from all the other connected components of $R$.}
}
\endinsert
The above expression for $R\veew T$ can be applied to the more general case
when $R=\{\CR^i\}$ and $T_\oplus=(T,\omega)=(\{\CT_j\},\{w_j\})$ have
more that one connected component, with the result that each connected
component $\CT_j$ of $T$ is fused to the connected component $\CR^i$ of
$R$ which contains its root $w_j$, and cut out from all the other connected
components of $R$ which it intersects (see \runometf ).
Note that the operation $\veew $ crucially depends on the position of
the roots of the diagram $T$ on the right with respect to the connected
components of the diagram $R$ on the left, but that these roots are not
retained
as roots of the resulting diagram $R\veew T$ which {\it by definition} is
unrooted.
The product of Taylor operators associated with the nest
${\tilde \CN}_\oplus$ as given by \newnestd\
still corresponds in this case to the combined result of the Taylor operation
$\Tay_{T_\oplus}$ followed by the coalescence of the ${\rm Card}(R)$  connected
components of $R$ toward arbitrary points.

\medskip
Finally, we return to the original question of finding the nests ${\CN
'}_\oplus$ which
give the same UV behavior as $T_\oplus$ when components of $R$ coalesce.
These are the rooted nests which build the same factorized integrand
\exempa\ or \exempb\ (possibly generalized to several connected
components). They are characterized by $\CN_\oplus \subset {\CN '}_\oplus
\subset {\tilde \CN}_\oplus $. We therefore get the four nests:

\eqn\newnests{\eqalign{{\CN_\oplus}&=\Big\{
(T,\omega) \Big\}\cr
{\CN_\oplus}_2&=\Big\{ (R\wedge T, \bullet ),
(T,\omega) \Big\}\cr
{\CN_\oplus}_3&=\Big\{
(T,\omega),(R\veew T ,\bullet) \Big\}\cr
{\tilde \CN}_\oplus&=\Big\{ (R\wedge T, \bullet ),
(T,\omega),(R\veew T ,\bullet) \Big\} .\cr }}
One can check (see Appendix D) that the $(-1)$ and symmetry factors associated
with these four nests sum up to give zero exactly
(this includes a sum over the unspecified compatible roots $\bullet $).
As a consequence, the divergences induced
in the contributions of the four nests above by the coalescence of the points
in the subset $\CR$ cancel exactly.
This property can be generalized to nests $\CN_\oplus$ with an arbitrary
number of diagrams as well as to successive coalescences associated with
a saturated nest $\CS$. Indeed, from the nest $\CN_\oplus$,
we can build a family of
nests ${\CN '}_\oplus$ giving the same divergences when points coalesce
successively according to the nested structure of $\CS$; we then can check
that these divergences cancel exactly within the obtained family.
The details of this construction will be discussed in the next section.

\subsec{Equivalence classes of nests: general construction}
In this section, we present a general procedure for classifying nests
according to the diverging behavior of the associated counterterm in
a given sector. Our construction is inspired by a construction
by Berg\`ere and Lam in \BergLam\
in the context of local
field theories in the Schwinger representation.
Extensive modifications are however necessary in order to make this
construction
applicable in our context.

We denote by $\CS=\{R^0,R^1\ldots ,R^{N-1}\}$ a saturated nest of $\CG$,
which will be kept fixed throughout this section. We are going to regroup
all rooted nests into equivalence classes, associated with $\CS$.
\par\noindent{\it $\diamond$ Tableau construction}

 From now on and until the end of the article,
the only rooted nests which we shall consider will be {\it compatibly rooted
nests}.

Let us thus consider an arbitrary compatibly rooted nest
$\CN_\oplus=\{T_{\sJ \oplus} ;\SJ=0,\ldots ,\ST\}$ where
$T_{\sJ \oplus}=(T_\sJ,\omega_\sJ)$.
For this compatibly rooted nest, we define the (unrooted) complete diagram
\eqn\RIJ{R_\sJ^\sI\equiv R^\sI\veewj T_\sJ\equiv \big( R^\sI \setminus (T_\sJ
\setminus \omega_\sJ) \big)\vee T_\sJ}
and build the tableau
\eqn\omegaconst{\encadre{\hbox{$\displaystyle \matrix{
T_0&R_0^1\wedge T_1&R_0^2\wedge T_1&\ldots&R_0^\sI\wedge T_1&\ldots&R_0^{N-2}
\wedge T_1&R_0^{N-1}\wedge T_1 \cr
 & & & & & & &\cr
T_1&R_1^1\wedge T_2&R_1^2\wedge T_2&\ldots&R_1^\sI\wedge T_2&\ldots&R_1^{N-2}
\wedge T_2 &R_1^{N-1}\wedge T_2\cr
\vdots&\vdots&\vdots&\ddots&\vdots&\ddots&\vdots&\vdots\cr
T_\sJ&R_\sJ^1\wedge T_{\sJ+1}&R_\sJ^2\wedge T_{\sJ+1}&\ldots&R_\sJ^\sI\wedge
T_{\sJ+1}&\ldots&R_\sJ^{N-2} \wedge T_{\sJ+1} &R_\sJ^{N-1} \wedge T_{\sJ+1}\cr
\vdots&\vdots&\vdots&\ddots&\vdots&\ddots&\vdots&\vdots \cr
T_\sT&R_\sT^1\wedge T_{\sT+1}&R_\sT^2\wedge T_{\sT+1}&\ldots&R_\sT^\sI\wedge
T_{\sT+1}&\ldots&R_\sT^{N-2} \wedge T_{\sT+1} &R_\sT^{N-1}\wedge T_{\sT+1}\cr
}$}}
}
where by convention $T_{\sT +1}\equiv G=\big\{ \CG \big\}$.
Notice that for $R^0=G_\odot$, we have $R_\sJ^0=R_\sJ^0\wedge T_{\sJ+1}=T_\sJ$.
Hence the first column $T_\sJ=R_\sJ^0\wedge T_{\sJ+1}$ of the tableau can
be seen as being build from $R^0$, with the same structure as the other
columns.
Notice also that since $R^{N-1}=G$, $R_\sJ^{N-1}=G$ for any ${\scriptstyle J}$,
hence $R_\sJ^{N-1}\wedge T_{\sJ +1}=T_{\sJ +1}$. Therefore the last element
of a given line of the tableau is identical to the first element of
the following line.
Finally, since $R^\sI \prec R^{\sI +1}$, then $R_\sJ^\sI\prec R_\sJ^{\sI +1}$
and
\eqn\ordre{R_\sJ^\sI\wedge T_{\sJ +1} \prec R_\sJ^{\sI +1}\wedge T_{\sJ +1} .}
Therefore, reading
the tableau in the natural order, {\it i.e.}
reading successive lines from the left to the right, we get a totally
nested structure, which defines an unrooted nest $\tilde \CN$.
This nest ${\tilde \CN}(\CS,\CN_\oplus)$  depends on
both the sector nest $\CS$ and the subtraction nest $\CN_\oplus$.
By construction, ${\tilde \CN}$ contains all the diagrams of $\CN$.
Of course, it may happen that two successive elements of the tableau
are identical (this is for instance the case for the last element of a line
and the first element of the next line), hence the tableau contains
redundant information.
\medskip
\noindent The nest ${\tilde \CN}$ is a generalization of the one
constructed in the previous section (Eq. \newnestd ). Indeed, if we
consider the nest ${\CN_\oplus}=\big\{(G_\odot,G_\odot),(T,\omega )\big\}$
and set $R^{\sI_0}=R$ at level ${\scriptstyle I}_0$ of the nest $\CS$,
we obtain in this case the simple tableau
\eqn\omegaex{\matrix{&&&\cr
T_0=G_\odot\hfill &\ldots&R_0^{\sI_0}\wedge T_1
=(R\veegodot G_\odot )\wedge T= R\wedge T\hfill &\ldots\cr & & & \cr
T_1=T\hfill &\ldots&R_1^{\sI_0}\wedge
T_2=(R\veew T)\wedge G=R\veew T\hfill &\ldots\cr
&&&\cr}}
where only columns $1$ and ${\scriptstyle I}_0$ are
specified. The general construction
\omegaconst\ therefore reproduces in this simple case exactly the largest nest
${\tilde \CN}$ (here unrooted) of \newnests\ .
\medskip
\noindent {\it $\diamond$ Reduction of the tableau}

Going back to the general case, we are now interested in
finding the {\it smallest} rooted nest $\CN^0_\oplus$ which,
under a construction similar to
\omegaconst , gives the same nest $\tilde \CN$ (that is
${\tilde \CN}(\CS,\CN^0_\oplus)={\tilde \CN}(\CS,\CN_\oplus)$).
More precisely, we must remove from $\CN_\oplus$ the diagrams $T_\sJ$ which
are not necessary to build $\tilde \CN$. Since $T_\sJ$ is involved is the
construction of the two lines ${\scriptstyle J}-1$ and $\scriptstyle J$,
removing $T_\sJ$ from the nest $\CN_\oplus$ amounts to replace
these two lines by a single line, which will be built directly
from $T_{\sJ-1}$ and $T_{\sJ +1}$. In this process, $N$
diagrams will be lost. Therefore, removing $T_\sJ$ will be possible
if the tableau contains $N$ redundant diagrams, which happens when
at least $N+1$ successive diagrams of the two lines ${\scriptstyle J}-1$ and
$\scriptstyle J$ are identical. This implies that there exists an
${\scriptstyle I}_0$ such that the two vertically adjacent elements of the
column ${\scriptstyle I}_0$ {\it coincide} at levels ${\scriptstyle J}-1$
and $\scriptstyle J$:
\eqn\supcond{R_{\sJ -1}^{\sI_0}\wedge T_{\sJ}=R_\sJ^{\sI_0} \wedge T_{\sJ +1}}
that is, on the tableau:
\eqn\omegared{\matrix{
T_0&R_0^1\wedge T_1&\ldots&R_0^{\sI_0}\wedge T_1&\ldots&R_0^{N-1}
\wedge T_1 \cr
\vdots&\vdots&\ddots&\vdots&\ddots&\vdots\cr
\hbox{\vbox{\hbox{$\displaystyle T_{\sJ-1}$}\hbox{}\hbox{$\displaystyle
T_\sJ$}}}&
\hbox{\vbox{\hbox{$\displaystyle R_{\sJ-1}^1\wedge
T_\sJ$}\hbox{}\hbox{$\displaystyle R_\sJ^1\wedge T_{\sJ+1}$}}}&
\hbox{\vbox{\hbox{$\ldots$}\hbox{}\hbox{$\ldots$}}}&
\raise -10 pt \encadre{\hbox{\vbox{
\hbox{$\displaystyle R_{\sJ-1}^{\sI_0}\wedge T_\sJ$}\hbox{$\qquad \parallel
$}\hbox{$\displaystyle
R_\sJ^{\sI_0}\wedge T_{\sJ+1}$}}}}&
\hbox{\vbox{\hbox{$\ldots$}\hbox{}\hbox{$\ldots$}}}&
\hbox{\vbox{\hbox{$\displaystyle R_{\sJ-1}^{N-1}\wedge
T_\sJ$}\hbox{}\hbox{$\displaystyle R_\sJ^{N-1}\wedge T_{\sJ+1}$}}}\cr
\vdots&\vdots&\ddots&\vdots&\ddots&\vdots\cr
T_\sT&R_\sT^1\wedge T_{\sT+1}&\ldots&R_\sT^{\sI_0}\wedge
T_{\sT+1}&\ldots&R_\sT^{N-1} \wedge T_{\sT+1} \cr
}}
Then, by the inclusion property \ordre , all the diagrams of ${\tilde \CN}$
between $R_{\sJ-1}^{\sI_0}\wedge T_\sJ$
and $R_\sJ^{\sI_0}\wedge T_{\sJ+1}$ are identical,
hence equal to $T_\sJ$ itself.
We thus don't loose any information by replacing the two lines
${\scriptstyle J}-1$ and $\scriptstyle J$ by the single line:
\eqn\singline{\matrix{T_{\sJ -1},&R_{\sJ-1}^1\wedge T_\sJ,&\ldots, &
\ \ \ R_{\sJ-1}^{\sI_0}\wedge T_\sJ\hfill& & \cr & & &=T_\sJ\hfill&  & \cr
& & &=R_\sJ^{\sI_0}\wedge T_{\sJ+1},\hfill &\ldots,&R_\sJ^{N-1}\wedge
T_{\sJ+1}.\cr}}
The important point is that this new line is precisely
the one which would have been constructed directly
by \omegaconst , when applied to the nest
$${\CN '}_\oplus=(T_{0\oplus},
T_{1\oplus},\ldots ,T_{\sJ -1\oplus},T_{\sJ +1\oplus}, \ldots, T_{T\oplus})$$
obtained from $\CN_\oplus$ by removing $T_{\sJ\oplus}$ (notice
that the induced rooting of this nest remains compatible).
Indeed, the construction
\omegaconst\ for ${\CN '}_\oplus$ simply corresponds to suppressing the
$\scriptstyle J$--line and to substituting to the $({\scriptstyle J}-1)$--line
the new line, constructed from $T_{\sJ-1\oplus}$ and $T_{\sJ +1\oplus}$:
\eqn\singliner{T_{\sJ -1}, \ \ \ R_{\sJ-1}^1\wedge T_{\sJ+1},\ \ \ \ldots,
\ \ \ R_{\sJ-1}^{\sI_0}\wedge T_{\sJ+1},\ \ \ \ldots,\ \ \ R_{\sJ-1}^{N-1}
\wedge T_{\sJ+1},}
the other lines staying unchanged.
It is the purpose of Appendix E to establish in detail the statement,
on which all our construction will rely, that the lines \singline\ and
\singliner\ are actually {\it identical} when \supcond\ is satisfied.
As a consequence, the nests
${\tilde \CN}(\CS, \CN_\oplus)$ and ${\tilde \CN}(\CS, {\CN '}_\oplus)$
are equal. In particular, we note
that $T_\sJ$, while absent from ${\CN '}_\oplus$, is still
present in ${\tilde \CN}(\CS,{\CN '}_\oplus)$ since
\eqn\survie{T_\sJ=R_{\sJ-1}^{\sI_0}\wedge T_{\sJ +1} .}
The ``suppression" of line
${\scriptstyle J}$ from \omegared\ when \supcond\ is satisfied,
consistent with the construction of ${\tilde \CN}(\CS,{\CN '}_\oplus)$,
can be visualized as follows:
\medskip
$$\matrix{{\scriptstyle J}-1&
\hbox to 60pt{\hrulefill}\hskip -60pt \raise 3.pt\hbox to 60 pt {\hrulefill}
\ {\scriptstyle I}_0\
\hbox to 60pt{\hrulefill}\hskip -60pt \raise .3pt\hbox to 60 pt {\hrulefill}
& & &\cr
& & & {\scriptstyle J}-1&
\hbox to 60pt{\hrulefill}\hskip -60pt \raise 3.pt\hbox to 60 pt {\hrulefill}
\ {\scriptstyle I}_0\
\hbox to 60pt{\hrulefill}\hskip -60pt \raise 2.5pt\hbox to 60 pt {\hrulefill}
\hskip -60pt \raise 5.pt\hbox to 60pt{\hrulefill}
\cr
{\scriptstyle J}&
\hbox to 60pt{\hrulefill}\hskip -60pt \raise .3pt\hbox to 60 pt {\hrulefill}
\ {\scriptstyle I}_0\
\hbox to 60pt{\hrulefill}\hskip -60pt \raise 2.5pt\hbox to 60 pt {\hrulefill}
\hskip -60pt \raise 5.pt\hbox to 60pt{\hrulefill}
&\longrightarrow & &\cr
&&&{\scriptstyle J}+1&\ldots \hfill\cr
{\scriptstyle J}+1&\ldots\hfill&&&\cr}
$$
where the double and triple lines represent successively nested	(in general
distinct) diagrams, while the single line represents a series of identical
diagrams.
\medskip
We therefore have at our disposal a reduction procedure, which allows
for the substitution to the nest $\CN_\oplus$ of the reduced nest
${\CN '}_\oplus$,
with one diagram less, which still generates the same nest ${\tilde \CN}$.
This process can be iterated to suppress all the diagrams $T_\sJ$ of the
original nest $\CN$ which are such that they satisfy the coincidence
property \supcond\ for
at least one ${\scriptstyle I}_0$ ($1\le {\scriptstyle I}_0\le N-1$).
When two successive lines possess this coincidence property, for
some ${\scriptstyle I}_0$ and ${\scriptstyle I}_1$, the reduction
is {\it associative}, that is its result is independent of the order of the
operations, as represented on the following picture:
$$\matrix{&&
\hbox to 20pt{\hrulefill}\hskip -20pt \raise 3.pt\hbox to 20 pt {\hrulefill}
\ {\scriptstyle I}_0\
\hbox to 20pt{\dotfill}\hskip -20pt \raise 3.pt\hbox to 20 pt {\dotfill}
\ {\scriptstyle I}_1\
\hbox to 20pt{\dotfill}
&&\cr
&&&&\cr
\hbox to 20pt{\hrulefill}\hskip -20pt \raise 3.pt\hbox to 20 pt {\hrulefill}
\ {\scriptstyle I}_0\
\hbox to 55pt{\hrulefill}\hskip -55pt \raise .3pt\hbox to 55 pt {\hrulefill}
&\rightarrow&
\hbox to 55pt{\dotfill}
\ {\scriptstyle I}_1\
\hbox to 20pt{\hrulefill}\hskip -20pt \raise 2.5pt\hbox to 20 pt {\hrulefill}
\hskip -20pt \raise 5.pt\hbox to 20pt{\hrulefill}
&&\cr
&&&&\cr
\hbox to 20pt{\hrulefill}\hskip -20pt \raise .3pt\hbox to 20 pt {\hrulefill}
\ {\scriptstyle I}_0\
\hbox to 20pt{\dotfill}\hskip -20pt \raise 3.pt\hbox to 20 pt {\dotfill}
\ {\scriptstyle I}_1\
\hbox to 20pt{\dotfill}
&&&\rightarrow&
\hbox to 20pt{\hrulefill}\hskip -20pt \raise 3.pt\hbox to 20 pt {\hrulefill}
\ {\scriptstyle I}_0\
\hbox to 20pt{\dotfill}\hskip -20pt \raise 3.pt\hbox to 20 pt {\dotfill}
\ {\scriptstyle I}_1\
\hbox to 20pt{\hrulefill}\hskip -20pt \raise 2.5pt\hbox to 20pt {\hrulefill}
\hskip -20pt \raise 5.pt \hbox to 20pt{\hrulefill}
\cr
&&&&\cr
\hbox to 55pt{\dotfill}
\ {\scriptstyle I}_1\
\hbox to 20pt{\hrulefill}\hskip -20pt \raise 2.5pt\hbox to 20 pt {\hrulefill}
\hskip -20pt \raise 5.pt\hbox to 20pt{\hrulefill}
&\rightarrow&
\hbox to 20pt{\hrulefill}\hskip -20pt \raise 3.pt\hbox to 20 pt {\hrulefill}
\ {\scriptstyle I}_0\
\hbox to 55pt{\hrulefill}\hskip -55pt \raise .3pt\hbox to 55 pt {\hrulefill}
&&\cr
&&&&\cr
&&
\hbox to 20pt{\hrulefill}\hskip -20pt \raise .3pt\hbox to 20 pt {\hrulefill}
\ {\scriptstyle I}_0\
\hbox to 20pt{\dotfill}\hskip -20pt \raise 3.pt\hbox to 20 pt {\dotfill}
\ {\scriptstyle I}_1\
\hbox to 20pt{\hrulefill}\hskip -20pt \raise 2.5pt\hbox to 20 pt{\hrulefill}
\hskip -20pt \raise 5.pt\hbox to 20pt{\hrulefill}
&&\cr
}$$
Notice furthermore
that a configuration like
$$\matrix{
{\scriptstyle J}-1&
\hbox to 55pt{\hrulefill}\hskip -55pt \raise 3.pt\hbox to 55 pt {\hrulefill}
\ {\scriptstyle I}_0\
\hbox to 20pt{\hrulefill}\hskip -20pt \raise .3pt\hbox to 20 pt {\hrulefill}
\cr
&&&&\cr
{\scriptstyle J}&
\hbox to 20pt{\hrulefill}\hskip -20pt \raise .3pt\hbox to 20 pt {\hrulefill}
\ {\scriptstyle I}_1\
\hbox to 20pt{\hrulefill}\hskip -20pt \raise .3pt\hbox to 20 pt {\hrulefill}
\ {\scriptstyle I}_0\
\hbox to 20pt{\hrulefill}\hskip -20pt \raise .3pt\hbox to 20pt {\hrulefill}
\cr
&&&&\cr
{\scriptstyle J}+1&
\hbox to 20pt{\hrulefill}\hskip -20pt \raise .3pt\hbox to 20pt {\hrulefill}
\ {\scriptstyle I}_1\
\hbox to 55pt{\hrulefill}\hskip -55pt \raise 2.5pt\hbox to 55 pt {\hrulefill}
\hskip -55pt \raise 5.pt\hbox to 55pt{\hrulefill}
\cr
}$$
which would cause obstruction to associativity, is actually forbidden
since it would imply $T_\sJ=T_{\sJ+1}$, which is ruled out by definition.
Notice finally that the ``suppression" of a line $\scriptstyle J$
does not create new coincidences (that is coincidences which did not exist
before suppression). Indeed, the only pairs of vertical neighbors which
are modified by the suppression are those of the lines ${\scriptstyle J}-2$ and
${\scriptstyle J}-1$ for ${\scriptstyle I}>{\scriptstyle I}_0$ one the one
hand,
and those of the lines ${\scriptstyle J}$
and ${\scriptstyle J}+1$ for ${\scriptstyle I}<{\scriptstyle I}_0$ on the other
hand,
as can be seen on the following picture:
$$\matrix{{\scriptstyle J}-2&\hbox to 75pt{}\hbox to 60pt{\dotfill}\hskip -60pt
\raise 3.pt \hbox to 60pt{\dotfill}& & & \cr
& & &{\scriptstyle J}-2&\hbox to 75pt{}
\hbox to 60pt{\dotfill}\hskip -60pt\raise 3.pt
\hbox to 60pt{\dotfill}\cr
{\scriptstyle J}-1&
\hbox to 60pt{\hrulefill}\hskip -60pt \raise 3.pt\hbox to 60 pt {\hrulefill}
\ {\scriptstyle I}_0\
\hbox to 60pt{\hrulefill}\hskip -60pt \raise .3pt\hbox to 60 pt {\hrulefill}
& & &\hskip 100.pt \ne \hfill \cr
& &\longrightarrow & {\scriptstyle J}-1&
\hbox to 60pt{\hrulefill}\hskip -60pt \raise 3.pt\hbox to 60 pt {\hrulefill}
\ {\scriptstyle I}_0\
\hbox to 60pt{\hrulefill}\hskip -60pt \raise 2.5pt\hbox to 60 pt {\hrulefill}
\hskip -60pt \raise 5.pt\hbox to 60pt{\hrulefill}
\cr
{\scriptstyle J}&
\hbox to 60pt{\hrulefill}\hskip -60pt \raise .3pt\hbox to 60 pt {\hrulefill}
\ {\scriptstyle I}_0\
\hbox to 60pt{\hrulefill}\hskip -60pt \raise 2.5pt\hbox to 60 pt {\hrulefill}
\hskip -60pt \raise 5.pt\hbox to 60pt{\hrulefill}
& & &\hskip 25pt \ne \hfill
\cr
&&&{\scriptstyle J}+1&
\hbox to 60pt{\dotfill}\hskip -60pt\raise 2.5pt
\hbox to 60pt{\dotfill}\hskip -60pt\raise 5.pt \hbox to 60pt{\dotfill}
\hbox to 75pt{}\cr
{\scriptstyle J}+1&\hbox to 60pt{\dotfill}\hskip -60pt \raise 2.5pt
\hbox to 60pt{\dotfill}\hskip -60pt \raise 5.pt \hbox to 60pt{\dotfill}
\hbox to 75pt{}&&&\cr
}
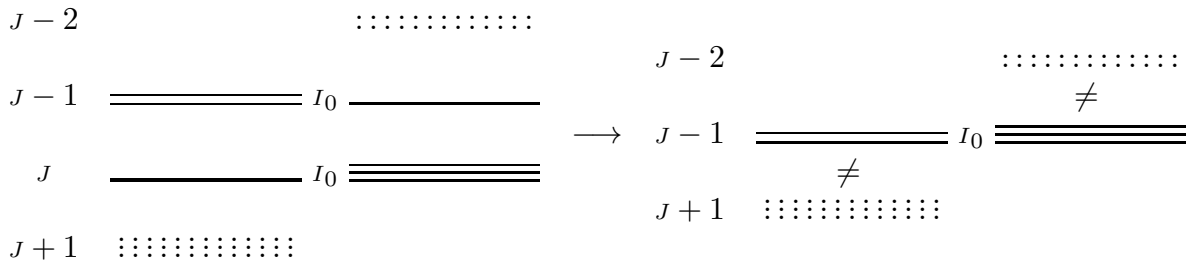
$$
A new coincidence would imply $T_{\sJ -1}=T_\sJ$ in the first (upper right)
case, and
$T_\sJ=T_{\sJ +1}$ in the second (lower left) case,
and is thus impossible. Therefore, after ``suppression" of all the lines
of the original nest which present a vertical coincidence with the preceding
line, we end up with a tableau which no longer contains any pair of
coinciding vertical neighbors.
We denote by $\CN^0_\oplus$ the nest resulting from this reduction procedure,
that is the subset of $\CN_\oplus$ made of the
diagrams $T_{\sJ\oplus}$ for values of $\scriptstyle J$ corresponding to lines
which {\it remain} after reduction.

\medskip
\noindent {\it $\diamond$ Equivalence classes of nests}

The above reduction allows to assign to any compatibly rooted nest
$\CN_\oplus$ a unique minimal nest $\CN^0_\oplus$,
which is a subset of the original nest $\CN_\oplus$
(and in particular whose compatible rooting is the restriction
of the original rooting of $\CN_\oplus$ to $\CN^0$),
such that ${\tilde \CN}(\CS,\CN^0_\oplus)={\tilde \CN}(\CS,\CN_\oplus)$,
and whose tableau
\omegaconst\ is ``minimal", {\it i.e.} has no vertically adjacent coinciding
elements\foot{In general, this tableau still contains series of identical
successive elements, but not more than that $N$ successive elements
can be identical.}.

We define the equivalence class $\CC_\CS(\CN^0_\oplus)$ of a minimal
(with respect to $\CS$) nest
$\CN^0_\oplus$ as the set of all compatibly rooted nests
$\CN_\oplus$ which lead by reduction of their $\CS$-tableau to that minimal
nest $\CN^0_\oplus$.
$$\CN_\oplus \in \CC_\CS(\CN^0_\oplus)\quad \Longleftrightarrow \quad
\CN_\oplus \buildrel {\rm
tableau} \over \longrightarrow {\tilde \CN}(\CS,\CN_\oplus) \buildrel
{\rm reduction}\over \longrightarrow \CN^0_\oplus\ .$$
Of course, if $\CN^0_\oplus$ is minimal with respect to $\CS$,
one has $\CN^0_\oplus \in \CC_\CS(\CN^0_\oplus)$.
For \break any $\CN_\oplus \in \CC_\CS(\CN^0_\oplus)$, one has
${\tilde \CN}(\CS,\CN_\oplus )={\tilde \CN}(\CS,\CN^0_\oplus)$.

We have the following characterization, for any compatibly
rooted nest $\CN_\oplus$ (with $\CN$ the corresponding unrooted nest):
\medskip
\noindent{THEOREM } {\it Characterization of $\CC_\CS (\CN^0_\oplus)$}
\eqn\Charcno{\CN_\oplus \in \CC_\CS(\CN^0_\oplus) \qquad \Longleftrightarrow
\qquad ({\rm a})\ \CN^0_\oplus \subset \CN_\oplus \quad {\rm and} \quad
({\rm b})\  \CN\subset
{\tilde \CN}(\CS,\CN^0_\oplus) .}
A nest of the equivalence class $\CC_\CS(\CN^0_\oplus)$ is
thus constituted of {\it all} the diagrams of $\CN^0$ plus {\it some} of the
diagrams of ${\tilde \CN}(\CS,\CN^0_\oplus)$ not in $\CN^0$.
Its rooting is constrained to be both
compatible and such that its restriction to $\CN^0$ is the rooting
of $\CN^0_\oplus$.
Conversely, one builds all the elements of $\CC_\CS(\CN^0_\oplus)$ by
completing $\CN^0_\oplus$ by an arbitrary number of diagrams of
${\tilde \CN}(\CS,\CN^0_\oplus) \setminus\CN^0$
(that is diagrams of ${\tilde \CN}(\CS,\CN^0_\oplus)$
not in $\CN^0$),
and assigning to these extra elements any roots compatible
with the roots of $\CN^0_\oplus$.
\par \noindent The direct implication ($\Longrightarrow$) is immediate since
\item {-} the reduced rooted nest is always a subset of the original
rooted nest, hence (a);
\item {-} any diagram of $T_\sJ$ of $\CN_\oplus$ belong to
${\tilde \CN}(\CS,\CN_\oplus)$ and the reduction process is defined
so as to leave $\tilde \CN$ invariant. Thus $T_\sJ\in {\tilde \CN}(
\CS,\CN^0_\oplus)$, hence (b).
\par \noindent
The reverse implication ($\Longleftarrow$) is not immediate and is
proven in Appendix F.
\medskip
\noindent Notice finally that the diagram $G=\{\CG\}$ is always a diagram
of ${\tilde \CN}(\CS,\CN^0_\oplus)$ since the last element
(${\scriptstyle I}=N-1$) of the
last line (${\scriptstyle J}={\scriptstyle T}$) of the tableau
of any nest is always equal to $G$. As a consequence, $G$ is never
a diagram of $\CN^0_\oplus$ since it can be rebuilt from $\CN^0_\oplus$
by the tableau construction. Actually, if a nest contains
the diagram $G$, the line of its tableau
built from $G$ has all its elements equal to $G$, while the preceding
line has its last element equal to $G$; this
leads to the coincidence property for this two lines for
${\scriptstyle I}_0=N-1$, indicating that $G$ is to be suppressed in
the construction of $\CN^0_\oplus$.
Therefore, for any minimal nest $\CN^0_\oplus$, one
has $G\in {\tilde \CN}(\CS,\CN^0_\oplus)\setminus \CN^0_\oplus$.

\subsec{Factorization of the ${\bf R}$ operator inside an equivalence class}
As we have seen before, the reason for classifying nests
into equivalence classes was to regroup nests
whose diverging contributions in a given sector $\CS$
in the ${\bf R}$ operator \Roperbis\
cancel exactly.
Given a sector nest $\CS$, it is therefore natural to rewrite
the ${\bf R}$ operator, which is a sum over all compatibly
rooted nests, as a sum of reduced
operators ${\bf R}_{\CC_\CS(\CN^0_\oplus)}$,
each of them involving all the nests
in the equivalence
class $\CC_\CS(\CN^0_\oplus)$ of a minimal (w.r.t. $\CS$) nest $\CN^0_\oplus$.
This reads:
\eqn\Rclass{{\bf R}=\sum_{\CN^0_\oplus\ {\rm minimal}\atop
\qquad {\rm w.r.t.} \CS}{\bf R}_{\CC_\CS(\CN^0_\oplus)}
\ ,\qquad {\bf R}_{\CC_\CS(\CN^0_\oplus)}=-\sum_{\CN_\oplus \in \CC_\CS
(\CN^0_\oplus)}W(\CN_\oplus)\prod_{T_\oplus \in \CN_\oplus}\Big(
-\Tay_{T_\oplus}\Big).}
Each operator $R_{\CC_\CS(\CN^0_\oplus)}$ can then be rewritten
as a sum of {\it factorized} contributions associated with different
rootings of the elements of the equivalence class, as explained
now.

We will need a lemma about partial sums over compatible rootings of nests.
Let us consider a nest $\CM=\{T_\sJ;\SJ=1,\ldots,\ST\}$.
We denote by $\oplus_\CM$ a {\it compatible rooting} of $\CM$,
that is simply the specification for each diagram $T_\sJ$ of $\CM$
of a root diagram $\omega_\sJ$ such that
$\CM_{\oplus_\CM}\equiv\{(T_\sJ,\omega_\sJ);\SJ=1,\ldots \ST\}$
is a compatibly rooted nest.
\medskip
\noindent {\it Lemma:}

Given a compatibly
rooted nest $\CN_\oplus$ and an unrooted nest $\CM$ such
that $\CN\subset\CM$ (that is all the diagrams of $\CN$ are
diagrams of $\CM$), we can consider all the compatible rootings
$\oplus_\CM$ of $\CM$ such that $\CN_\oplus \subset \CM_{\oplus_\CM}$,
that is the compatible rootings of $\CM$ whose restriction to $\CN$
is the rooting in $\CN_\oplus$; we then have the useful sum rule for the
weights \Wnest  :
\eqn\somme{\encadre{\hbox{$\displaystyle
\sum_{\oplus_\CM: \  \CN_\oplus \subset \CM_{\oplus_\CM}}
W(\CM_{\oplus_\CM})=W(\CN_\oplus)$}}. }
This lemma is proven in Appendix G.
\medskip
We can now use this property in the case of an arbitrary
 nest $\CN_\oplus \in
\CC_\CS(\CN^0_\oplus)$ if we choose:
\eqn\tableaumin{\CM={\tilde \CN }(\CS,\CN^0_\oplus)}
since, from \Charcno , we have $\CN\subset \CM$.
Inserting \somme\ in the formula \Rclass\ for ${\bf
R}_{\CC_\CS(\CN^0_\oplus)}$,
we get:
\eqn\factRoper{\eqalign{{\bf R}_{\CC_\CS(\CN^0_\oplus)}&=-
\sum_{\CN_\oplus \in \CC_\CS(\CN^0_\oplus)}\sum_{\quad \oplus_\CM:
\atop \CN_\oplus
\subset \CM_{\oplus}} W(\CM_{\oplus})\prod_{T_\oplus \in \CN_\oplus}
\big( -\Tay_{T_\oplus}\big)\cr&\cr  &=-
\sum_{\quad \oplus_\CM: \atop \CN^0_\oplus \subset \CM_{\oplus}}
W(\CM_{\oplus}) \sum_{\CN_\oplus : \atop \CN^0_\oplus \subset \CN_\oplus
\subset \CM_{\oplus}} \prod_{T_\oplus \in \CN_\oplus} \big(
-\Tay_{T_\oplus}\big)\cr &\cr &=-
\sum_{\quad \oplus_\CM: \atop \CN^0_\oplus \subset \CM_{\oplus}}
W(\CM_{\oplus}) \prod_{T^0_\oplus \in \CN^0_\oplus} \big( -\Tay_{T^0_\oplus}
\big) \prod_{{T}_\oplus \in (\CM_{\oplus}\setminus \CN^0_\oplus)}
\big( 1-\Tay_{{T}_\oplus}\big)\cr}}
where $\CM_\oplus$ stands here for $\CM_{\oplus_\CM}$. In the second
equation, we used the characterization \Charcno\ of $\CC_\CS(\CN^0_\oplus)$.
The sum rule \somme\ allows us to reconstruct all possible rootings
of the nests $\CN$ in $\CC_\CS(\CN^0_\oplus)$ with the appropriate weight,
by first fixing the roots of $\CM_\oplus$ by a compatible extension
of the roots of $\CN^0_\oplus$, and then restricting these roots
of $\CM_\oplus$ to all intermediate subnests $\CN$ between $\CN^0$ and $\CM$
(notice that a given rooting of such a nest $\CN$ can come from different
rootings of $\CM$).
In the last equation, we used the fact that the set of rooted
nests $\CN_\oplus$ such
that $\CN^0_\oplus\subset\CN_\oplus\subset\CM_\oplus$ is built by
taking necessarily, on the one hand all the diagrams $T^0_\oplus$ of the
minimal nest $\CN^0_\oplus$ and, for each diagram $T_\oplus$
of $\CM_\oplus\setminus
\CN^0_\oplus$ on the other hand, deciding whether to take it or not,
hence choosing
$1$ or $-\Tay_{T_\oplus}$ in the expansion of the product of Taylor operators.

Notice finally that the compatibly rooted
nests $\CM_\oplus$ involved in \factRoper\
can actually be characterized independently of the minimal nest $\CN^0_\oplus$
from which they are built, by the property:
\eqn\maxnest{{\tilde \CN}(\CS,\CM_\oplus)=\CM .}
A compatibly rooted nest satisfying \maxnest\ will be called {\it maximal}
with respect to $\CS$.
With this definition, the equations \Rclass\ and \factRoper\ can be
replaced by the single equation:
\eqn\Rdecomp{\encadre{\hbox{$\displaystyle {\bf R}=\sum_{\CM_\oplus\
{\rm maximal}\atop
\qquad {\rm w.r.t.} \CS}W(\CM_\oplus)\, {\bf R}_{\CM_\oplus}$}}}
with
\eqn\factRM{\encadre{\hbox{$\displaystyle
{\bf R}_{\CM_\oplus}=-
\prod_{T^0_\oplus \in \CN^0_\oplus}\Big( -\Tay_{T^0_\oplus}\Big)
\prod_{T_\oplus \in (\CM_\oplus\setminus \CN^0_\oplus)}\Big(
1 -\Tay_{T_\oplus}\Big)
$}}}
where $\CN^0_\oplus$ is now the minimal nest obtained by reducing the tableau
of the maximal nest $\CM_\oplus$.

\newsec{Proof of UV convergence}
\def\Tt{{\tilde T}}
\def\CTt{{\tilde \CT}}
We are now in a position to
prove the finiteness of subtracted correlation functions
${\CZ^{\bf R}}^{(M)}_N$ in \ZMRen\  when $\epsilon = 0$.
Our strategy is the following:
\item{(I)} First
we partition the domain of
integration over positions into {\it extended} Hepp sectors (as
defined in section 7.2), each of them being
characterized by a saturated nest $\CS$.
\item{(II)} In each sector $\CS$,
we reorganize the ${\bf R}$ operator by use of \Rdecomp\
as a sum of operators ${\bf R}_{\CM_\oplus}$
associated with the different nests $\CM_\oplus$ maximal with respect to
$\CS$.
\item{(III)}
At the end, one can write ${\CZ^{\bf R}}^{(M)}_N$ as
\eqn\lastfinite{{\CZ^{\bf R}}^{(M)}_N(X_a)=\sum_{\CS}
\sum_{\CM_\oplus\ {\rm  maximal} \atop {\rm w.r.t.} \CS}W(\CM_\oplus)
\int_{\CH^{\CS}}\prod_{i\in \CG}d^Dx_i \, {\bf R}_{\CM_\oplus}\,
[I_\CG (x_i,X_a)] .}
It is therefore sufficient to prove the finiteness of the
integral:
\eqn\tobefinite{\int_{\CH^{\CS}}\prod_{i\in \CG}d^Dx_i \, {\bf
R}_{\CM_\oplus}\,
[I_\CG (x_i,X_a)]}
where we integrate over the domain $\CH^{\CS}$ defined by \extHepp\ \foot{
More precisely, we integrate over the $x_i's$ such that the $y_i$'s
defined by $y_i=x_{i+1}-x_1$ are in $\CH^\CS$, since a Hepp sector is actually
defined in terms of relative positions.}
with the measure
\Compacext , and where $\CM_\oplus$ is any  nest maximal with respect
to $\CS$.
\item{(IV)} Using the factorized form \factRM\ for ${\bf R}_{\CM_\oplus}$,
we first apply the Taylor operators $\Tay_{T^0_\sJ}$ associated with
diagrams of the minimal nest $\CN^0_\oplus$. This results in factorizing
$I_\CG(x_i,X_a)$ into a product of amplitudes $I_\Tt=\prod\limits_{\CTt\in \Tt}
I_{{\tilde \CT}}$ for
suitable reduced diagrams $\Tt$ made of subsets ${\tilde \CT}$ of $\CG$.
\item{(V)} We show that the product of the remaining
$(1-\Tay_{T_\oplus})$ operators
acts independently on each subdiagram ${\tilde T}$, leading to a subtracted
integrand for ${\tilde T}$.
\item{(VI)} We show that this subtracted integrand, once integrated
over points in the Hepp sector $\CH^\CS$, yields
a finite result when $\epsilon =0$.
\medskip
Points (I), (II) and (III) have been already discussed in section 7.
We now show points (IV), (V) and (VI) precisely.
\subsec{ Factorization of ${\tit I_\CG (x_i,X_a)}$ }
In order to precise the action of ${\bf R}_{\CM_\oplus}$ on $I_\CG (x_i,X_a)$,
let us first have a closer look at the tableau
$\CM={\tilde \CN}(\CS,\CN^0_\oplus)$.
We denote by $T^0_\sJ$, ${\scriptstyle J}=0,\ldots,{\scriptstyle T}$ the
diagrams of $\CN^0$, and by $T^\sI_\sJ$
\eqn\TIJ{T^\sI_\sJ=(R^\sI \veewzeroj T^0_{\sJ})\wedge T^0_{\sJ+1} \qquad
0\le {\scriptstyle I}\le N-1;\quad \SJ=0,\ldots,\ST}
the diagrams of $\CM$.
By convention, we have set $T^0_{\sT+1}=G$.
Starting from the factorized form \factRM\ for ${\bf R}_{\CM_\oplus}$
(and using the fact that the $\Tay$\ 's commute),
we first apply the Taylor operators $\Tay_{T^0_{\sJ\oplus}}$ associated
with diagrams
of the minimal nest $\CN^0_\oplus$. This results in factorizing
$I_\CG(x_i,X_a)$
into:
\eqn\tayfac{
\prod_{T^0_{\sJ\oplus} \in \CN^0_\oplus}\left(\Tay_{T^0_{\sJ\oplus}}\right)
\, I_\CG(x_i,X_a)=\ I_{{\tilde T}_{\sT+1}}(x_i,X_a)\,
\prod_{\sJ=1}^{\sT} I_{{\tilde T}_{\sJ}}(x_i)
}
where
\eqn\Ttild{{\tilde T}_\sJ\equiv T^0_\sJ\setminuswjmin T^0_{\sJ-1}}
is the (uncomplete) diagram
obtained from $T^0_\sJ$ by replacing by its root each
component of the preceding diagram $T^0_{\sJ-1}$ in $\CN^0_\oplus$.
Each $\Tt_\sJ$ is made of ${\rm Card}(T^0_\sJ)$ connected components
${\tilde \CT}_{\sJ,j}$
and in \tayfac\ the amplitude for $\Tt_\sJ$ is by definition equal to
\eqn\TtildI{I_{{\tilde T}_\sJ}\equiv \prod_{j=1}^{{\rm Card} (T^0_\sJ)}
I_{{\tilde \CT}_{\sJ,j}}.}
By convention, if some connected
component is reduced to one single vertex, the corresponding amplitude
is $1$. Eq. \tayfac\ establishes point (IV).
\medskip

Similarly to \Ttild , it is convenient to define $\Tt^\sI_{\sJ-1}$ as
the (uncomplete)
diagram obtained by reducing in
some diagram of the tableau $T^\sI_{\sJ-1}$ the preceding minimal diagram
$T^0_{\sJ-1}$ to its root diagram $\omega^0_{\sJ-1}$:
\eqn\TtildIJ{{\tilde T}^\sI_{\sJ-1}\equiv T^\sI_{\sJ-1}\setminuswjmin
T^0_{\sJ-1}.}
Notice that $\Tt^{0}_{\sJ-1}=\omega^{0}_{\sJ-1}$ and that
$\Tt^{N-1}_{\sJ-1}=\Tt_{\sJ}$.

 From \TIJ ,
the connected components of the diagram $\Tt^{\sI}_{\sJ-1}$ are made
out of the intersection of
the connected components $\CTt_{\sJ,j}$ of $\Tt_\sJ$ and of the connected
components $\CR^{\sI,i}$ of $R^\sI$
\eqn\TtildeIiJj{
\CTt^{\sI,i}_{\sJ-1,j}\,\equiv \,\CR^{\sI,i}\cap\CTt_{\sJ,j}
}
Furthermore, from the compatibility requirement for $\CM_\oplus$, the root in
$\CM_\oplus$
of any connected component of the diagram $T^\sI_{\sJ-1}$ automatically
belongs to the corresponding reduced connected component of the
reduced diagram ${\tilde T}^{\sI}_{\sJ-1}$. Therefore, the rooting of
$\CM_\oplus$ naturally induces a rooting for the diagrams
${\tilde T}^{\sI}_{\sJ-1}$.
We denote by $w^{\sI,i}_{\sJ-1,j}$ the root of ${\tilde \CT}^{\sI,i}_{\sJ-1,j}$
and by $\Tt^\sI_{\sJ-1\oplus}$ the set of all $(\CTt^{\sI,i}_{\sJ-1,j},
w^{\sI,i}_{\sJ-1,j})$ for varying $i$ and $j$.
\medskip \noindent
Let us for a while
concentrate on what happens inside some given subset $\CTt_{\sJ,j}$
which we shall assume to have at least two vertices
(${\rm Card}(\CTt_{\sJ,j})>1$).
We can consider the family of different (and non empty) rooted subsets
$(\CTt^{\sI,i}_{\sJ-1,j},w^{\sI,i}_{\sJ-1,j})$ for
all $i=1,\ldots,{\rm Card}(R^\sI)$ (with ${\scriptstyle J}$ and $j$ fixed) as a
complete rooted diagram $\Tt^\sI_{\sJ-1,j\oplus}$ of the subset
$\CTt_{\sJ,j}$ in which we are now working.
 From \TtildeIiJj , this is nothing but the restriction of the diagram $R^\sI$
to this
subset $\CTt_{\sJ,j}$, together with a set of roots.
The family of distinct $\Tt^\sI_{\sJ-1,j\oplus}$ for varying
${\scriptstyle I}$
forms a {\it compatibly rooted and saturated nest},
$\CS_{\sJ,j\oplus}$,
of $\CTt_{\sJ,j}$, which is nothing but
the restriction of the saturated nest $\CS$ to $\CTt_{\sJ,j}$, with
a given rooting.
We define:
\eqn\IbarJj{{\overline{\rm Ind}}({\scriptstyle J},j)=\left\{ {\scriptstyle I}
\ge 1:\
\Tt^\sI_{\sJ-1,j}\ne \Tt^{\sI-1}_{\sJ-1,j}
\right\}}
as the set of indices ${\scriptstyle I}$ (of the sector $\CS$) such that
inside $\CTt_{\sJ,j}$, a new element $\Tt$ appears at level
${\scriptstyle I}$ in the saturated nest
$\CS_{\sJ,j}$.
\medskip
We now again consider the whole diagram $\Tt_{\sJ}$ and define, in a way
similar to \IbarJj :
\eqn\IbarJ{{\overline{\rm Ind}}({\scriptstyle J})
=\left\{ {\scriptstyle I} \ge 1:\
\Tt^\sI_{\sJ-1}\ne \Tt^{\sI-1}_{\sJ-1}
\right\}
=\left\{ {\scriptstyle I} \ge 1:\
T^\sI_{\sJ-1}\ne T^{\sI-1}_{\sJ-1}
\right\}
}
as the set of ${\scriptstyle I}$ such that a new diagram appears in the
tableau at level $\scriptstyle I$ between $T^0_{\sJ-1}$ and  $T^0_\sJ$.
Of course, if $\Tt^{\sI}_{\sJ-1}\ne \Tt^{\sI-1}_{\sJ-1}$, there exists
at least one $j$ such that $\Tt^{\sI}_{\sJ-1,j}\ne \Tt^{\sI-1}_{\sJ-1,j}$
and thus:
\eqn\fJjtJ{{\overline{\rm Ind}}(\SJ)=\bigcup_{j=1}^{{\rm Card}(T^0_\sJ)}
{\overline{\rm Ind}}(\SJ,j)}
We moreover denote by
\eqn\iminmaxJ{
{\scriptstyle I}^{\rm min}({\scriptstyle J})=
\min \Big( {\overline{\rm Ind}}({\scriptstyle J})\Big) \ ,\qquad
{\scriptstyle I}^{\rm max}({\scriptstyle J})=
\max \Big( {\overline{\rm Ind}}({\scriptstyle J})\Big)
}
with the property that
\eqn\imaxJprop{
{\scriptstyle I}^{\rm max}({\scriptstyle J})=
\min\left\{{\scriptstyle I}\,:\,
\Tt^\sI_{\sJ-1}=\Tt_{\sJ}\right\}
= \min\left\{{\scriptstyle I}\,:\,
T^\sI_{\sJ-1}=T^{0}_{\sJ}\right\}
}
is the index $\SI$ such that $T^0_\sJ$ appears at first in the tableau
\noindent We set:
\eqn\IsetJ{{\rm Ind}({\scriptstyle J})={\overline{\rm Ind}}(\SJ)
\setminus \{ \SI^{\rm max}({\scriptstyle J}) \} }
(which may be empty).

Finally, we define
\eqn\Ibar{{\overline{\rm Ind}}=\bigcup_{\sJ =1}^{\sT+1} {\overline{\rm Ind}}
({\scriptstyle J}).}

With these notations, the $(1-\Tay )$ operators in \factRoper\
act independently on each amplitude $I_{\Tt_{\sJ}}$. The operator
$(1-\Tay_{T_{{\scriptscriptstyle K}\oplus}^{\sI}})$ acts on $I_{\Tt_{\sJ}}$
only if ${\scriptstyle K}={\scriptstyle J}-1$, and results
in this case in $(1-\Tay_{\Tt^{\sI}_{\sJ-1\oplus}})[I_{\Tt_{\sJ}}]$.
We thus can express ${\bf R}_{\CM_\oplus}[I_\CG ]$ as a product of
subtracted amplitudes for each reduced diagram $\Tt_{\sJ}$.
The subtracted amplitude
for $\Tt_{\sJ}$ is obtained by the successive action on $I_{\Tt_{\sJ}}$
of a $(1-\Tay_{\Tt^{\sI}_{\sJ-1\oplus}})$ operator for each $\SI\in
{\rm Ind}(\SJ)$. The case $\SJ=\ST+1$ is special since, since in addition
to the $(1-\Tay_{\Tt^{\sI}_{\sT\oplus}})$ operator for each $\SI\in
{\rm Ind}(\ST+1)$, a $(1-\Tay)$ operator is also associated with
$\Tt^{\sI^{\rm max}(\sT+1)}_{\sT\oplus}$
\foot{Notice that $\Tt_{\sT+1}$ has only one connected component.}.
The factorization of ${\bf R}_{\CM_\oplus}[I_\CG ]$
is then expressed in the following equation:
\eqn\temporary{\eqalign{
{\bf R}_{\CM_\oplus}[I_\CG (x_i,X_a)]=&
\prod_{\sI\in {\overline{\rm Ind}}(\sT+1)}(1-\Tay_{{\tilde
T}_{\sT\oplus}^{\sI}})
I_{{\tilde T}_{\sT+1}}(x_i,X_a)\cr &\quad \times
\prod_{\sJ=1}^{\sT}\Big[
\prod_{\sI\in {\rm Ind}(\sJ)}(1-\Tay_{{\tilde T}_{\sJ-1\oplus}^{\sI}})
[I_{{\tilde T}_{\sJ}}(x_i)]\Big] .\cr}
}
We recall that
\eqn\Rappel{\Tay_{\Tt^\sI_{\sJ-1\oplus}}=\prod_{j=1}^{{\rm Card}(T^0_\sJ)}
\Tay_{\Tt^\sI_{\sJ-1,j\oplus}}=\prod_{j=1}^{{\rm Card}(T^0_\sJ)}
\prod_{(\CTt^{\sI,i}_{\sJ-1,j},w^{\sI,i}_{\sJ-1,j})\in \Tt^\sI_{\sJ-1,j\oplus}}
\Tay_{(\CTt^{\sI,i}_{\sJ-1,j},w^{\sI,i}_{\sJ-1,j})} .}
We have thus achieved point (V).
It remains to show that the subtractions
associated with the $(1-\Tay)$'s are sufficient to make \temporary\
integrable in the sector $\CS$.

\subsec{Appropriate tree variables}

In section 5, in order to prove the convergence of the original (unsubtracted)
integral in some Hepp sector (for $\epsilon>0$), we found useful to express
the measure in terms of tree variables for some specific tree
(which defined the sector).  Those tree variables are no longer
adapted to deal with the subtracted amplitude
${\bf R}_{\CM_\oplus}[I_\CG (x_i,X_a)]$ since they do not take into account
the factorization \temporary\ of
${\bf R}_{\CM_\oplus}[I_\CG (x_i,X_a)]$. Instead,
we shall look for tree variables associated with
a tree that, inside each subset $\CTt_{\sJ,j}$, forms a subtree compatible
with the sector.  \medskip The basic idea is that, since the nest
$\CS_{\sJ,j\oplus}$, which has been defined above as the restriction
of the sector nest $\CS$ to $\CTt_{\sJ,j}$, is both saturated in
$\CTt_{\sJ,j}$ and rooted, it naturally defines a unique oriented ordered
tree ${\bf T}_{\sJ,j}$ spanning
the vertices of $\CTt_{\sJ,j}$, as discussed in section 7.2.
The corresponding line vectors are naturally ordered by increasing
values of ${\scriptstyle I}$ in
${\overline {\rm Ind}}(\SJ, j) $
and denoted by
\eqn\lambdaIJj{\lambda^\sI_{\sJ,j};\quad {\scriptstyle I}\in
{\overline{\rm Ind}}(\SJ,j). }
 From the nested structure of $\CN^0_\oplus$, we deduce that the union
of the trees ${\bf T}_{\sJ,j}$ for varying ${\scriptstyle J}$ and $j$
(including ${\scriptstyle J}={\scriptstyle T}+1$)
forms a tree of the set $\CG$:
\eqn\Aptree{{\bf T}=\bigcup {\bf T}_{\sJ,j}=\Big( \lambda^\sI_{\sJ,j};\quad
{\scriptstyle J}=1,\ldots,
{\scriptstyle T}+1; \ j=1,\ldots, {\rm Card}(T^0_{\sJ});\
{\scriptstyle I}\in {\overline{\rm Ind}}(
{\scriptstyle J},j)
\Big)}
\midinsert
\figinsert{13.truecm}{6.5truecm}{%\hsize}{
\figcap\apptreevf{Appropriate tree variables.
At level $\SJ$, inside a connected
component $\CTt_{\sJ,j}$ of $\Tt_{\sJ}$ (dashed circles), we build an oriented
ordered tree with line vectors $\lambda^{\sI}_{\sJ,j}$.
As shown in the framed box, this tree is built in a way similar to what was
done in \treetonestf\ (b), now from the
rooted saturated nest $\CS_{\sJ,j\oplus}$. This nest is here made of the three
diagrams $\Tt^{\sI_1}_{\sJ-1,j}$, $\Tt^{\sI_2}_{\sJ-1,j}$ and
$\Tt^{\sI_3}_{\sJ-1,j}$ ({\it i.e.} ${\overline{\rm Ind}}(\SJ,j)=
\{ \SI_1,\SI_2,\SI_3\}$) whose roots are represented  by the dashed squares.
At level $\SJ+1$, the connected components of $\Tt_{\sJ}$ are fully
contracted toward their roots (big black dots), which are the vertices
of $\CTt_{\sJ+1,k}$. An oriented ordered tree with line vectors
$\lambda^{\sI'}_{\sJ+1,k}$ is then built inside $\CTt_{\sJ+1,k}$. The trees
at level $\SJ$ and $\SJ+1$ can be fused into a single oriented (but
only partially ordered) larger tree contributing to \Aptree\ .}
}
\endinsert
\noindent (see \apptreevf ).
In particular, this tree has $N-1$ line vectors. We can therefore
use the corresponding
tree variables $\lambda^\sI_{\sJ,j}$ as integration variables, instead
of the $N-1$ relative positions $y_i=x_{i+1}-x_1$ in $\RR^{N-1}$ (or
$\RR^{N+M-1}$ when $M$ external points are present).
\medskip
Notice that the tree ${\bf T}$ is not in general compatible with
the sector $\CS$, that is in general, $\CS({\bf T})\ne \CS$.
Still, since $\CS_{\sJ,j}$ is the restriction of $\CS$ to the
subset $\CTt_{\sJ,j}$, the subtree ${\bf T}_{\sJ,j}$ of ${\bf T}$
remains compatible with $\CS$. By this we mean that one can find
ordered trees of $\CG$ compatible with $\CS$, and which contain
${\bf T}_{\sJ,j}$ as an ordered subtree.
We can therefore take advantage of the inequalities \raplamb\ and
get the following bounds for ratios of length of $\lambda^\sI_{\sJ,j}$
inside the domain $\CH^\CS$:
\eqn\bounds{\eqalign{{1\over {\scriptstyle I}}&\le {|\lambda^\sI_{\sJ,j} |
\over |{\lambda}^{\sI}_{\sJ',j'} |}
\le {\scriptstyle I} \cr&{|\lambda^\sI_{\sJ,j} |\over
|{\lambda}^{\sI'}_{\sJ',j'} |}\le {\scriptstyle I} \qquad {\rm for}\quad
{\scriptstyle I} < {\scriptstyle I}\,' .\cr }}
This means that two $\lambda$'s with the same index $\scriptstyle I$ are
of the same order, while the $\lambda$'s with higher index
${\scriptstyle I}\,'>{\scriptstyle I}$ cannot vanish more rapidly
than those with index $\scriptstyle I$.
\medskip
Finally, since the vectors $\lambda^{\sI}_{\sJ,j}$ defining
the subtree ${\bf T}_{\sJ,j}$ are built from the {\it rooted}
nest $\CS_{\sJ,j\oplus}$, whose roots are precisely the roots
$w^{\sI,i}_{\sJ-1,j}$ of the subsets $\CTt^{\sI,i}_{\sJ-1,j}$,
the action of dilation operations \xrhobis
\eqn\actionpos{\prod_{i}
\CD^\rho_{(\CTt^{\sI,i}_{\sJ-1,j},w^{\sI,i}_{\sJ-1,j})}}
(for some fixed $\SJ$ and $j$) on the positions of the vertices of
$\CTt_{\sJ,j}$ is exactly performed by the transformation
\eqn\actionmod{|\lambda^{\sI'}_{\sJ,j} | \to \rho |\lambda^{\sI'}_{\sJ,j}|
\qquad {\rm for} \qquad \SI\,' \le \SI }
on the {\it modules} of the $\lambda$ variables.
\medskip
In a way similar to what we did in section 5.1, it is natural to rewrite
the vectors $\lambda$'s in terms of real variables $\beta^\sI$ which
measure ratios of successive modules $|\lambda |$'s, together with angular
variables $\theta$.

\def\MAX{E}
For definiteness, we write the elements of ${\overline{\rm Ind}}$ (\Ibar ) as
\eqn\indelmts{{\overline{\rm Ind}}=
\{ \SI_1<\SI_2<\ldots\SI_E\} .}
For each $\SI$ in ${\overline{\rm Ind}}$, we choose one of the
$\lambda^\sI_{\sJ,j}$ of the tree ${\bf T}$ as a representative
of all the lines which appear at level $\SI$, and denote
it by $\lambda^\sI$.
We then define the $\beta$ variables as the ratios of these
representatives $\lambda^\sI$ for successive $\SI$ in ${\overline{\rm Ind}}$:
\eqn\thebetas{\eqalign{\beta^{\sI_k}&=
{|\lambda^{\sI_k}|\over |\lambda^{\sI_{k+1}}|}
\qquad k=1,\ldots,\MAX -1 \cr
& \beta^{\sI_{\MAX}}=|\lambda^{\sI_{\MAX}}| .
\cr }}
 From \bounds , they satisfy
\eqn\betabounds{\beta^{\sI_k} \le \SI_k \qquad k=1,\ldots,{\MAX}-1 .}
Eq. \thebetas\  defines $\beta^\sI$ for $\SI\in{\overline{\rm Ind}}$.
We shall also use the convention
\eqn\betaone{\beta^\sI=1\quad{\rm if}\quad \SI\notin{\overline{\rm Ind}} .}
In order to compare $\lambda^\sI_{\sJ,j}$ to its representative
$\lambda^\sI$, we define
\eqn\chiIJj{\chi^\sI_{\sJ,j}={|\lambda^\sI_{\sJ,j}|\over
|\lambda^\sI |},}
with of course $\chi^\sI_{\sJ,j}=1$ if $\lambda^\sI_{\sJ,j}$
has been chosen as a representative.
We thus have
\eqn\lambbetchi{|\lambda^\sI_{\sJ,j}|=\chi^\sI_{\sJ,j}\beta^\sI
\beta^{\sI+1}\ldots\beta^{N-1}}
and the bounds
\eqn\chibounds{{1\over \SI}\le \chi^\sI_{\sJ,j} \le \SI .}
Finally, in addition to their moduli, the line vectors $\lambda^\sI_{\sJ,j}$
are characterized by a set of relative angles $\theta^{\sI,n}_{\sJ,j}$
labeled by some extra index $n$. These angles can be constructed in
different ways, corresponding in particular
to different orderings of the tree ${\bf T}$. In any case, as in section 5.2,
these angle variables do not actually play any role in the proof
of the finiteness of the integral \tobefinite . Therefore, we shall
not make their construction more explicit.
\medskip
In terms of the $|\lambda |$ and $\theta$ variables, the measure term
can be written, up to a global numerical factor,
as in {\it e.g.} \measthetaext , as
\eqn\measlambdaIJ{
\prod_{\sJ=1}^{\sT+1}\prod_{j=1}^{{\rm Card}(T^0_{\sJ})}
\prod_{\sI\in {\overline{\rm Ind}}(\sJ,j)}\Big[
d|\lambda^\sI_{\sJ,j}|\, |\lambda^\sI_{\sJ,j}|^{D-1}
\Big(
\prod_{n}(\sin
\theta^{\sI,n}_{\sJ,j})^{p(D,n)}d\theta^{\sI,n}_{\sJ,j}\Big)\Big]
}
where $p(D,n)$ is some positive number (when $D\ge N+M-1$).

\subsec{Subtracted integrand}

The tree variables of the preceding section, together with
the factorization \temporary\ allow us to work separately
inside each reduced diagram $\Tt_{\sJ}$. Indeed, the amplitude
$I_{\Tt_{\sJ}}$ for $\SJ\le \ST$ is a function of the
variables $\lambda^\sI_{\sJ,j}$ for the same $\SJ$ only,
with $\SI \in {\overline{\rm Ind}}(\SJ,j)$
(the case $\SJ=\ST+1$ which is special since it also involves the set
of external points, will be discussed separately).
Going back to the definition of the amplitude $I$, we can write
\tobefinite\ in a form where the measure and the integrand
are factorized simultaneously. For each $\Tt_{\sJ}$ (with $\SJ\le \ST$),
we get

\eqn\thebigone{\eqalign{\prod_{j=1}^{{\rm Card}(T^0_\sJ)}
\prod_{\sI\in {\overline{\rm Ind}}(\sJ,j)}d|\lambda^\sI_{\sJ,j}|
|\lambda^\sI_{\sJ,j}|^{D-1}&\big(\prod_n\sin(\theta_{\sJ,j}^{\sI,n})^{p(D,n)}
\big)\cr &\times \prod_{\sI\in {\rm Ind}(\sJ)}\big(1-
\Tay_{\Tt_{\sJ-1\oplus}^\sI} \big)
\Big[\prod_{j=1}^{{\rm Card}(T^0_\sJ)}\det(\Pi^{{\bf T}_{\sJ,j}})
\Big]^{-{d\over 2}}\cr}
}
where $\Pi^{{\bf T}_{\sJ,j}}$ is the matrix defined by \Piquad ,\Pifig \
for the subtree ${\bf T}_{\sJ,j}$. Its elements
$\Pi^{{\bf T}_{\sJ,j}}_{\sI,\sI'}$ are labeled by elements $\SI,\SI'$
of ${\overline{\rm Ind}}(\SJ,j)$.
As in the convergence proof of section 5, we introduce the normalized
matrix:
\eqn\UpsilJj{\Avril^{{\bf T}_{\sJ,j}}_{\sI,\sI'}\equiv{1\over A_D(\nu)}
{\Pi^{{\bf T}_{\sJ,j}}_{\SI,\SI'}
\over |\lambda^\sI_{\sJ,j}|^\nu|\lambda^{\sI'}_{\sJ,j}|^\nu}}
which, inserted in \thebigone , gives
\eqn\theBigone{\eqalign{\prod_{j=1}^{{\rm Card}(T^0_\sJ)}
\prod_{\sI\in {\overline{\rm Ind}}(\sJ,j)}{d|\lambda^\sI_{\sJ,j}|\over
|\lambda^\sI_{\sJ,j}|^{1-\epsilon}}&
\big(\prod_n\sin(\theta_{\sJ,j}^{\sI,n})^{p(D,n)}
\big)\cr & \times
\prod_{\sI\in {\rm Ind}(\sJ)}\big(1-\Tay^{\ 0}_{\Tt_{\sJ-1\oplus}^\sI} \big)
\Big[\prod_{j=1}^{{\rm Card}(T^0_\sJ)}
\det(\Avril^{{\bf T}_{\sJ,j}})\Big]^{-{d\over 2}} .\cr}
}
Since through \UpsilJj\ we have extracted the most singular factor
of the $\Pi$ matrices under rescalings $\CD^\rho$ (see \xrhobis ),
the Taylor operators $\Tay^{\ 0}$ appearing in \theBigone\
are now defined as
\eqn\taylorzero{
\Tay^{\ 0}\ =\ \lim_{\rho\to 0} \CD^\rho .}
The properties of $\det(\Avril^{{\bf T}_{\sJ,j}})$ are the same as those
mentioned in section 5.2. In particular, $\det(\Avril^{{\bf T}_{\sJ,j}})$
is a function of the ratios of
$\lambda^\sI_{\sJ,j}$ for successive $\SI$ in ${\overline{\rm Ind}}(\SJ,j)$,
which play the role of the $\beta_\alpha$ variables of section 5, and
are now products of the $\beta$ and $\chi$ variables defined above.

\noindent Then $\det(\Avril^{{\bf T}_{\sJ,j}})$
is a bounded function of the
$\beta^\sI$ and $\chi^\sI_{\sJ,j}$ variables
on the domain $\CH^\CS$, and is equal to $1$ when
all the $\beta^\sI$ are set to zero.

\noindent Due to our choice for the $\lambda^\sI_{\sJ,j}$ variables,
the action of $\Tay^{\ 0}_{\Tt_{\sJ-1\oplus}^\sI}$ on
$\displaystyle \prod_{j=1}^{{\rm Card}(T^0_\sJ)}
\Big(\det(\Avril^{{\bf T}_{\sJ,j}})\Big)^{-{d\over 2}}$ simply
corresponds to set $\beta^\sI=0$ in all the $\Avril^{{\bf T}_{\sJ,j}}$
for different $j$ (see \actionpos\ and \actionmod ).
Therefore
$\displaystyle
(1-\Tay^{\ 0}_{\Tt^\sI_{\sJ-1\oplus}})\Big[\prod_{j=1}^{{\rm Card}(T^0_\sJ)}
\Big(\det(
\Avril^{{\bf T}_{\sJ,j}})\Big)^{-{d\over 2}}\Big]$
vanishes
when $\beta^\sI\to 0$.
This is the key property which will ensure the finiteness of the
subtracted integrals. First we have to generalize this fact to all the $\beta$
variables. This is contained
in the following stronger property, as
shown in Appendix H:
\medskip
\noindent PROPOSITION:
\eqn\GrandO{\encadre{\hbox{$\displaystyle
\prod_{\sI\in {\rm Ind}(\sJ)}
(1-\Tay^{\ 0}_{\Tt^\sI_{\sJ-1\oplus}})\Big[\prod_{j=1}^{{\rm Card}(T^0_\sJ)}
\Big(\det( \Avril^{{\bf T}_{\sJ,j}})\Big)^{-{d\over 2}}
\Big]
={\cal O}\Big(\prod_{\sI^{\rm min}(\sJ)\le \sI<\sI^{\rm max}(\sJ)}
(\beta^\sI)^\delta \Big)
$}}}
with $\delta=\min (\nu,1-\nu)$ as in \defdelt.

\medskip
The above discussion holds for $\SJ\le\ST$ only. The
case $\SJ=\ST+1$ (and $j=1$) requires a separate analysis.
We now have
\eqn\ITT{I_{\Tt_{\sT+1\oplus}}=
\Big(\det( \Avril^{{\bf T}_{\sT+1,1}})\Big)^{-{d\over 2}}
\exp\Big(-{1\over 2}\sum_{a,b}\kvec_a\cdot\kvec_b\Delta_{ab} \Big)
}
and a property similar to \GrandO :
\eqn\GrandOT{\encadre{\hbox{$\displaystyle
\prod_{\sI\in {\overline{\rm Ind}}(\sT+1)}\kern -15pt
(1-\Tay^{\ 0}_{\Tt^\sI_{\sT\oplus}})\Big[\Big(\det(
\Avril^{{\bf T}_{\sT+1,1}})\Big)^{-{d\over 2}}\kern -5pt
\exp\Big(-{1\over 2}\sum_{a,b}\kvec_a\cdot\kvec_b\Delta_{ab}\Big)
\Big]
={\cal O}\Big(\kern -20pt \prod_{\qquad \sI \ge \sI^{\rm min}(\sT+1) }
\kern -20pt (\beta^\sI)^\delta \Big)
$}}}
\medskip
\subsec{Proof of finiteness }

 From the above discussion, we arrive at the
following form for \tobefinite\ at $\epsilon=0$:
\eqn\thelastone{\eqalign{\int_{\CD^\CS}&
\prod_{\sJ=1}^{\sT+1}\prod_{j=1}^{{\rm Card}(T^0_{\sJ})}
\Bigg[ \prod_{\sI\in {\overline{\rm Ind}}(\sJ,j)}\Big(\prod_{n}
(\sin \theta^{\sI,n}_{\sJ,j})^{p(D,n)}d\theta^{\sI,n}_{\sJ,j}\Big)
{\prod_{\sI\in {\overline{\rm Ind}}(\sJ,j)}}\kern -10pt {\raise 5pt \hbox{$'$}
}\kern 10pt {d\chi^\sI_{\sJ,j}
\over \chi^\sI_{\sJ,j}}\Bigg]
\cr& \times \prod_{\sI\in {\overline{\rm Ind}}}{d\beta^\sI\over \beta^\sI}
\ \ \CO \Big( \prod_{\sI\ge \sI^{\rm min}(\sT+1)}(\beta^\sI)^\delta \Big)
\ \prod_{\sJ=1}^{\sT}\CO \Big( \prod_{\sI^{\rm min}(\sJ)\le \sI<\sI^{\rm
max}(\sJ)
}(\beta^\sI)^\delta \Big)
\cr}
}
where $\prod'$ means that we omit the values of $\SI$ such that
$\lambda^\sI_{\sJ,j}$ is a representative, and
where the domain of integration $\CD^\CS$
reproduces the domain of integration $\CH^\CS$ for
the relative positions of internal points.
Inside $\CD^\CS$, the variables $\chi^{\sI}_{\sJ,j}$ are bounded from below
according to \chibounds . Therefore, the integration over these
$\chi^{\sI}_{\sJ,j}$ variables and the integration over the
$\theta^{\sI,n}_{\sJ,j}$ variables do not produce any divergence.
For the integral to be convergent,
it is actually sufficient
that, for each $\SI\in {\overline{\rm Ind}}$, at least one $(\beta^\sI)^\delta$
is present in the product of $\cal O$'s appearing in \thelastone , thus
making the integration over $\beta^\sI$ UV convergent.
This will be true if
\eqn\prbeta{{\overline{\rm Ind}}\subset
\Big\{ \bigcup_{\sJ=1}^{\sT}\Big[\SI^{\rm min}(\SJ),\SI^{\rm max}(\SJ)\Big)
\Big\} \cup\Big[\SI^{\rm min}(\ST+1),N-1\Big] .}
Now, from their definition \iminmaxJ , all the diagrams $T^\sI_{\sJ-1}$
for $\SI\ge \SI^{\rm max}(\SJ)$ and the diagrams $T^\sI_{\sJ}$ for
$\SI<\SI^{\rm min}(\SJ+1)$ are equal and identical to $T^0_\sJ$. Since, by
hypothesis, the nest $\CN^0_\oplus$ is minimal and therefore
its tableau has no equal vertically adjacent diagrams, we deduce that
\eqn\inegi{\SI^{\rm max}(\SJ)\ge \SI^{\rm min}(\SJ+1) .}
Using this inequality for each $\SJ$, it is easy to check that the r.h.s of
\prbeta\ is actually equal to
\eqn\intermin{\Big[\min_{\sJ=1}^{\sT+1}\SI^{\rm min}(\SJ)\ ,N-1\Big]=\Big[
\min( {\overline{\rm Ind}}) \ , N-1\Big]}
and the required property \prbeta\ follows. This proves the convergence of
\thelastone , Q.E.D.

\newsec{Discussion}

\subsec{Analytic continuation and convergence at small ${\tit D}$ }

Up to now, the finiteness of $\CZ_N$ (resp. $\CZ_N^{(M)}$) at
$\epsilon >0$ and that of ${\CZ^{\bf R}}_N$ (resp. ${\CZ^{\bf R}}_N^{(M)}$)
at $\epsilon \ge 0$ were proven for large enough dimension $D$ only, that is
$D\ge N-1$ (resp. $D\ge N+M-1$), $\nu$ being fixed.
If we now want to recover the physical models with a fixed value of $k$
(typically $k=2$) and of $\nu$, this  requires a fixed value of $D=k-2\nu$
(satisfying \domdef\ so that $0<\nu<1$).
All the diagram contributions to $\CZ_N$ (resp. ${\CZ^{\bf R}}_N^{(M)}$)
with $N\ge D+1$ (resp. $D-M+1$)
then have to be defined by the analytic
continuation procedure discussed in section 3, in a regime where
the products of the
measure (in the distance or the tree variables) by the integrands
(resp. the subtracted integrands) that we have considered become
{\it distributions}.
This is the case for all the diagrams but a finite number of these.

To end this study, we have to make sure that, in this regime,
these integrals (resp. subtracted integrals) are still finite in the sense of
distributions for $\epsilon >0$ (resp. $\epsilon \ge 0$).
We shall not give a rigorous and complete proof of this fact,
but we shall rather
outline the main steps of the argument.
\medskip
First we have to check that the absolute convergence of the unsubtracted
amplitude $\CZ_N$ for $\epsilon >0$ given in section 5 extends to $D < N-1$.
Considering the integral representation \conve\ for the contribution to
$\CZ_N$ of a given generalized Hepp sector $\CH^{\bf T}$ , expressed in
spherical coordinates, and using \boundDet , we get for this integral:
\eqn\convebis{
\int_{{\cal D}^{{\bf T}}} \prod_{\alpha =1}^{N-1}
(\beta_\alpha)^{\alpha\epsilon-1}
d\beta_\alpha
\,\prod_{\alpha=2}^{N-1}\prod_{n =1}^{\alpha -1}
\left(\sin\left( \theta_{\alpha,n}\right)\right)^{D-1-n}
d\theta_{\alpha,n}\left(
\det\left[\Avril^{{\bf T}}_{\alpha \beta}(\beta\hbox{'s},\theta\hbox{'s})
\right]\right)^{-{d\over 2}}\ ;
}
one sees that the problem of UV convergence (which comes from the small
$\beta_\alpha$ behavior) is completely decoupled from the problem of
analytic continuation of the measure in $D$ (which comes from the behavior
of the integral when  $\theta_{\alpha ,n}\to 0{\hbox{ or }}\pi$ for $n>D$).
As already discussed in subsection 3.3, an explicit representation
of the analytically continued amplitude can be written, for non integer $D$,
by subtracting the divergent powers of $\theta$ and $\pi-\theta$ (this is the
standard finite part prescription).
The resulting integration over the $\theta$'s are convergent, for fixed
non-zero $\beta$'s.
 From the explicit form of the matrix $\Avril^{\bf T}_{\alpha\beta}$,
one can check that the subtractions in $\theta$ do not introduce dangerous
negative powers of the $\beta$'s (at least in the sector $\CH^{\bf T}$
{\it i.e.} $\CD^{{\bf T}}$),
so that the power counting argument in the $\beta$'s stays valid.
Finally one can check that (as already done in subsection 3.3),
the poles that occur at integer $D$ are cancelled by the corresponding
zeros of the global factor $S_D S_{D-1}\ldots S_{D-N+2}$ in the measure
\meastheta , so that the
unsubtracted amplitude $\CZ_N$ is finite for any $D>0$ and $\epsilon >0$.
\medskip
The same argument can be applied to the subtracted amplitude at
$\epsilon=0$.
Starting from the expression \thelastone\ for the part associated
with the maximal nest $\CM_\oplus$ of the subtracted amplitude
in an extended Hepp sector, some of the $p(D,n)$ exponents become
negative for $D<N+M-1$, and the integration over the corresponding
angular variables $\theta^{\sI, n}_{\sJ,j}$ requires a finite part
subtraction prescription.
Again, one can argue that these subtractions do not interfere with
the power counting in $\beta$'s  and $\chi$'s, and that the small $\beta$
estimates \GrandO\ and \GrandOT\ remain valid for the $\theta$-subtracted
integrands.
\medskip
Finally, one can extend this analysis to small negative $\epsilon$, and show
that for a subtracted amplitude of order $N$, no UV divergences
occur as long as ${\rm Re}(\epsilon)>-\delta/(N-1)$, with
$\delta=\min (\nu , 1-\nu )$, as in \defdelt .
Indeed, for $\epsilon \ne 0$, we must modify \thelastone\ by inserting in
the integrand
\eqn\glouk{
\prod_{\sJ=1}^{\sT+1}\prod_{j=1}^{{\rm Card}(T^0_{\sJ})}
{\prod_{\sI\in {\overline{\rm Ind}}(\sJ,j)}}\kern -10pt {\raise 5pt \hbox{$'$}
}\kern 10pt
\left(\chi^\sI_{\sJ,j}\right)^\epsilon
\times \prod_{\sI\in {\overline{\rm Ind}}}
\left(\beta^\sI\right)^{n(\sI)\epsilon}\ ,
}
where $n(\SI)$ is the number of line vectors $\lambda_{\sJ, j}^{\sI'}$ with
an index ${\SI'}\le \SI$.
One has clearly $n(\SI)\le\SI\le N-1$.
Since the subtracted interaction term is (from \thelastone\ and
\prbeta ) a
$\displaystyle \CO \left( \prod_{\sI\in{\overline{\rm Ind}}}\left(
\beta^\sI\right)^\delta\right)$, the convergence at small $\beta$'s
is guaranteed for
${\rm Re}(\epsilon)>-\delta/(N-1)$.

\medskip
Finally, we have not discussed the problem of the convergence or summability
of the perturbative series for our model. Since the
model is expected to make sense for both $b>0$ and $b<0$ (with a finite
free energy proportional to the internal volume in the latter case),
we expect that
the radius of convergence of these series will be non-zero, and in fact
infinite for the unrenormalized series (which exists for $\epsilon >0$, thus
defining entire functions of $b$).

\subsec{Universal scaling properties of the manifold}

In this subsection, we shall derive some physical implications of the
existence of a renormalized theory, well defined at $\epsilon = 0$.
We shall consider here explicitly the case of elastic membranes with
$k=2$ in \Ham .

The main result of the preceding
sections is that the subtracted amplitudes \ZMRen\ for the correlation
functions
remain finite at $\epsilon = 0$. In terms of these, the full correlation
functions
\eqn\RenExpbis{\CZ^{(M)} (X_a,\kvec_a;b)=
{\CZ^{\bf R}}^{(M)}(X_a,\kvec_a;b_R)\ =\ \sum_{N=0}^\infty \,{(-b_R)^N \over
N!}\,
{\CZ^{\bf R}}_N^{(M)}(X_a,\kvec_a)
}
have a series expansion in terms of the effective
excluded volume parameter:
\eqn\brenbis{b_R={1\over \CV_{\CS_D}}\left(
V_{\RR^d} - \CZ \right)\ ,}
which represents the resummed one-point interaction of the manifold
with the impurity. As functions of $b_R$ and $\epsilon$, these correlation
functions thus stay finite at $\epsilon =0$.
\medskip
\noindent {\it $\diamond$ Existence of a Wilson function}

Our renormalization operation involves a peculiar renormalized coupling
constant $b_R$ \brenbis , which is a function:
\eqn\brbr{b_R\equiv b_R(b,X;\epsilon )}
where $X$ is the internal linear size of the manifold, defined by
\eqn\Xsize{\CV_{\CS_D}\equiv X^D .}
As usual, since the renormalization operator ${\bf R}$ deals only with
local counterterms, other choices of the renormalized coupling constant
are possible,
keeping the correlation functions finite as in \RenExpbis .
In particular, the theory describing the manifold of a given size $X$
remains finite when expressed in terms of
the parameter
\eqn\brlam{b_R(\lambda ) \equiv b_R (b, \lambda X;\epsilon ) \ ,}
which corresponds to the renormalized coupling constant of a (reference)
manifold with different size $\lambda X$.
In particular, the original $b_R(b,X;\epsilon)$ itself can be expressed
in terms of $b_R(\lambda )$ (and $\lambda $):
\eqn\bofb{b_R(b,X;\epsilon)=B_R(b_R(\lambda),\lambda,X;\epsilon)}
where $B_R$ stays finite at $\epsilon = 0$.
This information is best expressed by writing
\eqn\totolo{0=\lambda {{\rm d}{\phantom{\lambda}} \over {\rm d}\lambda }
b_R(b,X;\epsilon)
=\lambda {{\rm d}{\phantom{\lambda}}\over {\rm d}\lambda } b_R(\lambda) \ \
{\partial {\phantom {b_R}}\over \partial b_R} \Big|_{\lambda,X}B_R
+ \lambda {\partial {\phantom{\lambda}} \over \partial \lambda}
\Big|_{b_R(\lambda),X}
B_R\ ,}
from which we deduce that the quantity:
$\displaystyle {\lambda {{\rm d} {\phantom {\lambda}} \over {\rm d} \lambda}
b_R (\lambda)}$
remains finite at $\epsilon =0$ when expressed in terms of $b_R(\lambda)$,
$X$ and $\lambda$. This ensures in particular the finiteness at $\epsilon =0$
of the Wilson function:
\eqn\Wils{X {\partial {\phantom {X}}\over \partial X}\Big|_b b_R\equiv
\lambda {{\rm d} {\phantom {\lambda}}\over {\rm d}{\lambda }}b_R(\lambda)
\Big|_{\lambda =1}\ .}
As in \defofg , it is convenient to introduce the dimensionless coupling
constants
\eqn\defofgbis{
\eqalign{ g&\equiv \Big( 2\pi\, A_D(\nu) \Big)^{-{d/ 2}}b_R
X^\epsilon \ ,\cr
z&\equiv \Big( 2\pi\, A_D(\nu) \Big)^{-{d/ 2}}b X^\epsilon \ ,\cr }
}
with $A_D(\nu )=(S_D(2-D)/2)^{-1}$ for $k=2$.
The associated Wilson function then does not depend on $X$
explicitly and reads:
\eqn\Wilso{W(g,\epsilon)\equiv X{\partial {\phantom {X}}\over \partial X}
\Big|_b g = \epsilon z {{\rm d} g \over {\rm d}z} \  .}
It is finite at $\epsilon =0$, to all orders in $g$, and has the first
order expansion \RGW :
\eqn\RGWbis{
W(g)\,= \,\epsilon\,g-{1\over 2}S_D\,g^2
\,+\,{\cal O}(g^3,g^2\epsilon )\ ,}
with a fixed point at
\eqn\fixedbis{g^\star={2\epsilon\over S_D}+\CO (\epsilon^2 )\ .}
\medskip
\noindent{\it $\diamond$ Universality for the excluded volume and
the osmotic pressure}

Let us consider the quantity
\eqn\excvol{\aa = V_{\RR^d}-\CZ=b_R\CV_{\CS_D}\ ,}
which has the dimension of a $d$-volume.
For $b>0$ (repulsive interaction) it is positive and
represents an effective hard-sphere like excluded volume for the
manifold around the impurity.

According to the definition \defofgbis\ of $g$, we have explicitly
\eqn\atog{\aa=g\ \big( 2\pi\, A_D(\nu)\big)^{d/2} \CV_{\CS_D}^{d \nu /D}\ .}
The internal volume of the manifold, $\CV_{\CS_D}$,
is not directly observable, but,
according to \msqext\ and \ADnu , it is
related to the geometrical extension of the membrane in
bulk $d$-dimensional space, when no impurity is present ($b=0$).
This extension can be measured, for instance, by the radius of
gyration $\rr_G$ of the noninteracting manifold, defined as
\eqn\GyrRat{\eqalign{
\rr_G^2\,&\equiv \,{1\over 2 {\CV}_{{\CS}_D}^2}{\Big\langle \int_{{\CS}_D}
d^Dx\int_{{\CS}_D} d^Dy
\,[\rvec (x)-\rvec (y)]^2 \Big\rangle}_0\cr
&=\,{\rm Tr'}\left({1\over -\Delta}\right)\ ,\cr}
}
where ${\rm Tr'}$ means the sum over the non-zero eigenvalues of the
Laplacian $\Delta$ on the closed manifold. Consequently we have
\eqn\GyrVol{\rr_G^2\,=\,{\bf c}\,{\CV_{\CS_D}}^{2\nu/D}}
where the dimensionless
constant {\bf c} depends on the geometrical shape of the manifold
(it will be different for a sphere, an ellipsoid, a torus, etc...), and
requires
the knowledge of the true massless propagator $G(x,y)$ on the manifold ${\CV}$,
solution of
\eqn\propV{-\Delta_x G(x,y)\,=\,\delta^D(x,y)-{1\over {\CV_{\CS_D}}}}

We consider explicitly the case where the external space
dimension is lower than $d^\star$, so that a repulsive interaction ($b>0$)
is relevant.
When the size of the membrane becomes large,
$g$ then reaches its (IR stable) fixed point
value $g^\star$ in \atog , and we get the universal scaling law:
\eqn\univlaw{
\aa\,=\,{\bf a}^\star\,\rr_G^d\ ,
}
where the dimensionless constant
${\bf a}^\star=g^\star(2\pi A_D(\nu))^{d/2}{\bf c}^{-d/2}$
depends on the intrinsic
geometrical shape of the manifold, but neither on its size, nor on the
details and the amplitude of the repulsive interaction, and is therefore,
in this restricted sense, universal.

An ideal solution of $\nn$ identical
membranes interacting with one impurity, with concentration
$\cc=\nn / V_{\RR^d}$ in a box of volume $V_{\RR^d}$, presents a shift
of the osmotic pressure $\pp$ from
its ideal gas value. Owing to its relation \excvol\ to
the one-manifold partition function, the excluded volume $\aa$ directly
yields, by standard rules of thermodynamics,
\eqn\osmot{\pp / k_BT= {\cc \over 1-\aa / V_{\RR^d} } = \cc\,(1+\aa /
V_{\RR^d} + \ldots )\ .}
This law expresses the increase of the pressure due to the presence
of the impurity in the solution with finite volume, and can be thought of
as a finite size effect. The thermodynamic limit can be reached for
a finite concentration $\cc_I$ of impurities. One then gets the virial
expansion of the osmotic pressure:
\eqn\osmo{\pp / k_BT\,=\,\cc +\cc\,\cc_I \aa + \ldots\,=
\,\cc + {\bf a}^\star \cc\, \cc_I\,\rr_G^d\,+\,\ldots }

Let us stress that the dimensionless quantity ${\bf a}^\star$, which is
independent of the microscopic parameters and appears in the
expression for the osmotic pressure, is directly related to the fixed point
value $g^\star$ with the choice \defofgbis\ for the renormalized constant $g$.
This is entirely similar to the case of a polymer solution
with excluded volume \desClJan
\ref\Noda{J. des Cloizeaux and I. Noda, Macromolecules {\bf 15} (1982) 1505.}.

\medskip
\noindent{\it $\diamond$Pinned manifold}

Let us introduce the partition function of a manifold pinned at the origin
at one of its points $X_1$:
\eqn\pinnedfunc{\CZ^\diamond \equiv \int {\cal D}[\rvec]\exp(-{\cal H})\
\delta^d (\rvec (X_1)) \ .}
Owing to the internal spherical symmetry of the manifold, $\CZ^\diamond$
is independent of $X_1$ and actually equals:
\eqn\pinfuncint{\CZ^\diamond =\int {\cal D}[\rvec]\exp(-{\cal H})\
{1\over \CV_{\CS_D}}
\int_{\CS_D} d^Dx\, \delta^d (\rvec (x))\ . }
 From \partfunc , one has clearly:
\eqn\czbow{\CZ^\diamond = - {1\over \CV_{\CS_D}} {\partial \hfill \over
\partial b}\Big|_X \CZ(b,X)={\partial b_R \over \partial b}\Big|_X \ .}
Notice that, while the unrestricted partition function $\CZ$ has the
dimension of a $d$-volume, the pinned-manifold partition function
$\CZ^\diamond$ is dimensionless and is thus a function
$\CZ^\diamond (z;\epsilon)$ of $z$ (and $\epsilon $) only.
According to \defofgbis\ and \Wilso , we have
\eqn\dbR{\CZ^\diamond ={\partial \hfill \over \partial b}\Big|_X b_R =
{ {\rm d} g \hfill \over {\rm d} z} = {1\over \epsilon z} W (g(z); \epsilon )
\ .}
Notice that $\CZ^\diamond $ itself is not renormalized, {\it i.e.}
not finite at $\epsilon =0$ as a function of $g$, but that $\epsilon z
\CZ^\diamond =W(g,\epsilon )$ is renormalized.
When the size $X$ becomes large (for $\epsilon$ and $b$ positive)
$z$ becomes large and $g(z)$ tends to its limit $g^\star$,
the Wilson function vanishing as:
\eqn\Wivan{W(g(z);\epsilon)=(g(z)-g^\star)W'(g^\star)+\ldots\ , }
with
\eqn\glimit{g(z) - g^\star \sim {\hbox{const}}\ z^{W'(g^\star)/ \epsilon}\ ;}
(Notice that $W'(g^\star)<0$; see \RGflows ). This finally leads to the
scaling law for $\CZ^\diamond$:
\eqn\scaldiam{\CZ^\diamond \sim {\hbox{const}}\ z^{-1+W'(g^\star)/\epsilon}
\sim {\hbox{const}}\ (b^{1/ \epsilon}X)^{W'(g^\star)-\epsilon}\ .}
At first order in $\epsilon$, $W'(g^\star)=-\epsilon + \CO (\epsilon^2)$,
whence
\eqn\scaldiambis{\CZ^\diamond \sim {\hbox{const}}\ (b^{1/ \epsilon}
X)^{-2\epsilon}\ .}

\medskip
\noindent{\it $\diamond$ Universal $1/r$ repulsion law}

The pinned-manifold partition function $\CZ^\diamond$ is
a particular case of a more general restricted partition function
to which we now turn.
We introduce:
\eqn\czcirc{\CZ^\diamond (X_1, \rvec ; X, b\, ;\epsilon )
=\int {\cal D}[\rvec]\exp(-{\cal H})\
\delta^d (\rvec (X_1)- \rvec ) }
which describes the partition function of a manifold held by one
of its points at the position $\rvec$ relative to the origin.
It is the Fourier transform of the
one-point correlation function \zvertex\ for $M=1$, that is:
\eqn\tranz{\CZ^\diamond (X_1, \rvec; X , b\, ;\epsilon )=\int d^d \kvec_1
\ \exp ( - i \ \kvec_1\cdot \rvec )\  \CZ^{(1)}(X_1, \kvec_1 ; b, X, \epsilon )
\ .}
As above, for a closed manifold,
$\CZ^\diamond (X_1, \rvec ; X , b\, ; \epsilon )$
is actually independent of $X_1$ and equal to
\eqn\tobeused{\CZ^\diamond (\rvec ; X, b\, ; \epsilon ) = \int {\cal D}
[\rvec]\exp(-{\cal H})\
{1\over \CV_{\CS_D}}
\int_{\CS_D} d^Dx\, \delta^d (\rvec (x) -\rvec ) \ .}
The relations of this partition function to the former ones are
\eqn\relz{\eqalign{ \CZ^\diamond ({\vec 0}) &= \CZ^\diamond \ ,\cr
\int_{\RR^d} d^d \rvec \  \CZ^\diamond (\rvec ) &= \CZ \ .\cr }}
By rotational symmetry, the quantity $\CZ^\diamond $ depends only
on $r\equiv |\rvec |$. It is furthermore dimensionless, and thus can
be written as a function of $z$ and $r/X^\nu $ (and $\epsilon $):
\eqn\zredva{\CZ^\diamond (\rvec;X,b\,;\epsilon)\equiv \CZ^\diamond [r/X^\nu,z\,
;\epsilon ]\ .}
As we have seen for $\CZ^\diamond$ \dbR , $\CZ^\diamond [r/X^\nu,z\,
;\epsilon]$
is not exactly renormalized, when expressed in terms of $g$, but
$\epsilon z \CZ^\diamond [r/X^\nu,z\,;\epsilon ]$ is.
It is interesting to consider the limit when the
interaction parameter $b$ goes to
infinity, while keeping the size $X$ of the manifold finite. We expect
$\CZ^\diamond [r/X^\nu,z\,;\epsilon]$ to reach a finite limit
\eqn\finlim{\CZ_\infty^\diamond [r/X^\nu\,;\epsilon ]\equiv
\lim_{z\to \infty}\CZ^\diamond [r/X^\nu,z\,;\epsilon ]\ .}
According to \relz\ and \brenbis , we have
\eqn\tesre{\int_{\RR^d} d^d\rvec \ (\CZ^\diamond [r/X^\nu,z\,;\epsilon ] -1)=
-b_R \CV_{\CS_D}=- g \Big( 2\pi\, A_D(\nu)\Big)^{d/2}\, X^{\nu d}\ .}
In the limit $z\to \infty$, $g$ tends to $g^\star$, and we therefore have
\eqn\tesrelim{\int_{\RR^d} d^d{\vec u}\
(\CZ_\infty^\diamond [u\,;\epsilon ] -1)=-
g^\star \Big( 2\pi\, A_D(\nu)\Big)^{d/2}\ ,}
which is consistent with the assumption that the limit in \finlim\
actually exists.

In the scaling regime $r/X^\nu \ll 1$, we expect the marked point
to be strongly
repelled from the origin, and thus $\CZ^\diamond_\infty$ to vanish as
a power law:
\eqn\scalcirc{\CZ_\infty^\diamond [r/X^\nu\,; \epsilon ] \sim \hbox{const}
\Big({r\over X^\nu}
\Big)^\theta \ .}
This vanishing of $\CZ^\diamond[r/X^\nu,z\,;\epsilon ]$
in the successive limits $z\to \infty $ and
$r\to 0$ is consistent with that obtained in the reversed double
limit $r= 0$, and
$z\to \infty$, which corresponds to the vanishing of $\CZ^\diamond $ at
infinite
$z$ according to \scaldiam .

The contact exponent $\theta$ can be obtained as follows. For finite $b$ and
large $X$, we expect a universal $X$-dependence of $\CZ^\diamond [r/X^\nu,z
\,;\epsilon]$, irrespective of the particular value given to $r$. This
dependence is in particular known exactly when $r=0$, according to \scaldiam .
It must also be the same for $r\ne 0$ fixed and $b\to \infty$, that is a
behavior which is given by \scalcirc . This leads to identifying the contact
exponent with:
\eqn\thetval{\theta = {\epsilon -W'(g^\star ) \over \nu}\ .}
Notice that the argument above, intuitively clear on physical grounds,
is usually mathematically justified in field theory from the
existence of a short-distance operator product expansion. A rigorous
proof of the existence of
such a short-distance expansion in our case is beyond the scope
of this paper. The repeated appearance of $W'(g^\star)$ in \scaldiam\ and
\thetval\ suggests that all scaling behaviors in this theory are
controlled by a single scaling anomalous dimension, {\it i.e.} the universal
slope of the Wilson function at the fixed point.
\medskip
Equation \scalcirc\ allows us to derive a universal expression for the
repulsive force exerted by the impurity on the membrane,
\eqn\unifor{{\bf {\vec f(\rvec)}}/k_B T =\nabla_{\rvec}\log \CZ^\diamond
(\rvec) =
\theta {\rvec \over r^2} \ .}
According to the discussion above, this force law is valid in the
scaling regime $b^{-\nu/\epsilon} \ll r \ll X^\nu $, where $b^{-1/\epsilon}$
plays the same physical role as an ultraviolet cut-off for internal distances.

\medskip\noindent{\it $\diamond$ Scaling laws for the delocalization
transition}

Finally, we have seen in subsection 2.1 that for $d>d^\star$ (that is
$\epsilon <0$), the non-trivial fixed point $g^\star$ is now negative and
IR repulsive, and corresponds to a delocalization transition with non-trivial
critical exponents, for a particular negative critical value $b^\star$
of the bare coupling constant $b$.
In the localized phase ($b<b^\star$), the correlation functions
such as $\langle \rvec (x) \rvec (y) \rangle$ and the associated correlation
length $\xi_{\|}$ (in the
internal $D$-dimensional space) should be finite, as well as the average
distance ${r}=\langle|\rvec |\rangle$
of the manifold to the attractive impurity.
At the transition these quantities should diverge as
\eqn\delocscal{
\xi_{\|}\propto (b^\star-b)^{-\nu_\|}\qquad;
\qquad {r}\propto (b^\star-b)^{-\nu_\perp}\ .
}
Standard arguments lead to
\eqn\nuparal{\nu_\| = {1\over W'(g^\star)} = -\,{1\over\epsilon}\,+\,\ldots}
and
\eqn\nuperp{\nu_\perp = \nu_\|\, \nu\qquad.}
Indeed, $\rvec$ has no anomalous dimension and therefore, $r$ scales as
${\xi_\|}^\nu$ with $\nu=(2-D)/2$ from \scaldim .

\newsec{Conclusion}
\subsec{Summary}

In this last section, we would like to summarize the main steps of our
construction and outline the main ingredients which
ensure the renormalizability of the theory. We then discuss some
possible extensions of our results.
\medskip
\noindent{ \it $\diamond$ Existence of a perturbative expansion analytically
continued in $D$}

\item{(I)} The first ingredient is the existence, for integer dimension
$D$ of the manifold, of a formal perturbative expansion for the model.
The diagrams present an invariance under global
Euclidean motions in $\RR^D$ of the interaction points
(or under the group ${\rm SO}(D+1)$ for finite volume manifolds with
the internal geometry of the sphere $\CS_D$).
The interaction
terms, which are determinants involving the internal Green functions
between interaction
points, can then be expressed in terms of mutual squared distances only.
On the other hand, the external dimension $d$ appears only in the power
$(-d/2)$ of the interaction determinant.

\item{(II)} The second step is the construction of a measure term,
analytic in $D$, in terms of the above set of internal mutual squared
distances. One
can then use for convenience any equivalent measure, for instance in terms
of Cartesian or spherical coordinates in a space with a given integer
dimension (typically $\RR^{N-1}$ for a diagram of order $N$), $D$ itself
appearing as an analytic variable. This measure has in general to be understood
as a distribution. \par

\noindent Points {(I)} and {(II)} allow us to define a
perturbative expansion for the model, analytically continued in $D$.
Its main features are the following:
\itemitem{-} It can be viewed as a generalization of the Schwinger parametric
representation of Feynman amplitudes for local field theory,
with the one-dimensional $\alpha$-parameters
replaced by $D$-dimensional parameters.
\itemitem{-} It appears as a string-like theory, in the sense that it presents
only one diagram to each order in perturbation.
\itemitem{-} It reduces to the expansion of a local field theory when $D=1$,
expressed in the Schwinger $\alpha$-representation.
The field theoretic diagrammatic contributions are recovered in the limit
$D \to 1$ through the
analytic continuation of the measure term.
\par

\noindent{\it $\diamond$ Renormalizability}

The essential properties which are key to renormalizability are
the following:

\item{(III)} Schoenberg's theorem: this property of the interaction
determinants ensures that divergences in the
integrals of the diagrammatic expansion occur only at short-distances (UV),
as in ordinary local field theories. Infrared (IR)
divergences also can occur if the internal space is
infinite, a problem
which is dealt with by considering a finite membrane, {\it e.g.}
the sphere $\CS_D$ with finite volume $\CV_{\CS_D}$.

\item{(IV)} Factorization of the interaction term: this property states
that, when a subset of interaction points contracts toward a vertex,
the interaction determinant factorizes into the product of the interaction term
of the contracting subset by that simply obtained by replacing the whole subset
by its contraction vertex. The possibility of replacing a set of coalescing
points by a single contraction vertex, and of factorizing out the
corresponding divergence
is the key for renormalizability. Mathematically,
it allow us to make the theory finite by letting
a subtraction operator act on the integrand.
This operator essentially subtracts factorized equivalents so as to remove
the UV divergences. It is constructed from elementary Taylor operators
associated with subsets of points, then organized in forests or nests,
corresponding to the hierarchical structure of the divergences.

\item{(V)} Factorization of the measure: this property, obviously satisfied
for integer $D$, is preserved by the analytic continuation
of the measure
to non integer $D$. It allows us to integrate separately the factorized
determinants which are to be subtracted from the original amplitude,
and thus to interpret them as counterterms: the
subtraction operation is then a simple reexpression of the partition function
(or correlation functions) in terms of an effective (renormalized) coupling
constant.  \par

\noindent Points (III) and (IV) are properties of the interaction determinant
themselves, while point (V) is a general property of the measure.

\subsec{Prospects}

Let us finally discuss possible outcomes of our results.
As already discussed, the model \Ham\ of a manifold interacting
with a single point serves indeed as a laboratory
for studying the renormalizability of more general models of
interacting crumpled manifolds.
A prominent model of this class is of course the Edwards model \EdwardsM\
of a self-avoiding manifold interacting via a short range two-body
pseudopotential. Its perturbative expansion is similar in
structure to the one studied here. We indeed believe that the mathematical
techniques developed in this article can be applied and generalized to
the Edwards model, and provide both conceptually and practically a framework
for a similar proof of its renormalizability.

When reviewing the general scheme above, we note that point (I) is already
known for the self-avoiding model
\DHK\ .
Points (II) and (V) are actually valid for any manifold Hamiltonian.
The specificity of a given model is actually
encoded in its interaction determinants, for which
properties similar to those of (III) and (IV) have to be analyzed
in each case, and established in order
to eventually build a subtraction procedure and prove renormalizability
\ref\DDG{F. David, B. Duplantier and E. Guitter, work in progress.}.

This scheme should be directly applicable to a series of
manifolds theories with interactions, such as many-body or long-range
interactions ... These models generalize to arbitrary internal dimension
$D$ models of interacting polymers ($D=1$). All the latter models are
known to be
equivalent to some $n$-component field theories in the limit $n=0$, with
standard Feynman diagram expansions. When extended to manifolds of
arbitrary internal dimension, these models become theories
with a single diagram to each order in perturbation (a property which
is shared with string theories, although in our case the manifold has a fixed
internal metric). Interestingly enough,
the topological complexity of the usual Feynman diagrams is encoded in
the $D$-measure on the manifold, and arises in the limit $D=1$
from the ordering constraints along the one-dimensional (polymer) line.
More generally, it would be interesting to try and express field theories
with an arbitrary number $n$ of components as $D=1$ limits of ``manifolds"
string-like models, yet to be invented.

\bigbreak\bigskip\bigskip\centerline{{\bf Acknowledgements}}\nobreak
We thank M. Berg\`ere for helpful discussions
and for a careful reading of the manuscript.

\vfill\eject

\appendix{A}{From vectors to scalar products}
In this appendix we derive \intu\ \measuij.
First we insert the relation $u_{ij}=x_i\cdot x_j$ in the l.h.s. of
\intu\
\eqn\firstA{
\int \prod_{i=1}^N d^Dx_i\, f(x_i\cdot x_j)
\ =\
\int \prod_{i\le j} d u_{ij}\,\int \prod_{i}d^Dx_i\,
\prod_{i\le j}\delta (u_{ij}-x_i\cdot x_j) \ f([u_{ij}])
\ .}
Second we use the fact that the function
\eqn\secondA{
\sigma^{(D)}_N(u_{ij})\ =\ \int \prod_{i}d^Dx_i\,
\prod_{i\le j}\delta (u_{ij}-x_i\cdot x_j)
}
is invariant under SO($N$) rotations $R$ ($u\to R^tuR$) to diagonalize
$u_{ij}$ and express \secondA\ in terms of the $N$ eigenvalues $\lambda_i$
$i=1,\ldots ,N$ of $u_{ij}$
\eqn\thirdA{
\sigma^{(D)}_N(u_{ij})\ =\ \int \prod_{i} d^Dx_i\,\prod_{i\le j} \delta
(\lambda_i\delta_{ij}-x_i\cdot x_j)
\ .}
Third we perform the change of variables $x_i\to \sqrt{\lambda_i}x_i$
and get
\eqn\forthA{
\sigma^{(D)}_N(u_{ij})\ =\ \prod_{i=1}^{N} \lambda_i^{D-N-1\over 2}\,
\int \prod_{i}d^Dx_i\,\prod_{i\le j}\delta (\delta_{ij}-x_i\cdot x_j)
\ .}
The remaining integral over the $x_i$'s gives the volume of SO($D$)/SO($D-N$)
and we
obtain finally \measuij
\eqn\fifthA{\eqalign{
\sigma^{(D)}_N(u_{ij})\ &=\ \left({\scriptscriptstyle \prod\limits_i}
\lambda_i\right)^{D-N-1\over 2}\cdot
{{\rm Vol}({\rm SO}(D))\over{\rm Vol}({\rm SO}(D-N))}\cr
&=\ (\det[u_{ij}])^{D-N-1\over 2}\ {S_D\over 2}\ldots{S_{D-N+1}\over 2}\ .\cr
}}

\appendix{B}{Factorization of the measure}
To prove \factorid\ let us decompose the $N\times N$ symmetric positive
definite
scalar product matrix $[u]_N$ into blocks of size $P$ and $Q$ ($P+Q=N$):
\eqn\firstB{
[u]_N\ =\ \left(\matrix{[u]_P\hfil&[v]\hfil\cr[v]^t\hfil&[u]_Q\hfil\cr}\right)
\ .}
Equation \factorid\ is equivalent to the fact that, given the positive definite
matrices
$[u]_P$ and $[u]_Q$, when integrating over all $P\times Q$ matrices $[v]$
such that $[u]_N$ (defined by \firstB ) is positive definite, we have
for arbitrary non-integer $D$
\eqn\secondB{
\int d[v]\,\sigma^{(D)}_N([u]_N)\ =\ \sigma^{(D)}_P([u]_P)\,
\sigma^{(D)}_Q([u]_Q)
\ .}
Since $[u]_P$ and $[u]_Q$ are positive definite we can take their square root
$[u]_P^{1\over 2}$ and $[u]_Q^{1 \over 2}$ and write $\det([u]_N)$ in the
expression \measuij\ for $\sigma^{(D)}_N$ as
\eqn\thirdB{
\det [u]_N\,=\ \det ([u]_P)\,\det ([u]_Q)\,\det({\bf 1}- [u]_P^{-{1\over 2}}
[v] [u]_Q^{-1} [v]^t [u]_P^{-{1\over 2}})
\ .}
Now, one can perform the change of variable
$[v]\rightarrow [u]_P^{1\over 2} [v] [u]_Q^{1\over 2} $
which induces a Jacobian $J=\det([u]_P)^{{Q\over 2}} \det ([u]_Q)^{{P\over 2}}$
in \secondB .
We thus obtain finally that the l.h.s. of \secondB\ is equal to the r.h.s. of
\secondB , up to a constant $C$ which depends on $D$, $P$ and $Q$, but not
on $[u]_P$ and $[u]_Q$, and which is given
by
\eqn\forthB{
C\ =\
{{\rm Vol}({\rm SO}(D))\over {\rm Vol}({\rm SO}(D-N))}
{{\rm Vol}({\rm SO}(D-P))\over {\rm Vol}({\rm SO}(P))}
{{\rm Vol}({\rm SO}(D-Q))\over {\rm Vol}({\rm SO}(Q))}
\,\int d[v] \left( \det ({\bf 1}-[v][v]^t)\right)
^{D-N-1\over 2}
\ .}
(The domain of integration for $[v]$ is now such that
$\left(\matrix{{\bf 1}_P & [v]\cr[v]^t & {\bf 1}_Q \cr}\right)$ is positive
definite)
\medskip
It remains to prove that $C=1$.
This can be done in a simple way by proving that the factorization identity
\factorid\ holds for some particular function $f([u])$. As an example we can
take
the exponential
\eqn\fifthB{
f([u]_N)\ =\ \exp (-{\rm tr}[u]_N)
\ ,}
since we can easily calculate explicitly (see below)
\eqn\sixthB{
I_N\ =\ \int_{{\cal U}_N} d[u]_N\,\sigma^{(D)}_N([u]_N)\,\exp(-{\rm tr}([u]_N))
\ =\ (\pi)^{N{D\over 2}}\ ,
}
and therefore factorization holds in this case since:
\eqn\seventhB{
f([u]_N)\,=\,f([u]_P)f([u]_Q)\qquad{\rm and}\qquad I_N\,=\,I_P I_Q
\ .}
The direct computation of $I_N$ (\sixthB\ ) for any $D$ proceeds as follows.
The set ${\cal U}_N$ is the set of symmetric positive matrices. By ${\rm
SO}(N)$
orthogonal transformations, it can be reduced to the set of diagonal matrices
with positive eigenvalues $\lambda_i$ ($i=1,\ldots,N$),
with the new measure:
\eqn\seventhB{d[u]_N=\ {\rm Vol}({\rm SO}(N)) {1\over N!} \prod_{i=1}^{N}
d\lambda_i \ \Delta(\lambda) \ ,}
where the $\lambda$'s are integrated from $0$ to $\infty$ and $\Delta(\lambda)$
is the Jacobian
\ref\mehta{M. L. Mehta, {\sl Random matrices},
second edition, Academic Press, New York (1991).}
\eqn\eighthB{\Delta(\lambda)=\prod_{1\le j<l \le N} |\lambda_j-\lambda_l|\ .}
In terms of these variables, $I_N$ reads explicitly:
\eqn\ninethB{I_N={ {\rm Vol}({\rm SO}(D)) {\rm Vol}({\rm SO}(N))\over
N!\,{\rm Vol}({\rm SO}(D-N))}\int_0^\infty \prod_{j=1}^N\, d\lambda_j
\ \Delta(\lambda)
\, \exp(-{ \textstyle\sum\limits_{j=1}^{N} \lambda_j})
\, \Big({\textstyle\prod\limits_{j=1}^{N} \lambda_j}\Big)^{{D-N-1\over 2}}. }
The calculation is completed by using the Selberg integral formula \mehta\ ,
\ref\Selberg{A. Selberg, Norsk Matematisk Tidsskrift {\bf 26} (1944) 71-78.}:
\eqn\tenthB{\int_0^\infty (\Delta(\lambda))^{2\gamma}
\prod_{j=1}^N \big[ \lambda_j^{\alpha-1} \ \exp (-\lambda_j)\  d\lambda_j\big]
=\prod_{j=0}^{N-1}{\Gamma(1+\gamma +j\gamma )\Gamma (\alpha + j\gamma )\over
\Gamma(1+\gamma)} }
for $\gamma={1\over 2}$ and $\alpha ={D-N+1 \over 2}$, which leads finally
to \sixthB\ .

\appendix{C}{Factorization of $ {\bf det}\tit {\left(\left[\Pi^{{\bf
T}}_{\alpha\beta}\right]\right)}$}

Let us consider an ordered tree ${\bf T}$ and the corresponding vectors
$\lambda_1, \ldots , \lambda_{N-1}$ with $|\lambda_1|\le \ldots
\le |\lambda_{N-1}|$.
We have by definition
\eqn\firstC{\Pi^{{\bf T}}_{\alpha\beta}=-{A_D(\nu)\over 2}\Big\{
|R_{\alpha\beta}
+\lambda_\beta-\lambda_\alpha|^{2\nu}-|R_{\alpha\beta} +\lambda_\beta|^{2\nu}
-|R_{\alpha\beta}-\lambda_\alpha|^{2\nu}+|R_{\alpha\beta}|^{2\nu}\Big\}\ ,}
where $R_{\alpha\beta}$ is one ``basis" of the quadrilateral
\eqn\secondC{R_{\alpha\beta}=x_{i_\beta}-x_{i_\alpha}
\ .}
\midinsert
\figinsert{5.truecm}{8.truecm}{%\hsize}{
\figcap\quadRf{The quadrilateral picturing the matrix element
$\Pi^{{\bf T}}_{\alpha\beta}$ and its ``basis" vector $R_{\alpha\beta}$.}
}
\endinsert
The vector $R_{\alpha\beta}$ is a linear combination of the $\lambda$'s
\eqn\thirdC{R_{\alpha\beta}=\sum_{\gamma=1}^{N-1}c_{\alpha\beta}^\gamma\,
\lambda_\gamma}
where $c_{\alpha\beta}^\gamma=0,\pm 1$.
Suppose we make the following rescaling
\eqn\forthC{\lambda_\alpha \to \lambda_\alpha(\rho)=\left\{
\matrix{\rho \lambda_\alpha \hfill&\quad {\rm if} \quad {\alpha \le P-1}\hfill
\cr &\cr \lambda_\alpha \hfill&\quad{\rm if}\quad {\alpha \ge P}\hfill \cr}
\right.}
for some $P$, $2\le P\le N$ and with a contraction factor $\rho$,
$0\le \rho \le 1$.
Under this rescaling, $R_{\alpha\beta}$ becomes
\eqn\fifthC{\eqalign{
R_{\alpha\beta}(\rho)&=\sum_{\gamma=P}^{N-1}c_{\alpha\beta}^\gamma
\lambda_\gamma+\rho\,\sum_{\gamma=1}^{P-1}c_{\alpha\beta}^\gamma
\lambda_\gamma \cr &=R_{\alpha\beta}^0+\rho R_{\alpha\beta}^1\ .\cr}
}
We therefore have two possibilities:

\item{(a)} $R_{\alpha\beta}^0=0$. This means that $R_{\alpha\beta}$
is formed only of vectors $\lambda_\gamma$ with $\gamma \le P-1$, which
are all contracted, hence $R_{\alpha\beta}$ itself is contracted.
By definition, this is also the case when $R_{\alpha\beta}$ is $0$, that is
when $x_{i_\alpha}=x_{i_\beta}$.

\item{(b)} $R_{\alpha\beta}^0\ne 0$. This occurs when $R_{\alpha\beta}$
is spanned by at least one $\lambda_\gamma$ which is not contracted,
that is with $\gamma \ge P$.\par
\midinsert
\figinsert{11.truecm}{7.truecm}{%\hsize}{
\figcap\treesTif{Classification of the line vectors of the tree ${\bf T}$
into subtrees ${\bf T}_i$. The dashed lines in (a) correspond to contracting
branches of the tree ${\bf T}$, and are organized into two connected subtrees
${\bf T}_1$ and ${\bf T}_2$ in (b) . The full lines in (a) correspond
to non-contracting branches and are organized into a single connected subtree
${\bf T}_3$ in (b), by fully contracting the dashed lines in (a).}
}
\endinsert

\noindent This allows us to classify the $\lambda$'s into subtrees as follows
(see \treesTif ):

\item{-} We regroup the $\lambda_\alpha$'s with $\alpha\le P-1$
({\it i.e.} corresponding to contracted lines) into equivalence
classes by deciding that $\lambda_\alpha$ and $\lambda_\beta$ are equivalent
if $R^0_{\alpha\beta}=0$. The equivalence classes ${\bf T}_1,\ldots ,{\bf
T}_{m-1}$
(with $2\le m \le P$ depending on ${\bf T}$)
correspond to the $m-1$ distinct connected subtrees
which build the subset of the contracted lines.
Case (a) above thus corresponds to $\lambda_\alpha$ and $\lambda_\beta$
in the same
equivalence class, that is
in the same connected subtree of contracted lines.
Case (b) corresponds to $\lambda_\alpha$ and $\lambda_\beta$ in two distinct
equivalence classes, that is
in two distinct connected subtrees of contracted lines.

\item{-} We regroup the $\lambda_\alpha$ with $\alpha\ge P$ into a single
connected tree ${\bf T}_m$ obtained by setting $\lambda_\beta=0$ for $\beta \le
P-1$
in the original tree ${\bf T}$.\par
\medskip
We will now show that, for $\rho \to 0$:
\eqn\sixthC{\det\left(\left[\Pi^{{\bf T}}(\rho)\right]\right)=
\rho^{2\nu (P-1)}\prod_{i=1}^{m}\det\left(\left[\Pi^{{{\bf T}}_i}\right]
\right)\left\{1+{\cal O}(\rho^{2\delta})\right\}
\ .}
Let us consider two lines $\lambda_\alpha$ and $\lambda_\beta$.

\noindent {\bf Case 1:} $\alpha\le P-1$, $\beta \le P-1$

\noindent {Case 1(a):} $R^0_{\alpha\beta}=0$

This case corresponds to two $\lambda$'s in the same contracting connected
subtree ${\bf T}_i$ for some $i\le m-1$. In \firstC\ , $\lambda_\alpha$,
$\lambda_\beta$ and $R_{\alpha\beta}$ all get a factor $\rho$, hence
\eqn\seventhC{\Pi^{{\bf T}}_{\alpha\beta}(\rho)=\rho^{2\nu}\Pi^{{\bf
T}}_{\alpha\beta}\ .}
It is furthermore clear that $R_{\alpha\beta}$ is spanned only by $\lambda$'s
in ${\bf T}_i$, hence
\eqn\eighthC{\Pi^{{\bf T}}_{\alpha\beta}(\rho)=\rho^{2\nu}\Pi^{{{\bf
T}}_i}_{\alpha\beta}\ .}
\noindent {Case 1(b):} $R^0_{\alpha\beta}\ne 0$

This case corresponds to two $\lambda$'s in two distinct contracting
connected subtrees ${\bf T}_{i_1}$ and ${\bf T}_{i_2}$. Since $R_{\alpha\beta}$
does not
contract to zero,
we can formally expand \firstC\ in power of $\lambda_\alpha$ and
$\lambda_\beta$. The matrix element $\Pi^{{\bf T}}_{\alpha\beta}$ is by
definition
the interaction between two dipoles $\lambda_\alpha$, $\lambda_\beta$
separated by $R_{\alpha\beta}$. It is therefore clear that the first
term in the multipolar expansion is of order
\eqn\ninethC{\Pi^{{\bf T}}_{\alpha\beta}\propto |R_{\alpha\beta}|^{2\nu-2}
\lambda_\alpha\cdot\lambda_\beta +\ldots \ .}
Therefore, expanding in $\rho$ yields immediately
\eqn\tenthC{\eqalign{
\Pi^{{\bf T}}_{\alpha\beta}(\rho)&\propto \rho^2\,|R^0_{\alpha\beta}|^{2
\nu-2} \lambda_\alpha\cdot\lambda_\beta +\ldots \cr &={\cal O}(\rho^2)
=\rho^{2\nu}{\cal O}(\rho^{2\delta})\ ;\cr}}
(See \defdelt ).
As we shall see below, this element, which mixes several subtrees ${\bf T}_i$,
is vanishing sufficiently fast as to disappear in the limit $\rho \to 0$.
\medskip
\noindent {\bf Case 2:} $\alpha\le P-1$, $\beta \ge P$

In this case, we have
\eqn\eleventhC{\Pi^{{\bf T}}_{\alpha\beta}(\rho) \propto
\matrix{(1)&(2)&(3)&(4)\cr |R_{\alpha\beta}(\rho)
+\lambda_\beta-\rho\lambda_\alpha|^{2\nu}&-|R_{\alpha\beta}(\rho)
+\lambda_\beta|^{2\nu}&
-|R_{\alpha\beta}(\rho)-\rho\lambda_\alpha|^{2\nu}&
+|R_{\alpha\beta}(\rho)|^{2\nu}\cr&&&\cr }}

\noindent{Case 2(a):} $R^0_{\alpha\beta}=0$

Substituting $R_{\alpha\beta}(\rho)=\rho R^1_{\alpha\beta}$
in \eleventhC\ , the last two terms $(3)$ and $(4)$ are homogeneous
to $\rho^{2\nu}$, while the expansion of $(1)-(2)$ in power of $\rho$
gives a leading term linear in $\rho$. On the whole, we can write
\eqn\twelfthC{\Pi^{{\bf T}}_{\alpha\beta}=\rho^\nu\, {\cal O}(\rho^\delta)\ .}
\noindent {Case 2(b):} $R^0_{\alpha\beta}\ne 0$

This time, the expansion of $(1)-(2)$ on the one hand, and $-(3)+(4)$
on the other hand, in formal powers of $\rho\lambda_\alpha$ leads
immediately to a matrix element of order $\rho$, hence
\eqn\thirteenC{\Pi^{{\bf T}}_{\alpha\beta}={\cal O}(\rho)=
\rho^\nu\, {\cal O}(\rho^\delta)\ .}
\medskip
\noindent {\bf Case 3:} $\alpha\ge P$, $\beta\ge P$

In this case, $\lambda_\alpha$ and $\lambda_\beta$ are not contracted
and belong to ${\bf T}_m$. In the limit $\rho \to 0$, $R_{\alpha\beta}$
is simply replaced by $R^0_{\alpha\beta}$. Whatever the value of
$R^0_{\alpha\beta}$, this corresponds precisely to
\eqn\fifteenthC{\eqalign{\Pi^{{\bf T}}_{\alpha\beta}(\rho)&=
\Pi^{{{\bf T}}_m}_{\alpha\beta}+\rho^\nu\, {\cal O}(\rho^\delta)\cr
&= \Pi^{{{\bf T}}_m}_{\alpha\beta}+ {\cal O}(\rho^{2 \delta})\ .\cr}}
We can summarize all these cases by writing the synoptic table
\eqn\lastC{\det\left(\Pi^{{\bf T}}(\rho)\right)=\det \matt}
where we have permuted the $P-1$ first lines and columns so as
to regroup the $\lambda$'s according to their equivalence classes.
Therefore, each of the first $m-1$ blocks corresponds to a connected
fully contracting subtree, while the last block corresponds to ${\bf T}_m$.
This rearrangement leaves the determinant invariant.
The factorization property \sixthC\ can now be read from the block structure
of the matrix in \lastC\ .
\medskip
Considering the reduced matrix $\Avril^{{\bf T}}$ defined in \Upsil\ , we
have a similar block structure
\eqn\lasttC{\det\left(\Avril^{{\bf T}}(\rho)\right)=\det \mattt}
and we can now let $\rho\to 0$ and get
\eqn\leastC{\det\left(\left[\Avril^{{\bf T}}(\rho\to 0)\right]\right)=
\prod_{i=1}^{m}\det\left(\left[\Avril^{{{\bf T}}_i}\right]
\right)
\ ,}
which means that, in this limit, the tree has been disconnected into
several components on which its determinant is exactly factorized.
\medskip
Let us know turn to the variables $\beta$'s defined in \lamtobeta\ .
Notice that due to the rescaling \Upsil\ , $\det(\Avril^{{\bf T}})$ is
actually independent of the global scale factor $\beta_{N-1}$\foot{
This homogeneity property holds only for the choice \propS\ for the
propagator, even on the sphere. Otherwise, both $\beta_{N-1}$ and the IR
regulator $R$ would appear and lead to a slightly more complicated
discussion.}.
Each variable $\beta_\gamma$ can be associated with a contracting factor
$\rho=\beta_\gamma$.
Therefore, once expressed in term of the $\beta$'s , $\det(\Avril^{{\bf T}})$
is such that, if we let one $\beta$ tend to zero (say $\beta_\gamma$),
keeping the others non zero, we have
\eqn\leasttC{\eqalign{\det\left(\Avril^{{\bf
T}}\left(\beta_1,\ldots,\beta_{\gamma-1},
\beta_\gamma\to 0,\beta_{\gamma+1},\ldots,\beta_{N-2};\theta^{{\bf T}}
\right)\right)
&=\prod_{i=1}^{m-1}\det\left(\Avril^{{{\bf
T}}_i}\left(\beta_1,\ldots,\beta_{\gamma-1}
;\theta^{{{\bf T}}_i}\right)\right)\cr &
\times \det\left(\Avril^{{{\bf T}}_m}\left(\beta_{\gamma+1},\ldots,\beta_{N-2}
;\theta^{{{\bf T}}_m}\right)\right)\cr}
}
where the $m-1$ first determinants in the r.h.s. involve $\beta_\alpha$ with
$\alpha < \gamma$ only, while the last determinant involves $\beta_\alpha$
with $\alpha>\gamma$ only. The angular parameter set $\theta^{{\bf T}}$
associated
with ${\bf T}$ is left untouched by the rescaling, but simply decomposed into
subsets $\theta^{{{\bf T}}_i}$ associated with the line vectors of the distinct
subtrees ${\bf T}_i$ (see \treesTif ).
We are now interested in values of $\beta$ and $\theta$ varying
inside the domain ${\cal D}^{{\bf T}}$ and look at the possible zeros
 of $\det(\Avril^{{\bf T}})$ inside ${\cal D}^{{\bf T}}$. We already know that
such
zeros can be reached only
when one $\beta$ at least goes to zero.
We thus fix all the variables $\theta$, and
all the variables $\beta$ non zero except for one of them, $\beta_\gamma$.
The quantity $\beta_\gamma^{\rm min}$ in \domainT\
is therefore fixed, either strictly positive or zero.
If it is strictly positive, this means that
$\beta_\gamma$ cannot reach $0$ within the domain ${\cal D}^{{\bf T}}$
for this particular configuration of the other variables.
This happens when the tree ${\bf T}_m$, obtained by fully contracting the lines
$\lambda_1,\ldots,\lambda_\gamma$ of
${\bf T}$, is not compatible with the definition of the sector ${\cal D}^{{\bf
T}}$.
The only relevant case is therefore $\beta_\gamma^{\rm min}=0$.
When $\beta_\gamma \to 0$, we can use equation \leasttC\ .
The trees ${\bf T}_i$, $1\le i \le m-1$, were already subtrees of ${\bf T}$,
hence
the associated determinants $\det(\Avril^{{{\bf T}}_i})$, which involve only
non vanishing $\beta$'s, do not vanish.
The new tree ${\bf T}_m$, which appears in the contraction process,
is now {\it compatible} with the sector, which
again implies that no fortuitous coincidence of its vertices can
occur, and $\det(\Avril^{{{\bf T}}_m})$ itself cannot vanish.
Thus $\det(\Avril^{{\bf T}})$ cannot vanish in this limit $\beta_\gamma \to 0$.
This process can be iterated on the remaining determinants in \leasttC\
for successive $\beta$'s going to zero. This shows that $\det(\Avril^{{\bf
T}})$
does not vanish
for any number of $\beta$'s going to zero. Hence we reach the important
result that $\det(\Avril^{{\bf T}})$ cannot vanish inside
the whole sector ${\cal D}^{{\bf T}}$. Since ${\cal D}^{{\bf T}}$ is bounded
(excluding
the variable $\beta_{N-1}$ which does not enter in $\det(\Avril^{{\bf T}})$),
$\det(\Avril^{{\bf T}})$ is moreover bounded from below by a strictly
positive number.

\appendix{D} {Example of cancellation of symmetry factors}
Let us consider as in \newnests\ the four compatible nests:
\eqn\firstD{\eqalign{{\CN_\oplus}&=\Big\{
(T,\omega) \Big\}\cr
{\CN_\oplus}_2&=\Big\{ (R\wedge T, \bullet ),
(T,\omega) \Big\}\cr
{\CN_\oplus}_3&=\Big\{
(T,\omega),(R\veew T ,\bullet) \Big\}\cr
{\tilde \CN}_\oplus&=\Big\{ (R\wedge T, \bullet ),
(T,\omega),(R\veew T ,\bullet) \Big\}\cr }}
where $R=\{\CR\}$, $T=\{\CT\}$ and $\omega=\big\{ \{w\} \big\}$ with $w\in
\CT$.
We want to show that the sum of the $(-1)$ and symmetry factors associated
with these nests (taking into account the degeneracy coming from the
unspecified compatible roots $\bullet$) is equal to $0$.
We recall that to a compatible nest ${\CN '}_\oplus$ is associated the factor
in front of the associated Taylor operators
(here we forget about the first diagram
$T_0=(G_\odot,G_\odot)$ implicit in all the nests of \firstD ,
and the corresponding
global $(-1)$ factor):
\eqn\SecondD{
(-1)^{{\rm Card}({\CN '}_\oplus)}W({\CN '}_{\oplus})\,=\,
(-1)^{{\rm Card}({\CN '}_\oplus)}\prod_{
       {w'\ {\rm root}}\atop
       {{\rm of}\,{\CN '}_{\oplus}}
      }
{1\over |\CT_{w'}|}
}
with $\CT_{w'}$ being the largest connected component (among
all connected components of all diagrams of ${\CN '}_\oplus $) whose root is
$w'$.
\medskip The factor associated with $\CN_\oplus$ in \firstD\
is thus $(-){1\over |\CT |}$. Let us now discuss the three remaining
nests in \firstD .

\medskip \noindent {Case (a):} $w\in \CR$ (see \largnestf )

The root of the connected component $\CR\cap \CT$ of $R\wedge T$
must be equal to $w$. The factor associated with ${\CN_{\oplus }}_2$ is then
${1\over |\CT |}$. In ${\CN_\oplus}_3$, the root of the connected component
$\CR\cup\CT$ of $R\veew T$
is either equal to $w$, or belongs to $\CR\setminus \CT$. The factor
associated with ${\CN_\oplus}_3$ is therefore
${1\over |\CR\cup \CT |}$ in the first case, and ${1\over |\CT |}{ 1
\over |\CR \cup \CT |}$ in the second case, with degeneracy $|\CR \setminus
\CT|$.
Hence, the global factor associated with ${\CN_\oplus}_3$ and its possible
rootings is
${1\over |\CR\cup \CT |}+ {1\over |\CT |}{|\CR \setminus \CT | \over
|\CR \cup \CT |}$ which, using $|\CR\setminus\CT |+|\CR \cap \CT |=
|\CR \cup \CT |$, is nothing but ${1\over |\CT |}$.
The factor
associated with ${\tilde \CN}_\oplus$ is similarly equal to
$(-)\Big[ {1\over |\CR\cup \CT |}+ {1\over |\CT |}{|\CR \setminus \CT | \over
|\CR \cup \CT |}\Big]=-{1\over |\CT |}$.
By summing up all these factors for all elements of \firstD\ ,
we get zero as expected.

\medskip \noindent {Case (b):} $w\notin \CR$ (see \largnestf )

The root of the connected component $\CR\cap \CT$ of $R\wedge T$
can now be any vertex of $\CR\cap \CT$. The factor associated with
${\CN_{\oplus}}_2$
is in this case ${|\CR \cap \CT |\over |\CR \cap \CT |}{1\over |\CT |}=
{1\over |\CT |}$, since ${\CN_\oplus}_2$ has now two distinct
roots. In ${\CN_\oplus}_3$, the two roots of the two connected components $\CT$
and $\CR \setminus
\CT $ of $R\veew T$ are respectively $w$ and any vertex in $\CR \setminus \CT$.
The factor associated with ${\CN_{\oplus }}_3$ is then ${1\over |\CT |}
{|\CR \setminus \CT | \over |\CR \setminus \CT |}={1\over |\CT |}$, while
the factor associated with ${\tilde \CN}_\oplus $ is $
(-) {|\CR \cap \CT | \over |\CR \cap \CT |}{1\over |\CT |}{|\CR\setminus \CT
|\over |\CR \setminus \CT |}=(-){1 \over |\CT |}$.
Here too the sum of these factors gives zero as expected.

\appendix{E} {``Suppression" of a reducible line from the tableau nest}
We assume here that the coincidence \supcond\ holds in the tableau
\omegared . We therefore have as a starting point the set of identities:
\eqn\firstE{\eqalign{R_{\sJ-1}^\sI\wedge T_\sJ=T_\sJ &\qquad {\scriptstyle I}
\ge {\scriptstyle I}_0\cr
T_\sJ=R_\sJ^\sI\wedge T_{\sJ+1} &\qquad {\scriptstyle I}\le {\scriptstyle I}_0
 . \cr }}
We want to prove that the lines \singline\ and \singliner\ are then identical,
{\it i.e.}
that $T_\sJ$ can be skipped in the construction of the tableau.
We thus have to prove the two following sets of identities:
\item{{\bf (1)}} For ${\scriptstyle I}\ge {\scriptstyle I}_0$:
\eqn\secondE{R_\sJ^\sI\wedge T_{\sJ+1}=R_{\sJ-1}^{\sI}\wedge T_{\sJ+1} .}
\par
\item{{\bf (2)}} For ${\scriptstyle I}\le {\scriptstyle I}_0$:
\eqn\thirdE{R_{\sJ-1}^\sI\wedge T_{\sJ+1}=R_{\sJ-1}^{\sI}\wedge T_\sJ .}
\par
\noindent These two sets of indentities are consequences of the stronger
equality:
\eqn\fourthE{R_{\sJ}^\sI=R_{\sJ-1}^\sI\ ,\quad \forall {\scriptstyle I} \ge
{\scriptstyle I}_0 .}
Indeed \fourthE\ clearly implies \secondE\ for case {\bf (1)}.
Furthermore, for case {\bf (2)}, we make the following argument:

\noindent We use \firstE\ to write $T_\sJ$ as:
\eqn\fifthE{T_{\sJ}=R_\sJ^{\sI_0}\wedge T_{\sJ+1} .}
Hence, we have:
\eqn\sixthE{\eqalign{R_{\sJ-1}^\sI\wedge T_\sJ
&=(R_{\sJ-1}^\sI\wedge R_{\sJ}^{\sI_0}) \wedge T_{\sJ+1}\cr
&=R_{\sJ-1}^\sI \wedge T_{\sJ+1}\ ,\ {\rm QED}\cr}}
where we have made use of $R_{\sJ-1}^\sI\prec R_{\sJ-1}^{\sI_0}$ since
${\scriptstyle I}\le {\scriptstyle I}_0$, together with $R_{\sJ-1}^{\sI_0}
=R_\sJ^{\sI_0}$ as a particular case of \fourthE .
We are thus left with proving \fourthE .
\medskip
\noindent {\it Proof of \fourthE :}
\midinsert
\figinsert{8.truecm}{7.truecm}{%\hsize}{
\figcap\Riijjf{An application of the operation of \runometf\ :
the figure describes a connected component $\CR^{\sI,i}$ of $R^\sI$ and the
corresponding connected component $\CR^{\sI,i}_{\sJ-1}$ of
$R^\sI\veewjmin T_{\sJ-1}$, obtained by fusing to $\CR^{\sI,i}$
those connected components of $T_{\sJ-1\oplus}$ which have their root
inside $\CR^{\sI,i}$ and cutting out those which do not have their root
inside $\CR^{\sI,i}$, but still intersect $\CR^{\sI,i}$.}
}
\endinsert
It is first useful to characterize the connected components
of $R_{\sJ-1}^{\sI}$ (or of $R_\sJ^\sI$). Denoting by $\CR^{\sI,i}$
the connected components of $R^\sI$, a generic component $\CR_{\sJ-1}^{\sI,i}$
of $R_{\sJ-1}^\sI$ is of the form:
\eqn\seventhE{ \CR_{\sJ-1}^{\sI,i}=\Bigg[
\CR^{\sI,i}
\bigcup_{w_{\sJ-1,k}\in \CR^{\sI,i} }\CT_{\sJ-1,k}
\Bigg] \setminus
\Bigg(
\bigcup_{w_{\sJ-1,k'}'\notin \CR^{\sI,i}
}\CT_{\sJ-1,k'}'
\Bigg)\ ,}
which simply states that a connected component
$\CR_{\sJ-1}^{\sI,i}$ of $R_{\sJ-1}^\sI=R^\sI\veewjmin T_{\sJ-1}$
is obtained from a connected component $\CR^{\sI,i}$ of $R^\sI$
by (see \Riijjf ):
\item{-} considering all the connected components of $T_{\sJ-1}$ ;
\item{-} making the union with $\CR^{\sI,i}$ of those $\CT_{\sJ-1,k}$
which share their root with $\CR^{\sI,i}$;
\item{-} cutting out from $\CR^{\sI,i}$ those $\CT_{\sJ-1,k'}'$
which do not.  \par
\noindent Since the connected components of $T_{\sJ-1}$ are all
disjoint, the order of the union and cutting operations in \seventhE\ is
indifferent. Notice also that the connected components of $T_{\sJ-1}$
which do not intersect $\CR^{\sI,i}$ do not affect $\CR^{\sI,i}_{\sJ-1}$ in the
operation $\veewjmin$. Of course,
it may happen that $\CR_{\sJ-1}^{\sI,i}$ is empty and $R_{\sJ-1}^\sI$
has in general less connected components than $R^\sI$.
\par \noindent
For convenience, we introduce the notations:
\eqn\eighthE{A_{\sJ-1}^{\sI,i}=
\bigcup_{w_{\sJ-1,k}\in \CR^{\sI,i} }\CT_{\sJ-1,k}}
\eqn\ninethE{B_{\sJ-1}^{\sI,i}=
\bigcup_{w_{\sJ-1,k'}'\notin \CR^{\sI,i}
}\CT_{\sJ-1,k'}'}
which are {\it complementary sets} in $\CG$ since $T_{\sJ-1}$ is a complete
diagram.
With these notations, we have:
\eqn\tenthE{\CR_{\sJ-1}^{\sI,i}=\Big( \CR^{\sI,i} \cup A_{\sJ-1}^{\sI,i}\Big)
\setminus \Big( \CR^{\sI,i}\cap B_{\sJ-1}^{\sI,i}\Big)}
and a similar equation for the connected components $\CR_{\sJ}^{\sI,i}$
of $R_{\sJ}^{\sI}$. Therefore, to prove $R_{\sJ-1}^\sI=R_{\sJ}^\sI$, it
is enough to prove:
\eqn\eleventhE{\encadre{\hbox{$\displaystyle
\eqalign{A^{\sI,i}_{\sJ}&= A^{\sI,i}_{\sJ-1}\cr
B^{\sI,i}_{\sJ} &= B^{\sI,i}_{\sJ-1}\cr}$}}}
\medskip
\noindent The main ingredient comes from the property in \firstE :
$$T_\sJ=R_{\sJ-1}^\sI\wedge T_\sJ, \quad \forall {\scriptstyle I}\ge
{\scriptstyle I}_0\ ,$$
which implies
\eqn\twelfthE{\encadre{\hbox{$\displaystyle T_\sJ\prec R_{\sJ-1}^\sI$}}}
This means that {\it any connected component $\CT_{\sJ,j}$ of $T_\sJ$ which
intersects a connected component $\CR_{\sJ-1}^{\sI,i}$ is actually
entirely included it the latter}.

\medskip \noindent $\diamond$ We first prove $A_{\sJ-1}^{\sI,i}
\subset A_{\sJ}^{\sI,i}$:
\par \noindent Let us consider a connected component $\CT_{\sJ-1,k}$
of $T_{\sJ-1}$, such that $w_{\sJ-1,k}\in \CR^{\sI,i}$. From the nest
property, this connected component is included in a connected component
$\CT_{\sJ,j}$ of $T_\sJ$. By definition,
$\CT_{\sJ-1,k}\subset\CR_{\sJ-1}^{\sI,i}$
and therefore $\CT_{\sJ,j}$ intersects $\CR_{\sJ-1}^{\sI,i}$.
 From \twelfthE ,
$\CT_{\sJ,j}$ is necessarily included in $\CR_{\sJ-1}^{\sI,i}$
and in particular its root $w_{\sJ,j}$ belongs
to $\CR_{\sJ-1}^{\sI,i}$, thus to $\CR^{\sI,i}\cup A_{\sJ-1}^{\sI,i}$.
One has either $w_{\sJ,j}\in \CR^{\sI,i}$, or $w_{\sJ,j}\in \CT_{\sJ-1,l}$
for some connected component $\CT_{\sJ-1,l}$ (with $l\ne k$ in
general) of $T_{\sJ-1}$ such that
$w_{\sJ-1,l}\in \CR^{\sI,i}$. In the latter case, from the compatibility
condition for the roots, we have $w_{\sJ,j}=w_{\sJ-1,l}\in \CR^{\sI,i}$.
Therefore, in any case, $w_{\sJ,j}\in \CR^{\sI,i}$ and
$\CT_{\sJ,j}\subset A_{\sJ}^{\sI,i}$.
This implies
$\CT_{\sJ-1,k}\subset A_{\sJ}^{\sI,i}$, which leads to:
\eqn\thirteenthE{A_{\sJ-1}^{\sI,i}\subset A_{\sJ}^{\sI,i} .}
\medskip \noindent $\diamond$ We now prove $B_{\sJ-1}^{\sI,i}
\subset B_{\sJ}^{\sI,i}$:
\par\noindent We can use the fact that $R^\sI$ is a complete diagram,
thus each root $w_{\sJ-1,k}$ of a connected component $\CT_{\sJ-1,k}$
belongs to one and only one connected component $\CR^{\sI,i}$ of $R^{\sI}$.
The set $B_{\sJ-1}^{\sI,i}$ can therefore be expressed as:
\eqn\fourteenthE{B_{\sJ-1}^{\sI,i}=\bigcup_{i'\ne i}A_{\sJ-1}^{\sI,i'} .}
A similar equation holds for $B_{\sJ}^{\sI,i}$. Making use of
\thirteenthE\ for each $i'$ in the r.h.s. of \fourteenthE , we directly
arrive at:
\eqn\fifteenthE{B_{\sJ-1}^{\sI,i}\subset B_{\sJ}^{\sI,i} .}
\medskip The inclusion properties \thirteenthE\ and \fifteenthE ,
together with the fact that $A_{\sJ-1}^{\sI,i}$ and $B_{\sJ-1}^{\sI,i}$
on the one hand,
and $A_{\sJ}^{\sI,i}$ and $B_{\sJ}^{\sI,i}$ on the other hand,
are pairs of complementary sets of $\CG$, imply \eleventhE , hence \fourthE .

\appendix{F} {Addition of reducible lines in the tableau nest}
\noindent $\diamond$ We want to
prove first that, if we consider a compatibly rooted
nest $\CN_{\oplus}=\{T_{0\oplus},\ldots,T_{\sT\oplus}\}$ and build
the larger nest $\CN_\oplus'=\{T_{0\oplus},\ldots,T_{\sJ-1\oplus},
T_{\sJ-1\oplus}^{\sI_0},T_{\sJ\oplus},\ldots,T_{\sT\oplus}\}$ by inserting
between the levels ${\scriptstyle J}-1$ and $\scriptstyle J$ of
$\CN_\oplus$ an extra rooted diagram $T_{\sJ-1\oplus}^{\sI_0}=
(T_{\sJ-1}^{\sI_0},\omega_{\sJ-1}^{\sI_0})$
with:
\eqn\firstF{T_{\sJ-1}^{\sI_0}=R_{\sJ-1}^{\sI_0}\wedge T_\sJ\equiv T\,'}
and $\omega_{\sJ-1}^{\sI_0}\equiv \omega'$ an arbitrary set
of roots compatible with
the rooting of $\CN_\oplus$ (making $\CN_\oplus'$ compatibly rooted),
then the tableau of $\CN_\oplus'$ can be reduced to that of $\CN_\oplus$.
For convenience, we denote ${T_{\sJ-1}^{\sI_0}}_\oplus$ by $T_\oplus\,'=(T\,',
\omega')$.
\par \noindent More precisely, the tableau built from $\CN_\oplus'$ is:
\eqn\secondF{\matrix{
\vdots&&&&&\cr
\hbox{\vbox{\hbox{$\displaystyle T_{\sJ-1}$}\hbox{}\hbox{$\displaystyle T\,'
$}}}&
\hbox{\vbox{\hbox{$\displaystyle R_{\sJ-1}^1\wedge
T\,'$}\hbox{}\hbox{$\displaystyle {R\,'}^1\wedge T_{\sJ}$}}}&
\hbox{\vbox{\hbox{$\ldots$}\hbox{}\hbox{$\ldots$}}}&
\raise -10 pt \encadre{\hbox{\vbox{
\hbox{$\displaystyle R_{\sJ-1}^{\sI_0}\wedge T\,'$}\hbox{$\qquad
$}\hbox{$\displaystyle
{R\,'}^{\sI_0}\wedge T_{\sJ}$}}}}&
\hbox{\vbox{\hbox{$\ldots$}\hbox{}\hbox{$\ldots$}}}&
\hbox{\vbox{\hbox{$\displaystyle R_{\sJ-1}^{N-1}\wedge
T\,'$}\hbox{}\hbox{$\displaystyle {R\,'}^{N-1}\wedge T_{\sJ}$}}}\cr
&&&&&\cr
T_\sJ&&&&& \cr
\vdots&&&&&\cr
}}
where
\eqn\thirdF{{R\,'}^\sI\equiv R^{\sI}\veewprim T\,' .}
We want to prove that this tableau has the coincidence property for
${\scriptstyle I}={\scriptstyle I}_0$:
\eqn\fourthF{R_{\sJ-1}^{\sI_0}\wedge T\,'={R\,'}^{\sI_0}\wedge
T_\sJ\ ,}
and therefore can be reduced to the tableau of $\CN_\oplus$.
 From the definition of $T\,'$, the term on the l.h.s. of
\fourthF\ is nothing
but $R_{\sJ-1}^{\sI_0}\wedge T_\sJ$ and the coincidence property is equivalent
to:
\eqn\fifthF{R_{\sJ-1}^{\sI_0}\wedge T_\sJ={R\,'}^{\sI_0}\wedge
T_\sJ . }
This last equation is actually a consequence of the stronger indentity
\eqn\sixthF{R_{\sJ-1}^{\sI_0}={R\,'}^{\sI_0} }
which we prove now.
\medskip
\noindent {\it Proof of \sixthF }:
\par As in Appendix F, we consider a typical connected component
$\CR_{\sJ-1}^{\sI_0,i_0}$ of $R_{\sJ-1}^{\sI_0}$, defined by:
\eqn\seventhF{ \CR_{\sJ-1}^{\sI_0,i_0}=\Bigg[
\CR^{\sI_0,i_0}
\bigcup_{w_{\sJ-1,k}\in \CR^{\sI_0,i_0} }\CT_{\sJ-1,k}
\Bigg] \setminus
\Bigg(
\bigcup_{w_{\sJ-1,l}\notin \CR^{\sI_0,i_0}
}\CT_{\sJ-1,l}
\Bigg)\ ,}
or by the equivalent equation:
\eqn\eighthF{\CR_{\sJ-1}^{\sI_0,i_0}=\Big( \CR^{\sI_0,i_0} \cup
A_{\sJ-1}^{\sI_0,i_0}
\Big)
\setminus \Big( \CR^{\sI_0,i_0}\cap B_{\sJ-1}^{\sI_0,i_0}\Big)}
where
\eqn\ninethF{A_{\sJ-1}^{\sI_0,i_0}=
\bigcup_{w_{\sJ-1,k}\in \CR^{\sI_0,i_0} }\CT_{\sJ-1,k}\ ,}
\eqn\tenthF{B_{\sJ-1}^{\sI_0,i_0}=
\bigcup_{w_{\sJ-1,l}\notin \CR^{\sI_0,i_0}
}\CT_{\sJ-1,l} . }
The sets $A_{\sJ-1}^{\sI_0,i_0}$ and $B_{\sJ-1}^{\sI_0,i_0}$ are
complementary subsets of $\CG$ and, as in Appendix E:
\eqn\eleventhF{B_{\sJ-1}^{\sI_0,i_0}=\bigcup_{i\ne i_0}A_{\sJ-1}^{\sI,i} .}
We then can write for ${R\,'}^{\sI_0}$ an equation similar to
\eighthF\ with $A_{\sJ-1}^{\sI_0,i_0}$ and $B_{\sJ-1}^{\sI_0,i_0}$ replaced by:
\eqn\twelfthF{{A'}^{\sI_0,i_0}=
\bigcup_{{w'}_k^i\in \CR^{\sI_0,i_0} }{\CT'}_k^i}
\eqn\thirteenthF{{B'}^{\sI_0,i_0}=
\bigcup_{{w'}_l^i\notin \CR^{\sI_0,i_0}
}{\CT'}_l^i}
which are complementary subsets of $\CG$ and satisfy an equation similar
to \eleventhF .
In \twelfthF\ and \thirteenthF ,
${\CT'}_k^i$ is the generic connected component of
$T\,'$ given by:
\eqn\fourtenthF{{\CT'}_k^i=\CR_{\sJ-1}^{\sI_0,i}\cap\CT_{\sJ,k}}
and ${w'}_k^i$ is its root.
\par
In order to prove \sixthF , it is sufficient to prove that
$A_{\sJ-1}^{\sI_0,i_0} \subset {A'}^{\sI_0,i_0}$ . Indeed,
from \eleventhF\ and the similar equation for ${B'}^{\sI_0,i_0}$,
this inclusion will imply $B_{\sJ-1}^{\sI_0,i_0}\subset {B'}^{\sI_0,i_0}$. From
the complementarity property of $A_{\sJ-1}^{\sI_0,i_0}$ and
$B_{\sJ-1}^{\sI_0,i_0}$ on the one hand, and that of ${A'}^{\sI_0,i_0}$ and
${B'}^{\sI_0,i_0}$ on the other hand, the
two equalities:
\eqn\fifteenthF{A_{\sJ-1}^{\sI_0,i_0}= {A'}^{\sI_0,i_0}}
\eqn\sixteenthF{B_{\sJ-1}^{\sI_0,i_0} ={B'}^{\sI_0,i_0}}
follow, leading to \sixthF .
\medskip
\noindent $\diamond$ We are thus left with proving
$A_{\sJ-1}^{\sI_0,i_0}\subset {A'}^{\sI_0,i_0}$:
\par \noindent Let us consider a connected component $\CT_{\sJ-1,k}$
of $T_{\sJ-1}$, such that $w_{\sJ-1,k}\in \CR^{\sI_0,i_0}$. From the nest
property, this connected component is included in a connected component
$\CT_{\sJ,j}$ of $T_\sJ$. By definition,
$\CT_{\sJ-1,k}\subset\CR_{\sJ-1}^{\sI_0,i_0}$
and therefore
$\CT_{\sJ-1,k}\subset\CR_{\sJ-1}^{\sI_0,i_0}\cap\CT_{\sJ,j}\equiv
{\CT'}_j^{i_0}$.
The root ${w'}_j^{i_0}$ of ${\CT'}_j^{i_0}$ belongs
to $\CR_{\sJ-1}^{\sI_0,i_0}$, thus to
$\CR^{\sI_0,i_0}\cup A_{\sJ-1}^{\sI_0,i_0}$.
One has either ${w'}_j^{i_0}\in \CR^{\sI_0,i_0}$ or ${w'}_j^{i_0}\in
\CT_{\sJ-1,l}$
for some connected component $\CT_{\sJ-1,l}$ (with $l\ne k$ in
general) of $T_{\sJ-1}$ such that
$w_{\sJ-1,l}\in \CR^{\sI_0,i_0}$. In the latter case, from the compatibility
condition (in the nest $\CN_\oplus'$) between the root
${w'}_j^{i_0}$ and the roots of
$T_{\sJ-1}$, one has ${w'}_j^{i_0}=w_{\sJ-1,l}\in \CR^{\sI_0,i_0}$.
Therefore, in any case, ${w'}_j^{i_0}\in \CR^{\sI_0,i_0}$ and
${\CT'}_j^{i_0}\subset {A'}^{\sI_0,i_0}$.
This implies
$\CT_{\sJ-1,k}\subset {A'}^{\sI_0,i_0}$, which leads to:
\eqn\seventeenthF{A_{\sJ-1}^{\sI_0,i_0}\subset {A'}^{\sI_0,i_0}\ ,}
which completes the proof.
\medskip
\noindent $\diamond$ The above property generalizes to a nest $\CN_\oplus'$
obtained from $\CN_\oplus$ by inserting between the levels
${\scriptstyle J}-1$ and ${\scriptstyle J}$ an arbitrary number of
diagrams $T_{\sJ-1\oplus}^{\sI_0},T_{\sJ-1\oplus}^{\sI_1},\ldots,
T_{\sJ-1\oplus}^{\sI_K}$
with $1\le {\scriptstyle I}_0\le{\scriptstyle I}_1\le\ldots\le
{\scriptstyle I}_K\le N-2$, where as before:
\eqn\eighteenthF{T_{\sJ-1}^{\sI}=R_{\sJ-1}^{\sI}\wedge T_\sJ\ ,}
and where the roots of these extra diagrams are such that $\CN_\oplus'$
is compatibly rooted.
Indeed, one can proceed by recursion by adding first $T_{\sJ-1}^{\sI_K}=
R_{\sJ-1}^{\sI_K} \wedge T_\sJ$ between $T_{\sJ-1}$ and $T_{\sJ}$ .
Then one can add
$R_{\sJ-1}^{\sI_{K-1}}\wedge T_{\sJ-1}^{\sI_K}$ between
$T_{\sJ-1}$ and $T_{\sJ-1}^{\sI_K}$. From the nest property
of the sector nest $\CS$, we have $R_{\sJ-1}^{\sI_{K-1}}\prec
R_{\sJ-1}^{\sI_K}$ and this second added diagram is
nothing but $R_{\sJ-1}^{\sI_{K-1}}\wedge R_{\sJ-1}^{\sI_K}\wedge
T_{\sJ}=R_{\sJ-1}^{\sI_{K-1}}\wedge T_{\sJ}=T_{\sJ-1}^{\sI_{K-1}}$ as
wanted. This process can be repeated until the first diagram
$T_{\sJ-1}^{\sI_0}$ is inserted.
\medskip
\noindent $\diamond$ Finally, the
above property also generalizes to arbitrary insertions
between several pairs $({\scriptstyle J}-1,{\scriptstyle J})$, each
pair being actually decoupled from the other pairs.
\medskip
When applied to a minimal nest $\CN^0_\oplus$, this property means
that all the nests $\CN_\oplus'$ obtained from $\CN^0_\oplus$ by inserting
an arbitrary
number of diagrams of ${\tilde \CN}(\CS,\CN^0_\oplus)\setminus \CN^0$
(rooted with compatible roots) lead by reduction to $\CN^0_\oplus$, and
therefore belong to $\CC_\CS(\CN^0_\oplus)$.

\appendix{G}{Sum rule for the weights $\tit W$}
In this appendix, we prove \somme .
Given a nest $\CN$, we first give an alternative procedure to
construct all compatible rootings $\oplus_\CN$ of $\CN$, with
their weight factor $W(\CN_{\oplus_\CN})$ \Wnest .

Let $\sigma$ be a bijection from $\{1,2,\ldots,N\}$ into $\CG$
(it is nothing but an ordering of the $N$ vertices of $\CG$).
There are $N!$ such orderings.
To any subset $\CP$ of $\CG$, we assign  a root $p$ through $\sigma$
by the following definition:
\eqn\firstG{p=\sigma(k)\quad {\rm where}\quad k=\min (n\in\{1,\ldots,N\}\ :\
\sigma(n)\in \CP) .}
We denote this assignment procedure by:
\eqn\secondG{\CP\buildrel \sigma\over \longrightarrow p .}
It is easy to check that, when applied to all connected components
of all diagrams of $\CN$, this rooting procedure builds a compatible
rooting of $\CN$.
Moreover, all compatible rootings of $\CN$ can be built in that way.
Given such a rooting $\oplus_\CN$, the number of distinct orderings $\sigma$
which build $\oplus_\CN$ is:
\eqn\secbisG{K(\CN_{\oplus_\CN})\equiv {\rm Card}( \{ \sigma :\ \forall (\CT,w)
\ \hbox{rooted connected comp. of $\CN_{\oplus_\CN}$}\ ,\CT \buildrel
\sigma \over \longrightarrow w\}) .}
It is simply related to the weight $W(\CN_{\oplus_\CN})$ by
\eqn\thirdG{
{K(\CN_{\oplus_\CN})\over N!}
=W(\CN_{\oplus_\CN})=
\prod_w {1\over |\CT_w|}\ .
}
Indeed, given a subset $\CP$ of $\CG$ and a vertex $p$ in $\CP$, the number
of $\sigma$ which assign $p$ to $\CP$ is $N!/|\CP|$ (the probability for
$p$ to be the first vertex of $\CP$ to appear in the sequence $\sigma(1),
\ldots,\sigma(N)$ is $1/|\CP|$).
A compatible rooting $\oplus_\CN$ of $\CN$ is entirely known once one specifies
for each vertex $w$ the largest connected component of $\CN$, $\CT_w$, which
has $w$ as its root.
The above argument can then
be extended to all these largest connected components of $\CN$
containing the roots of $\oplus_\CN$, and leads to \thirdG .
\medskip
The proof of \somme is then straightforward.
Indeed, the r.h.s. of \somme\ is simply $1/(N!)$ times
\eqn\fourthG{{\rm Card} (\{ \sigma :\ \forall (\CT,w)
\ \hbox{rooted connected comp. of $\CN_{\oplus}$}\ ,\CT \buildrel
\sigma \over \longrightarrow w\}) ,}
while each term of the sum in the l.h.s of \somme\ is $1/(N!)$ times
\eqn\fifthG{\eqalign{{\rm Card} ( \{ \sigma :\ &\bullet  \forall (\CT,w)
\ \hbox{rooted connected comp. of $\CN_{\oplus}$}\ ,\CT \buildrel
\sigma \over \longrightarrow w\ ,\cr
&\bullet \forall (\CT,w)\ \hbox{rooted connected comp. of $\CM_{\oplus_\CM}$
not in $\CN_\oplus$}\ , \CT \buildrel
\sigma \over \longrightarrow w\}) .\cr}}
The sum over $\oplus_\CM$ in \somme\ relaxes the second constraint on $\sigma$
in \fifthG , and reproduces \fourthG . Hence \somme\ follows.

\appendix{H}{Estimates of subtracted integrands in a Hepp sector}
In this appendix, we prove \GrandO\ and \GrandOT .
We shall proceed in three steps:
\item{(I)} We first analyze the properties of the elements
of the matrix $\Avril^{{\bf T}_{\sJ,j}}$ in terms of the $\beta^\sI$
variables.
\item{(II)} We then write an integral representation of the $(1-\Tay \ )$
operators appearing in the l.h.s. of \GrandO\ or \GrandOT .
\item{(III)} We finally show \GrandO\ and \GrandOT .
\par
\vfill\eject
\medskip
\noindent{\hbox{$\diamond$ (I) {\it Properties of} $\Avril^{{\bf T}_{\sJ,j}}$}}

In this subsection, we shall work separately
inside each connected component $\CTt_{\sJ,j}$ of $\Tt_{\sJ}$.
As explained in section 8.2, the line vectors $\lambda^\sI_{\sJ,j}$
of the oriented ordered tree ${\bf T}_{\sJ,j}$ spanning $\CTt_{\sJ,j}$
are uniquely labeled by $\SI \in {\overline{\rm Ind}}(\SJ,j)$.
 From now on, we shall suppress the indices $(\SJ,j)$ and thus
denote $\lambda^\sI_{\sJ,j}$ by $\lambda^\sI$.
A typical element of the matrix $\Avril^{{\bf T}_{\sJ,j}}$ writes:
\eqn\firstH{\Avril^{{\bf T}_{\sJ,j}}_{\sK\sL}={-1\over 2
|\lambda^\sK |^\nu \  |\lambda^\sL |^\nu }
\Big\{ |R^{\sK\sL}
+\lambda^\sL-\lambda^\sK|^{2\nu}-|R^{\sK\sL} +\lambda^\sL|^{2\nu}
-|R^{\sK\sL}-\lambda^\sK|^{2\nu}+|R^{\sK\sL}|^{2\nu}\Big\}}
where $R^{\sK\sL}$ is the ``basis" of the quadrilateral
\eqn\secondC{R^{\sK\sL}=x_{i_\sL}-x_{i_\sK}
\ ,}
with $i_\sK$ and $i_\sL$ being the origins of $\lambda^\sK$ and
$\lambda^\sL$. The vector $R^{\sK\sL}$ is a linear combination
of the $\lambda^\sI$'s joining $x_{i_\sK}$ and $x_{i_\sL}$, and
since the tree ${\bf T}_{\sJ,j}$ has been built from the
rooted sector $\CS_{\sJ,j\oplus}$, this linear combination involves
only $\lambda^\sI$'s for $\SI > \min (\SK,\SL)$ (see section 7.2):
\eqn\thirdH{R^{\sK\sL}=\sum_{\sM>\min (\sK,\sL)}c^{\sK\sL}_\sM\, \lambda^\sM}
with $c^{\sK\sL}_\sM=0,\pm 1$.
\bigskip\noindent PROPOSITIONS:

\noindent$\bullet$ {\it Prop. 1:}
$\det(\Avril^{{\bf T}_{\sJ,j}})$ is a positive, non
vanishing continuous function on the compact domain $\CH^\CS$, and is therefore
bounded from below on $\CH^\CS$ by a strictly positive number.
In particular, the matrix $\Avril^{{\bf T}_{\sJ,j}}$ is invertible.

\noindent$\bullet$ {\it Prop. 2:}
$\Avril^{{\bf T}_{\sJ,j}}_{\sK\sL}$, as a function of the
$\beta$, $\chi$ and $\theta$ variables, depends on the $\beta^\sI$'s
for $\SI$ in some subset $\CJ_{\sJ,j}(\SK,\SL)$ only, defined as
\eqn\fourthH{\CJ_{\sJ,j}(\SK,\SL)=\Big\{ \SI : \min(\SK,\SL)\le
\SI < \max\big(\SK,\SL,\max(\SM : c^{\sK\sL}_\sM \ne 0)\big)\Big\} }
with the convention that:
$\max\big(\SK,\SL,\max(\SM : c^{\sK\sL}_\sM \ne 0)\big)
=\max(\SK,\SL)$,
if all the $c^{\sK\sL}$ are zero (that is if $R^{\sK\sL}=0$).

\noindent$\bullet$ {\it Prop. 3:}
Inside the sector $\CH^\CS$,
\eqn\fifthH{\Avril^{{\bf T}_{\sJ,j}}_{\sK\sL}=\CO\Big(
\prod_{\sI \in \CJ_{\sJ,j}(\sK,\sL)} (\beta^\sI)^\delta \Big) .}

\noindent$\bullet$ {\it Prop. 4:}
The matrix $\Avril^{{\bf T}_{\sJ,j}}$ is positive, and bounded from below
by a strictly positive constant. By this we mean that there exist a strictly
positive number $C$ such that \break $(\Avril^{{\bf T}_{\sJ,j}}-C\II)$ is a
positive
matrix on $\CH^\CS$.

\bigskip
\noindent Proposition 1 has already been proven in Appendix C, in the
restricted case of a generalized Hepp sector $\CH^{\bf T}$ attached
to some tree ${\bf T}$. The proof can be carried over to the whole
extended Hepp sector $\CH^\CS$ attached to the nest $\CS=\CS({\bf T})$.
Indeed, the spirit of the proof is that $\Avril^{{\bf T}}$ depends
only on ratios of successive $\lambda$'s ($\beta$ variables);
from the bounds on those
ratios inside $\CH^{{\bf T}}$, we deduce that if some points coincide
then one of these ratios at least must vanish, and
$\det(\Avril^{{\bf T}})$ factorizes and remains strictly positive.
Since, from Schoenberg's theorem, this is the only case when
$\det(\Avril^{{\bf T}})$ might have vanished, we deduce that it actually
never vanishes, and remains positive inside $\CH^{{\bf T}}$.
Inside $\CH^\CS$, we have weaker bounds on the ratios of $\lambda$'s
but one can check that this does not alter the proof.
\medskip
To prove Propositions 2 and 3, we first consider the trivial case
$\SK=\SL$. In this case $\CJ_{\sJ,j}(\SK,\SK)=\hbox{\O}$ but then
$\Avril^{{\bf T}_{\sJ,j}(\SK,\SK)}=1$, which satisfies these propositions.

\noindent
We can therefore assume that $\SK<\SL$. Four distinct situations may occur:
\item{(a)} $R^{\sK\sL}=0$:
then $\CJ_{\sJ,j}(\SK,\SL)=\{ \SI : \SK\le\SI<\SL\}$;
\par
If $R^{\sK,\sL} \ne 0$, we denote by:
\eqn\sixthH{\SP=\max (\SM :c^{\sK\sL}_\sM\ne 0 )\ .}
\item{(b)} If $\SP>\SL$,
then $\CJ_{\sJ,j}(\SK,\SL)=\{ \SI : \SK\le\SI<\SP\}$;
\item{(c)} If $\SK<\SP<\SL$,
then $\CJ_{\sJ,j}(\SK,\SL)=\{ \SI : \SK\le\SI<\SL\}$;
\item{(d)} If $\SP=\SL$,
then $\CJ_{\sJ,j}(\SK,\SL)=\{ \SI : \SK\le\SI<\SL\}$.
\medskip
We shall use the property that, if $\SA >\SB $ and $\SA > \SC$,
then the quantity:
\eqn\propgo{\displaystyle
{|\lambda^\sA+(\sum\limits_{\sB} \pm
\lambda^\sB) \pm \lambda^\sC|^{2\nu}-|\lambda^\sA
+(\sum\limits_{\sB} \pm \lambda^\sB) |^{2\nu}\over
|\lambda^\sA |^{2\nu-1}|\lambda^\sC |}
}
is bounded (in module) from above inside $\CH^\CS$.
This follows from the fact that, inside the sector $\CH^\CS$,
the ratios $|\lambda^\sB |/|\lambda^\sA |$,
$|\lambda^\sC |/|\lambda^\sA |$ and
$\displaystyle |\lambda^\sA |/|\lambda^\sA+\sum\limits_{\sB}\pm\lambda^\sB |$
are bounded. Eq. \propgo\
can then easily be obtained by use of the mean value theorem.

By a simple generalization of this property, one can show that,
if $\SA > \SB$, $\SA > \SC$ and $\SA > \SD$, then the quantity:
\eqn\propgh{\displaystyle {| \lambda^{\sA}\kern -3pt+(\sum\limits_{\sB} \pm
\lambda^\sB)
\kern -3pt\pm\kern -3pt\lambda^\sC\kern -5pt\pm\kern -3pt\lambda^\sD|^{2\nu}
\kern -5pt-\kern -3pt|
\lambda^{\sA}\kern -3pt+(\sum\limits_{\sB} \pm \lambda^\sB)
\kern -3pt\pm\kern -3pt\lambda^\sC|^{2\nu} \kern -5pt-\kern -3pt|
\lambda^{\sA}\kern -3pt+(\sum\limits_{\sB} \pm \lambda^\sB)
\kern -3pt\pm\kern -3pt\lambda^\sD|^{2\nu}\kern -5pt+\kern -3pt|
\lambda^{\sA}\kern -3pt+(\sum\limits_{\sB} \pm \lambda^\sB)
|^{2\nu}\over
|\lambda^{\sA}|^{2\nu-2}|\lambda^\sC ||\lambda^\sD |}}
is also bounded (in module) from above inside $\CH^\CS$.

Let us now consider cases (a)--(d) above.

\noindent Case (a):
We can write:
\eqn\caseaH{\eqalign{\Avril^{{\bf T}_{\sJ,j}}_{\sK\sL}&=-{1\over 2 }
\Big\{ \left({|\lambda^\sK |\over|\lambda^\sL |}\right)^{1-\nu}
{|\lambda^\sL-\lambda^\sK|^{2\nu}-|\lambda^\sL|^{2\nu}\over
|\lambda^\sL |^{ 2\nu -1}|\lambda^\sK |}
-\left({|\lambda^\sK|\over |\lambda^\sL |}\right)^{\nu}\Big\}\cr
&=\CO\Big\{\Big( {|\lambda^\sK| \over| \lambda^\sL |}\Big)^{1-\nu}\Big\}
+\CO\Big\{\Big( {|\lambda^\sK |\over |\lambda^\sL |}\Big)^{\nu}\Big\}
\cr &=\CO\Big(\prod_{\sK\le\sI<\sL}(\beta^\sI)^\delta\Big)\ ,\cr}}
which proves Proposition 3 in this case.
In \caseaH , we used \propgo\ and the fact that
$|\lambda^\sK|/|\lambda^\sL|$
is of the same order that $\displaystyle \prod_{\sK\le\sI<\sL}(\beta^\sI) $
since:
\eqn\sameorder{{|\lambda^\sK |\over |\lambda^\sL |}={\chi^\sK\over \chi^\sL}
\prod_{\sK\le\sI<\sL}(\beta^\sI) \ , }
and since the $\chi$ variables are bounded from above and from below.
 From \caseaH , we also deduce that
$\Avril^{{\bf T}_{\sJ,j}}_{\sK\sL}$ depends only on $|\lambda^\sK |
/ |\lambda^\sL | $, that is,
from \sameorder , depends only on $\beta^\sI$ for
$\SK \le \SI < \SL$, which precisely defines $\CJ_{\sJ,j}(\SK,\SL)$
in this case, whence Proposition 2.
\medskip\noindent
Case (b):
We can now write:
\eqn\casebH{\eqalign{&\Avril^{{\bf T}_{\sJ,j}}_{\sK\sL}=-{1\over 2}
\Big({|\lambda^\sK |\over |\lambda^{\sP}|}\Big)^{1-\nu}
\Big({|\lambda^\sL |\over |\lambda^{\sP}|}\Big)^{1-\nu}\times
\cr &
{| \lambda^{\sP}\kern -3pt+\sum c^{\sK\sL}_\sM \lambda^\sM
\kern -5pt+\kern -3pt\lambda^\sL\kern -5pt-\kern -5pt\lambda^\sK|^{2\nu}
\kern -5pt-\kern -3pt|
\lambda^{\sP}\kern -3pt+\sum c^{\sK\sL}_\sM \lambda^\sM
\kern -5pt+\kern -3pt\lambda^\sL|^{2\nu} \kern -5pt-\kern -3pt|
\lambda^{\sP}\kern -3pt+\sum c^{\sK\sL}_\sM \lambda^\sM
\kern -5pt-\kern -5pt\lambda^\sK|^{2\nu}\kern -5pt+\kern -3pt|
\lambda^{\sP}\kern -3pt+\sum c^{\sK\sL}_\sM \lambda^\sM
|^{2\nu}\over
|\lambda^{\sP}|^{2\nu-2}|\lambda^\sK ||\lambda^\sL |
}\cr&\qquad = \CO \Big\{
\Big({|\lambda^\sK |\over |\lambda^{\sP}|}\Big)^{1-\nu}
\Big({|\lambda^\sL |\over |\lambda^{\sP}|}\Big)^{1-\nu}\Big\}\cr
&\qquad =\CO \Big( \prod_{\sK\le\sI<\sL } (\beta^\sI)^{1-\nu }
\prod_{\sL\le\sI<\sP }(\beta^\sI)^{2-2\nu }
\Big) \cr&\qquad=\CO \Big(\prod_{\sK\le\sI<\sP}
(\beta^\sI)^\delta \Big) \ ,\cr
}}
by use of \propgh . This proves Proposition 3 in this case.
Moreover, from \casebH , $\Avril^{{\bf T}_{\sJ,j}}_{\sK\sL}$
can be written as a function of the ratios $|\lambda^\sK|/|\lambda^\sP|$,
$|\lambda^\sL|/|\lambda^\sP|$ and $|\lambda^\sM|/|\lambda^\sP|$. Since
$\SK<\SM,\SL<\SP$, these ratios involve $\beta^\sI$ for
$\SK\le\SI<\SP$ only. This again proves Proposition 2.
\medskip\noindent
Case (c): We now write:
\eqn\casecH{\eqalign{\Avril^{{\bf T}_{\sJ,j}}_{\sK\sL}&={1\over 2}
\Big({ |\lambda^\sK |\over |\lambda^\sP |}\Big)^{1-\nu}
\Big({ |\lambda^\sP |\over |\lambda^\sL |}\Big)^{\nu}
\Big\{ {|\lambda^\sP+\sum c^{\sK\sL}_\sM
\lambda^\sM -\lambda^\sK|^{2\nu}-|\lambda^\sP
+\sum c^{\sK\sL}_\sM \lambda^\sM |^{2\nu}\over
|\lambda^\sP |^{2\nu-1}|\lambda^\sK |} \Big\}\cr
&-{1\over 2}
\Big({| \lambda^\sK |\over | \lambda^\sL |} \Big)^{1-\nu}
\Big\{ {|\lambda^\sL +\lambda^\sP +\sum c^{\sK\sL}_\sM \lambda^\sM-
\lambda^\sK|^{2\nu}-|\lambda^\sL+\lambda^\sP+\sum c^{\sK\sL}_\sM
\lambda^\sM |^{2\nu}
\over |\lambda^\sL |^{2\nu -1}|\lambda^\sK |}\Big\}\cr
&=\CO\Big\{
\Big({ |\lambda^\sK |\over |\lambda^\sP |}\Big)^{1-\nu}
\Big({ |\lambda^\sP |\over |\lambda^\sL |}\Big)^{\nu}
\Big\}+\CO\Big\{
\Big({|\lambda^\sK |\over |\lambda^\sL |} \Big)^{1-\nu}
\Big\} \cr &=\CO\Big\{\prod_{\sK\le\sI<\sP}(\beta^\sI)^{1-\nu}
\prod_{\sP\le\sI<\sL}(\beta^\sI)^\nu \Big\}+\CO\Big\{
\prod_{\sK\le\sI<\sL}(\beta^\sI)^{1-\nu} \Big\}\cr&=\CO\Big(
\prod_{\sK\le\sI<\sL}(\beta^\sI)^\delta \Big) \ , \cr}}
which proves Proposition 2. Here again, we can write
$\Avril^{{\bf T}_{\sJ,j}}_{\sK\sL}$ as a function of
the ratios $|\lambda^\sK|/|\lambda^\sL|$,
$|\lambda^\sP|/|\lambda^\sL|$ and $|\lambda^\sM|/|\lambda^\sL|$. Since
$\SK<\SM<\SP<\SL$, we deduce Proposition 3.
\medskip\noindent
Case (d): In this case $R^{\sK\sL}=-\lambda^\sL+\sum c^{\sK\sL}_\sM
\lambda^\sM$ and the propositions can be obtained from case (c) by
simply interchanging $R^{\sK\sL}$ and $R^{\sK\sL}+\lambda^\sL$.
This achieves the proof of Propositions 2 and 3.
\medskip
\noindent Finally, Proposition 4 is a consequence of Propositions 1 and 3.
Indeed, from Proposition 3 and the fact that the $\beta^\sI$ are
bounded from above inside $\CH^\CS$, we obtain a uniform upper bound
for $|\Avril^{{\bf T}_{\sJ,j}}_{\sK\sL}|$ inside $\CH^\CS$. This
upper bound, together with the lower bound of Proposition 1
on $\det(\Avril^{{\bf T}_{\sJ,j}})$ gives a uniform upper bound for the
modules $|(\Avril^{{\bf T}_{\sJ,j}})^{-1}_{\sK\sL}|$ of the elements of
the inverse matrix. This then implies that $(\Avril^{{\bf T}_{\sJ,j}})^{-1}$
is bounded from above by a positive number $C^{-1}$ (that is
$(\Avril^{{\bf T}_{\sJ,j}})^{-1}-C^{-1}\II$ is a negative matrix), and, since
$(\Avril^{{\bf T}_{\sJ,j}})^{-1}$ is a positive matrix, that
$\Avril^{{\bf T}_{\sJ,j}} $ is bounded from below by the
strictly positive number $C$.
\medskip
\noindent{\hbox{ $\diamond$ (II) {\it Integral representation of} $(1-\Tay \
)$}}

 From now on, we shall work inside the whole diagram $\Tt_\sJ$ for fixed
$\SJ$ and treat in parallel its distinct connected components $\CTt_{\sJ,j}$
for varying $j$. This is achieved by introducing the block diagonal
matrix:
\eqn\bigmatY{\Avril^\sJ=\mattY}
with $j^{\rm max}={\rm Card} (T^0_\sJ)$. This matrix is such that
(for $\SJ\le\ST$):
\eqn\Ampdet{I_{\Tt_\sJ}=\Big( \det(\Avril^\sJ) \Big)^{-{d\over 2}} .}
Now we must consider the action of $(1-\Tay^{\ 0}_{\Tt^\sI_{\sJ-1\oplus}})$ on
$I_{\Tt_\sJ}$. For our particular choice of tree variables, the action of
$\Tay^{\ }0_{\Tt^\sI_{\sJ-1\oplus}}$ simply corresponds to set
$\beta^\sI=0$ in the matrix $\Avril^\sJ$. From the Propositions 2 and 3
of the preceding subsection, we know that an element
$\Avril^{{\bf T}_{\sJ,j}}_{\sK\sL}$ of the matrix $\Avril^\sJ$
either is independent of $\beta^\sI$ (if
$\SI\notin \CJ_{\sJ,j}(\SK,\SL)$ for this value of $j$), or
vanishes with $\beta^\sI$ at least as $(\beta^\sI)^\delta $ (if
$\SI\in \CJ_{\sJ,j}(\SK,\SL)$ ). Therefore the action of
$\Tay^{\ 0}_{\Tt^\sI_{\sJ-1\oplus}}$ simply corresponds to set to zero those
elements of $\Avril^\sJ$ which depend on $\beta^\sI$, leaving the other
elements
unchanged. In particular, this action is non trivial ({\it i.e.} non reduced
to the identity) when
\eqn\nontriv{\SI\in\CJ_\sJ
\equiv\bigcup\limits_{j,\sK,\sL}\CJ_{\sJ,j}(\SK,\SL).}
Conversely, if $\SI\notin\CJ_\sJ$, then $(1-\Tay^{0}_{\Tt^\sI_{\sJ-1\oplus}})
[I_{\Tt_{\sJ}}]=0$.
To perform the action of $\Tay^{0}_{\Tt^\sI_{\sJ-1\oplus}}$, it is convenient
to introduce an extra variable $t^\sI$ which multiplies
the elements $\Avril^{{\bf T}_{\sJ,j}}_{\sK\sL}$ such that
$\SI\in \CJ_{\sJ,j}(\SK,\SL)$.
We thus define
\eqn\Yoft{\Avril^{{\bf T}_{\sJ,j}}_{\sK\sL} (\{t\})\equiv
\Big( \prod_{\sI\in \CJ_{\sJ,j}(\sK,\sL)}t^\sI\Big) \ \Avril^{{\bf
T}_{\sJ,j}}_{\sK\sL}
\ ,}
and obtain a matrix $\Avril^\sJ(\{t\})$ which is a function of the
$t^\sI$'s for $\SI\in\CJ_\sJ$.
The action of $\Tay^{\ 0}_{\Tt^\sI_{\sJ-1\oplus}}$ then corresponds to set
$t^\sI=0$
(and set the other $t^\sI$'s equal to $1$).
We then have the following integral representation of a
$(1-\Tay_{\Tt^\sI_{\sJ-1\oplus}})$ operator with $\SI\in \CJ_\sJ$:
\eqn\Integraltone{
(1-\Tay^{\ 0}_{\Tt^\sI_{\sJ-1\oplus}})\Big[\prod_{j=1}^{{\rm Card}(T^0_\sJ)}
\Big(\det( \Avril^{{\bf T}_{\sJ,j}})\Big)^{-{d\over 2}}
\Big]
=\int^1_0\, dt^\sI\,{\partial\over\partial t^\sI}
\,\Big[ \det \big( \Avril^\sJ(\{t\}) \big)\Big]^{-{d\over 2}}
\ .}
Now we must apply a product of such Taylor operators for all the
$\SI\in{\rm Ind}(\SJ)$.
We can use the fact that $(1-\Tay \ )$ is a projector, and can thus be applied
several times to the same diagram. Since all the reduced diagrams
$\Tt^\sI_{\sJ-1}$ for
\eqn\aded{\SI\in {\overline{\CJ_\sJ}}\equiv \{\SI:\, \SI^{\rm min}(\SJ)\le\SI <
\SI^{\rm max}(\SJ)\}}
are equal to some $\Tt^\sI_{\sJ-1}$ for $\SI\in{\rm Ind}(\SJ)$, we have
\eqn\overTay{
\prod_{\sI\in {\rm Ind}(\sJ)}
(1-\Tay^{\ 0}_{\Tt^\sI_{\sJ-1\oplus}})
\,=\,
\prod_{\sI\in {\overline{\CJ_\sJ}}}
(1-\Tay^{\ 0}_{\Tt^\sI_{\sJ-1\oplus}})\ .}

 From their definitions and \fJjtJ\ and \iminmaxJ ,
it is clear that $\CJ_\sJ\subset
{\overline{\CJ_\sJ}}$. If $\CJ_\sJ\varsubsetneq{\overline{\CJ_\sJ}}$, then
the above product of $(1-\Tay \ )$, when acting on $I_{\Tt_\sJ}$, gives $0$,
as a consequence of the discussion above. \GrandO\ is then obviously
satisfied. We can therefore assume that $\CJ_\sJ={\overline{\CJ_\sJ}}$.
We then write the l.h.s of \GrandO\ as:
\eqn\Integralt{
\prod_{\sI\in {\rm Ind}(\sJ)}
(1-\Tay^{\ 0}_{\Tt^\sI_{\sJ-1\oplus}})\Big[\prod_{j=1}^{{\rm Card}(T^0_\sJ)}
\Big(\det( \Avril^{{\bf T}_{\sJ,j}})\Big)^{-{d\over 2}}
\Big]
=\int^1_0\,\prod_{\sI\in\CJ_\sJ}\,dt^\sI\,{\partial\over\partial t^\sI}
\,\Big[ \det \big( \Avril^\sJ(\{t\}) \big)\Big]^{-{d\over 2}}
\ ,}
with $\CJ_\sJ={\overline{\CJ_\sJ}}=\{\SI :\SI^{\rm min}(\SJ)\le\SI<
\SI^{\rm max}(\SJ)\}$.
\medskip
\noindent{\hbox{ $\diamond$ (III) {\it Proof of estimates
\GrandO\ and \GrandOT .}}}

First we use the fact that the property 4 of matrix
$\Avril^{{\bf T}_{\sJ,j}}$ extends to the matrix
$\Avril^{\sJ} (\{t\})$.
Specifically we have:
\smallskip\noindent
$\bullet$ {\it Prop. 5:} The matrix $\Avril^\sJ(\{t\})$ is positive and
bounded from below ({\it i.e.} $\Avril^\sJ(\{t\})-C\II>0$ for some positive
$C$) for all
$0\le t^\sI \le 1$, $\SI\in\CJ_\sJ$.
\medskip
Indeed, this property holds when each $t^\sI$ equals $0$ or $1$.
In this case, each block
$\Avril^{{\bf T}_{\sJ,j}}$ of the matrix is ``factorized" into a product of
sub-blocks $\Avril^{\bf T}$ for subtrees {\bf T}'s (see Appendix C) compatible
with the sector.
Each of these sub-matrices $\Avril^{\bf T}$ then satisfies Proposition 4,
as well as the matrix $\Avril^\sJ$.
To complete the proof of Proposition 5, we use the fact that the matrix
$\Avril^\sJ(\{t\})$ is a linear function of each $t^\sI$, and that it is
thus sufficient to have a lower bound at each corner of the hypercube
$0\le t^\sI\le 1$ ($\SI\in \CJ_\sJ$)
to have this bound inside the whole hypercube.
\medskip
A direct consequence of Proposition 5 is that $\Avril^\sJ(\{t\})$ is
invertible, and that $\big(\Avril^\sJ(\{t\})\big)^{-1}$ is positive
and bounded from above uniformly in the sector. In particular, the module
of all the
elements $\big(\Avril^\sJ(\{t\})\big)^{-1}_{\sK\sL}$ is also bounded
from above.
\par\noindent Another consequence of Proposition 5 is that
$\det\big(\Avril^\sJ(\{t\})\big)$ is uniformly bounded from below by a strictly
positive number.

Finally, if $\CE$ is some subset of $\CJ_\sJ$, it is clear from
Proposition 3 and the definition \Yoft\ of $\Avril^\sJ(\{t\})$ that, in the
considered Hepp sector, the partial set-derivative
\eqn\bopdoy{
\partial_\CE \Avril^\sJ_{\sK\sL}\equiv
\Big(\prod_{\sI\in\CE}{\partial\over\partial t^\sI}\Big)\,
\Avril^\sJ_{\sK\sL}
(\{t\})=\CO\Big(\prod_{\sI\in\CE}(\beta^\sI)^\delta\Big)}
for the $t^\sI$'s between $0$ and $1$.

To prove \GrandO , we now perform explicitly the derivatives with respect
to the $t^\sI$'s in the r.h.s. of \Integralt. This leads to an integral over
the
$t^\sI$ of a finite sum of terms of the form
\eqn\trucmuche{
{\rm Tr}\left(\partial_{\CE^1_1}\Avril\cdot \Avril^{-1}\ldots
\partial_{\CE^1_{k_1}}\Avril\cdot\Avril^{-1}\right)\ldots
{\rm Tr}\left(\partial_{\CE^n_1}\Avril\cdot \Avril^{-1}\ldots
\partial_{\CE^n_{k_n}}\Avril\cdot\Avril^{-1}\right)
\Big(\det(\Avril)\Big)^{-{d\over 2}}
\ ,}
made of a product of an arbitrary number $n$ of traces (the $i$th trace
involving a product of $k_i$ set-derivatives) and where the set of all
$\CE^l_m$'s form a partition of $\CJ_\sJ$
(here, $\Avril$ stands for $\Avril^\sJ(\{t\})$).
 From the estimates \bopdoy , from the upper bound on
$\big(\Avril^\sJ(\{t\})\big)^{-1}$ and
from the lower bound on $\det\big(\Avril^\sJ(\{t\})\big)$, we deduce that
in the Hepp sector:
\eqn\GrandOH{
\prod_{\sI\in {\rm Ind}(\sJ)}
(1-\Tay^{\ 0}_{\Tt^\sI_{\sJ-1\oplus}})\Big[\prod_{j=1}^{{\rm Card}(T^0_\sJ)}
\Big(\det( \Avril^{{\bf T}_{\sJ,j}})\Big)^{-{d\over 2}}
\Big]
={\cal O}\Big(\prod_{\sI\in \CJ_\sJ}
(\beta^\sI)^\delta \Big)
\ ,}
which is just the announced estimate \GrandO .
\medskip
It is not very difficult to extend the above analysis to the case of the
largest
diagram $\Tt_{\sT+1}$ which contains the external points.
Indeed, the action of the $\Tay^{\ 0}$'s on the extra
term $\displaystyle \exp(-{1\over 2}\sum_{a,b}\kvec_a\cdot\kvec_b\Delta_{ab})$
can also be implemented
through the $t$ variables, and one can check that the quadratic form
$\Delta_{ab}(\{t\})$ is still definite positive. This ensures that the estimate
\GrandOT\ is valid, Q.E.D.
\vfill\eject
\listrefs
\listfigs   %(if necessary)
\bye